\documentclass{article}

\usepackage[usenames]{color}
\usepackage[table]{xcolor}
\usepackage{soul}
\usepackage{textcomp}
\usepackage{multicol}  
\usepackage{multirow} 
\usepackage{xspace}
\usepackage{amsmath} 
\usepackage{amssymb}
\usepackage{graphicx}
\usepackage{breqn}
 \usepackage{hyperref}
\usepackage{algorithm}
\usepackage{algorithmic}
 \usepackage{cuted}
 \usepackage{subfigure}
 \usepackage{multirow}
 \usepackage{multicol}
 \usepackage{rotating}
 \usepackage{cancel}
 \usepackage[outdir=./]{epstopdf}
 
\usepackage{amsthm}
\theoremstyle{definition} 
\newtheorem{definition}{Definition}[section]
\newtheorem{proposition}{Proposition} 
\newtheorem{example}{Example}[section]

\newcommand*{\QEDB}{\hfill\ensuremath{\square}}%

\def\V{\mbox{$\mathcal{V}$}} % 

\newcommand{\mpara}[1]{\medskip\noindent{\bf #1}}

\def\fftwyt{FF-TW-YT}
\def\ghsotw{GH-SO-TW}
\def\greedy{M-EMCD\xspace}
\def\greedystar{M-EMCD$^*$\xspace}

 \usepackage{authblk}

\begin{document}
\title{Modularity in Multilayer Networks using \\Redundancy-based Resolution and\\ Projection-based Inter-Layer Coupling\thanks{Accepted for publication with \textit{IEEE Trans. on Network Science and Engineering}, April 2019. DOI: \texttt{10.1109/TNSE.2019.2913325} }\thanks{An abridged version of this work appeared in \cite{AmelioT17}.}}

\author[1]{Alessia~Amelio}
\author[2]{Giuseppe~Mangioni}
\author[1]{Andrea~Tagarelli\thanks{\em Corresponding author: Andrea Tagarelli.}} 

\affil[1]{\footnotesize Dept. of Computer Engineering, Modeling, Electronics and Systems Engineering (DIMES), University of Calabria, Rende (CS), Italy\\ Email: \{aamelio, tagarelli\}@dimes.unical.it.}

\affil[2]{\footnotesize Dept. of Electrical, Electronics and Computer Engineering, University of Catania, Catania, Italy\\ Email: giuseppe.mangioni@dieei.unict.it.}  

\date{}
  \maketitle

\begin{abstract}
The generalized version of modularity for multilayer networks, a.k.a.  multislice modularity, is characterized by two model parameters, namely   resolution factor and  inter-layer coupling factor. The former corresponds to  a notion of layer-specific relevance, whereas the inter-layer coupling factor represents the strength of node connections across the network layers. Despite the potential of this approach, the setting of both parameters   can be arbitrarily selected, without considering specific characteristics from the topology of the multilayer network as well as from an available community  structure. Also, the multislice modularity is not designed to explicitly model order relations over the layers, which is of prior importance for dynamic networks. 

This paper aims to overcome the main limitations of the multislice modularity by introducing a new formulation of modularity for multilayer networks. We revise the role and semantics of both the resolution and inter-layer coupling factors based on information available from the within-layer and inter-layer structures of the multilayer communities. Also, our proposed multilayer modularity is general enough to consider orderings of network layers and their constraints on layer coupling.  Experiments  were carried out on synthetic and  real-world multilayer networks using   state-of-the-art approaches for multilayer community detection. The obtained results have shown the meaningfulness of the proposed modularity, revealing the effects of different combinations of the resolution and inter-layer coupling functions. This work also represents a starting point for the development of new optimization methods for community detection in multilayer networks. 
\end{abstract}

%\begin{IEEEkeywords}
%modularity, multilayer networks, community detection
%\end{IEEEkeywords}

 % \thispagestyle{empty} 

 \section{Introduction}\label{s:introduction} 
Complex network systems, such as social networks, biological networks, and transportation networks, are inherently organized into \textit{communities}, a.k.a. clusters or modules,  with dense internal links and sparse external links.  
  Since members of a community tend to generally share common
properties, revealing the community structure in a network can provide a better understanding of the overall functioning of the network.

 The well-known    \textit{modularity}~\cite{Newman04,Newman04b} function was originally conceived to evaluate  a community structure in a network graph in terms of difference  between the  actual   number of edges linking nodes inside each community and the expected connectivity in the null model. Typically, the expected connectivity is expressed through a configuration graph model, having a certain degree distribution and randomized edges. Since this graph ignores any community structure, a large difference between actual connectivity and expected connectivity would indicate   the presence of a  community structure.   

Modularity has been widely utilized as objective function in several optimization methods designed for discovering communities  in networks~\cite{BrandesDGGHNW08,ChenKS14}, including greedy agglomeration~\cite{Newman04b,Clauset04},  spectral division~\cite{Newman05,WhiteS05}, simulated annealing~\cite{Reichardt06}, %~\cite{Guimera04,Reichardt06}
  or extremal optimization~\cite{Duch05}.

Traditional  approach to network analysis refers to the modeling of a real-world  system  as a single network of interacting entities. While this approach has been 
 %successfully 
widely used to study a variety of applications, % cases, 
%the world is full of examples 
 there are plenty of scenarios for which this methodology appears strongly limiting~\cite{BOCCALETTI20141}. 
In general, ties among entities could be induced by one or several types of  relations or interactions, or even be dependent on side-information 
based on specific dimensions or aspects of interest for the entities in the network. Within this view, \textit{multilayer networks}
~\cite{Magnanibook,Kivela+14} represent a powerful tool
to model systems interconnected by %several categories of connections: 
 multiple types of relations. In the multilayer network model, each type of connection is represented by a layer, and an entity may be present in 
different layers based on the type of relations it is connected to its neighbor  entities. 
 Just to mention one real example, nowadays online users usually have multiple accounts across different online social networks, and several online/offline relationships are likely to occur among the same group of individuals (e.g., following relations, like/comment interactions, working together, having lunch).
 It should be emphasized that  neglecting such a kind of complex 
organization by reducing the whole system to a single network (e.g., through some kind of projection,  or by aggregation), has been shown to be much less 
informative than the multilayer representation~\cite{e20120909}.  
For the above mentioned reasons, multilayer networks are experiencing an increasing interest from the scientific community, leading to an explosion of 
scientific papers in many areas of science, thus becoming one of the most used tools for interdisciplinary research \cite{Mucha10}, \cite{coscia2011}, \cite{BOCCALETTI20141}, \cite{Kivela+14}, \cite{KunchevaM15}, \cite{ZhangWLY16}, \cite{Magnanibook}, \cite{Tagar17}, \cite{Tagar18,torreggiani2018identifying}, \cite{mangioni2018multilayer}, \cite{alves2019nested}.
 
Clearly, the problem of community detection in such multilayer networks takes a central role to unveil meaningful patterns of node groupings into communities,  by leveraging the various interaction modes that involve all the entity nodes in the network.   
To address these needs, modularity has been  extended to the   general case of  multilayer networks. In particular, Mucha et al.~\cite{Mucha10} extend the modularity function to   arbitrary multilayer networks (also called \textit{multislice} in that work), by introducing two additional parameters w.r.t. classic modularity: a \textit{resolution} parameter  and an \textit{inter-layer coupling} factor. 
 The resolution parameter acts on the expected connectivity terms, thus controlling the effect on the size distribution of community due to the \textit{resolution limit} known in modularity~\cite{FortunatoB07}. 
 The inter-layer coupling factor focuses on the links across layers and hence impacts on the strength of the inter-layer connections of entities in the network. 
 While being important to enhance the ability of modularity in evaluating a community structure, the two parameters introduced in the multislice modularity are nonetheless subjected to   arbitrary choices, which raise a number of issues in the application of this modularity function.  In particular,  the resolution parameter can be  arbitrarily   set for each layer, but it discards any structure information at graph or community level. Moreover, the inter-layer coupling terms do not differentiate among the selected layers, and  all pairs of layers can in principle be considered, which  makes no sense in certain scenarios such as modeling of time-evolving networks.  
 
 The above considerations prompted us to revisit the notion of modularity in multilayer networks, and in particular to introduce  novel aspects to take into account in both the resolution and inter-layer coupling definitions.  First, the layer-specific resolution factor is also made dependent on  each particular community. We notice that, since a high-quality multilayer community should embed high information content among its nodes,  the resolution of a specific layer to control the expected connectivity of a given community in the modularity function should be lower as the contribution of that layer to the information content of the community is higher. 
 By relating the information content of a multilayer community to the amount and  variety of types of links   internal to the community, we  provide a new definition of resolution factor based on the concept of \textit{redundancy} of community. 
 Second, to determine the strength of coupling of nodes across layers, 
 we again  consider it at community level, such that for each pair of layers, the inter-layer coupling factor for nodes in a community depends on the relevance of the community projection on the two layers.    
 Moreover, we account for an available \textit{ordering of the layers}, and relating constraints on their coupling validity.

 Our main contributions are summarized as follows:
\begin{itemize}
\item We propose a novel definition of multilayer modularity, in which we reconsider  the role and semantics of its two key terms, that is, the  resolution factor and inter-layer coupling factor. We conceive parameter-free unsupervised approaches for their computation, which leverage information from the within-layer and across-layer structures of the   communities in the multilayer network. 
Moreover, our formulation of multilayer modularity is general enough to account for an available ordering of the layers, therefore is also well-suited to deal with temporal multilayer networks. 
\item 
We provide  theoretical insights into properties of the proposed multilayer modularity. More specifically, we investigate the effect of increasing the number of communities in the behavior of the multilayer modularity, and we analytically derive the lower and upper bounds in the values of the multilayer modularity. 
\item
We conduct an extensive experimental evaluation, primarily  to understand   how the proposed multilayer modularity behaves w.r.t. different settings regarding the resolution   and the inter-layer coupling terms.  Using 4  state-of-the-art methods for multilayer community detection (GL, PMM, LART,   and \greedystar), LFR synthetic multilayer networks and 10 real-world multilayer networks, 
results have shown the significance of our formulation and the different  expressiveness against the previously existing multislice    modularity.  

\end{itemize}

  %==========================================
  
\section{Related Work}   
%To uncover the community structure of a given network, classic algorithms for graph clustering cannot be used. In fact, such algorithms are focused on optimal subdivisions of graphs respect to min--flow cuts. To find communities we need a deeper analysis of link patterns and relations. For this reason, 
Community detection is a key-enabling task in network analysis and mining, with tons of methods developed in the last  ten years --- please refer to~\cite{FORTUNATO201075,2010Tang,coscia2011,FORTUNATO20161} for surveys on this topic.  
In addition, % to the development of new algorithms, 
   different metrics for community structure evaluation have been introduced. 
As discussed in Section~\ref{s:introduction}, the most popular and widely accepted  measure is the so-called ``modularity'', defined by Newman \cite{Newman04}.  Initially conceived for undirected networks, the modularity function has been subsequently extended to cover different cases. 
In~\cite{leicht2008}\cite{arenas2007}, modularity has been generalized to directed networks incorporating information contained in edge directions, while in~\cite{newman2004}   modularity is also adapted to capture communities in weighted networks.  
To overcome the well-known modularity resolution limit~\cite{FortunatoB07,XIANG20124995,Zhang2009}, 
in \cite{arenas2007-2} and \cite{XIANG2015127} modularity has been modified by incorporating a resolution parameter that helps reveal communities at different resolution scale.  
 A further step towards a generalization of the modularity refers to its extension to  signed networks~\cite{gomez2009,traag2009}.  %, i.e., graphs in which each edge has a positive or negative sign, such as in correlation networks. 
 %Bipartite networks (or two-modes networks) are a class of graphs whose nodes are divided into two sets, and only connections between two nodes in different sets are allowed. 
  Also, to deal with bipartite networks,  modifications have been proposed in~\cite{barber2007,barber2008,guimera2007}. 
 %There are several cases where nodes can belong to more communities or, in other words, communities overlap. 
To uncover overlapping communities, in~\cite{nicosia2009} the authors propose an extension to the modularity function that includes the notion of belonging (or membership) coefficient, which measures to which extent a node belongs to the various  communities. This approach is sometimes referred to   as fuzzy community discovering.   
  Finally, as introduced in Section~\ref{s:introduction}, modularity has been generalized by~\cite{Mucha10} to capture communities in multislice  networks. 
  Such a version of the modularity function is detailed in the next section.

\section{Background} 
\label{sec:background}

\subsection{Modularity}

Given an undirected graph $G=(\V,\mathcal{E})$, with $n=|\V|$ nodes and $m=|\mathcal{E}|$ edges,  let $\mathcal{C}$ be a community structure over $G$. For any  $v \in \V$,  we use   $d(v)$ to denote the degree of $v$, and for any community $C \in \mathcal{C}$, symbol $d(C)$ to denote the degree of   $C$, i.e.,  $d(C)=\sum_{v \in C} d(v)$; also,  the  total degree of nodes over the entire graph,     $d(\V)$, is defined as $d(\V)=\sum_{v \in \mathcal{V}} d(v)=\sum_{C \in \mathcal{C}} d(C)=2m$.  
Moreover, we denote with $d_{int}(C)$  the internal degree of $C$, i.e., the portion of $d(C)$ that corresponds to the number of edges linking nodes in $C$ to other nodes in $C$. % (i.e., twice the number of links internal to $C$). 
Newman and Girvan's \textit{modularity} is defined as follows~\cite{Newman04}:
\begin{equation}\label{eq:NGmodularity}
Q_{NG}(\mathcal{C})=  \sum_{C \in \mathcal{C}} \frac{d_{int}(C)}{d(\V)} - \left( \frac{d(C)}{d(\V)} \right)^2  
\end{equation}
 
 In the above equation, the first term is maximized when many edges are  contained in clusters, whereas the second term is minimized   by partitioning  the graph into many clusters with small total degrees. 
The value of $Q_{NG}$ ranges within  -0.5 and 1.0~\cite{BrandesDGGHNW08};  
it is  minimum for any bipartite network with canonic clustering, and  maximum when the network is composed by disjoint cliques.

\subsection{Multilayer network model}
Let $G_{\mathcal{L}} = (V_{\mathcal{L}}, E_{\mathcal{L}}, \V, \mathcal{L})$ be a \textit{multilayer network} graph, where $\V$ is a set of entities  and  $\mathcal{L}= \{L_1, \ldots, L_{\ell}\}$ is a set of layers. Each layer represents a specific type of relation between entity nodes. %, or edge-label. 
%We generalize the definition of multilayer network introduced in~\cite{Kivela+14} by also introducing a partial order relation $\prec_{\mathcal{L}}$ over the layers in order to model scenarios in which a particular ordering among layers is required; however, unless otherwise specified, we assume that $\prec_{\mathcal{L}}$ is an empty function by default (i.e., $\mathcal{L}$ is an unordered set). 
% 
Let $V_{\mathcal{L}} \subseteq \V \times \mathcal{L}$ be the set containing the entity-layer combinations, i.e., the occurrences of each   entity  in the   layers.   $E_{\mathcal{L}} \subseteq V_{\mathcal{L}} \times V_{\mathcal{L}}$ is the set of undirected links connecting the entity-layer elements. 
 For every  $L_i \in \mathcal{L}$, we define $V_i = \{v \in \V \ | \    (v, L_i) \in V_{\mathcal{L}}\} \subseteq \mathcal{V}$ as the set of nodes in the graph of $L_i$, and $E_i \subseteq V_i \times V_i$ as the set of edges in $L_i$. Each entity must be present in at least one layer, i.e., $\bigcup_{i = 1..\ell} V_i  = \V$, but each  layer is not required to contain all  elements of $\V$. 
We assume that the inter-layer links only connect the same entity in different layers, however  each entity in one layer could be  linked to the same entity in a few or all other layers.

\subsection{Multislice Modularity}
\label{sec:Qms}
Given a community structure   $\mathcal{C}$ identified over a multilayer network $G_{\mathcal{L}}$, the \textit{multislice modularity}~\cite{Mucha10} of   $\mathcal{C}$ is defined as:
\begin{eqnarray}\label{muchaMod}
Q_{\textrm{ms}}(\mathcal{C}) & = & \frac{1}{d(V_{\mathcal{L}})}  \sum_{\substack{u,v, \\ L_i,L_j}}   \left[  \left( A_{uvL_i} - \gamma_i \frac{d_{L_i}(u)d_{L_i}(v)}{2|E_i|} \right)  \times \right. \nonumber \\
  & \times &  \left. \delta_{L_i,L_j} +  \delta_{u,v}\mathrm{C}_{v,L_i,L_j} \right] \delta(g_{u,L_i}, g_{v,L_j})
\end{eqnarray}
where 
$d(V_{\mathcal{L}})$ is the total degree of the multilayer network graph, 
$d_{L_i}(u)$ denotes the degree of node $u$ in layer $L_i$, 
$A_{uvL_i}$ denotes a link between $u$ and $v$ in $L_i$,  
%$\mu$ is a normalization factor, 
   $2|E_i|$  is the total degree of the graph of layer $L_i$, $\gamma_i$ is the resolution parameter for layer $L_i$,   $\mathrm{C}_{v,L_i,L_j}$ quantifies the links of node $v$ across  layers $L_i$,  $L_j$.  
Moreover, the Dirichlet terms have the following meanings: 
$\delta_{L_i,L_j}$ is equal to 1 if $L_i=L_j$ and 0 otherwise, 
% (i.e., the classic modularity is evaluated on each layer graph), 
$\delta_{u,v}$ is equal to 1 if $u=v$ and 0 otherwise (i.e., the inter-layer coupling is allowed only for nodes corresponding to the same entity), 
and 
$\delta(g_{u,L_i}, g_{v,L_j})$ is equal to 1 if the community assignments of node $u$ in $L_i$ and node $v$ in $L_j$ are the same and 0 otherwise.

\mpara{Limitations of $Q_{\textrm{ms}}$.\ \ }
As previously  mentioned, a different   resolution parameter   $\gamma_i$ can be associated with each layer to express its relevance weight; however,  in~\cite{Mucha10}, there is no  specification of  any principled way to set a layer-weighting scheme, possibly including information from the available multilayer community structure.  
Moreover, neither the inter-layer coupling term  (i.e., $C_{v,L_i,L_j}$) or any constraint on the layer comparability are clearly  defined; actually,  all nonzero inter-layer edges are set to a constant value $\omega$, for all unordered pairs of layers.    
In general, both $\gamma_i$  and $\omega$ parameters can   assume any non-negative value, which further increases a clarity issue in the     properties of $Q_{\textrm{ms}}$.

%==========================================
 
\section{Proposed Multilayer Modularity}
In this section, we propose a new definition  of   modularity for multilayer networks that aims to overcome all of the issues of $Q_{\textrm{ms}}$ previously discussed. We pursue this goal by focusing on the role and semantics of the two key elements in multilayer modularity: the \textit{layer-specific resolution}   and the \textit{inter-layer coupling}.  

\begin{figure}[t!]
\centering  
\includegraphics[scale=0.42]{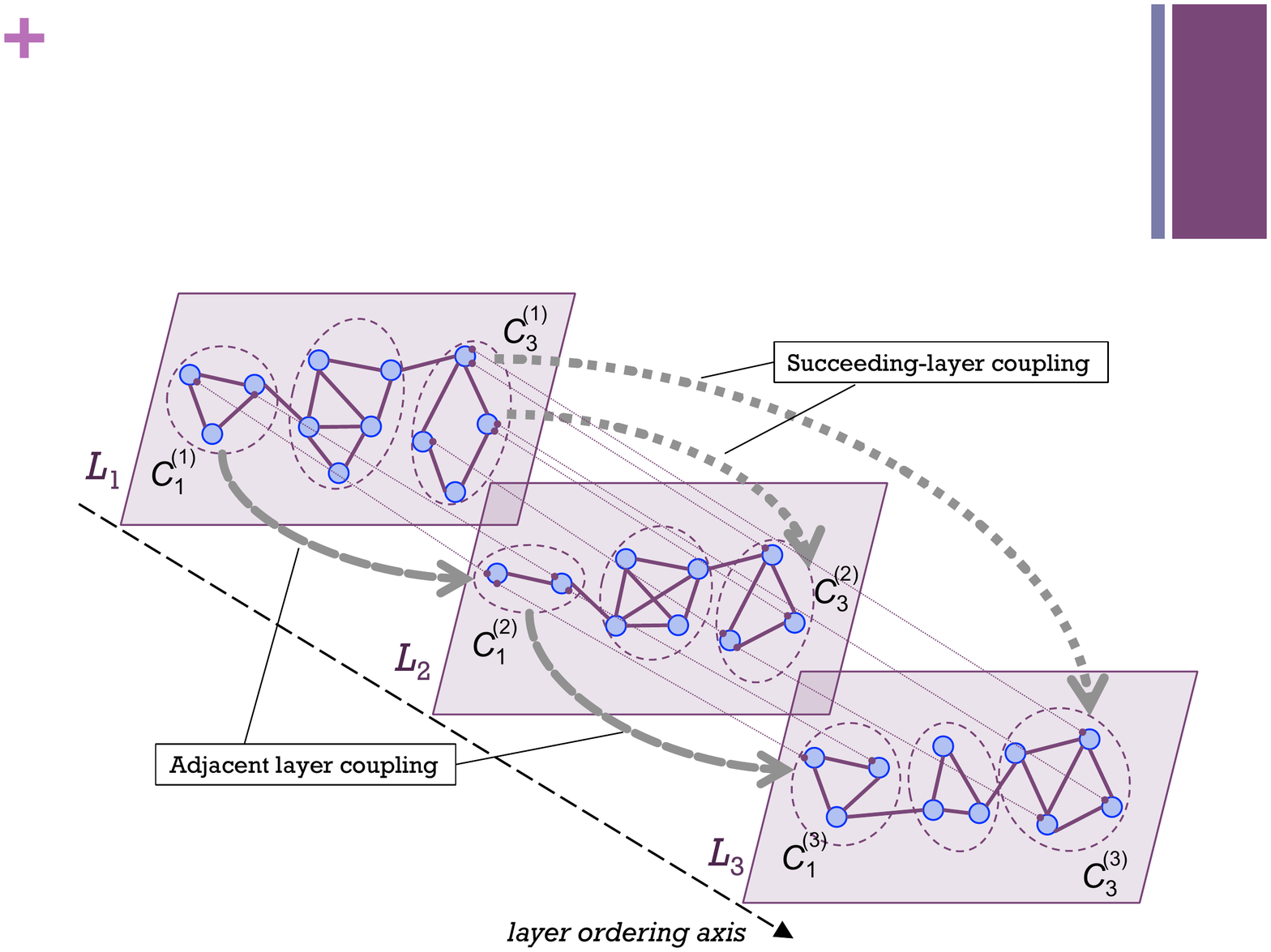}
\caption{Example multilayer network. The ordering over the set of layers enables the introduction of constraints to compare layers; for instance,   adjacent-layer coupling forces the comparison of adjacent layers only, while succeeding-layer coupling allows each layer to be compared with the subsequent ones.   
Note also that any community across the layers is visually represented by means of each projection from one layer to another valid layer; for instance,   
using succeeding-layer coupling, community $C_3$ in layer $L_1$ is projected onto layer $L_2$ and layer $L_3$.}
\label{fig:esempio}
\end{figure}

Our definitions of the two terms are independent on a-priori assumptions on the network and/or user-specified settings; by contrast, we  conceive  parameter-free unsupervised approaches for their computation, by leveraging information from the within-layer and inter-layer structures of the communities.  
 Our proposed   resolution factor is computed for pairs of layer and community, rather than  for each layer globally. Analogously, to define the inter-layer coupling term,  we   account for properties related to a community 
 on two layers; more in detail, 
 we evaluate the projections of a community  over any two comparable layers, i.e., the sets of nodes belonging to a community that lay on those layers.  
Remarkably, the comparability of layers is another key aspect of our definition of modularity:  we generalize   the inter-layer coupling term by admitting the existence of a \textit{partial order relation} $\prec_{\mathcal{L}}$ over the layers, in order  to properly represent  scenarios in which a particular ordering among layers is required. For instance, it may be the case that the network layers 
 have to be processed according to their natural order (e.g., lexicographic order of the network labels), or according to a temporal order; moreover, it may be required to compare   adjacent layers only, or each layer with any other  succeeding it in the ordering.  Figure~\ref{fig:esempio} provides an illustration of a multilayer network and the aforementioned key aspects we deal with in our proposed multilayer modularity, which is formally presented next.

 \vspace{2mm}
\begin{definition}[Multilayer Modularity]
\em
Let $G_{\mathcal{L}} = (V_{\mathcal{L}}, E_{\mathcal{L}}, \V, \mathcal{L})$ be  a multilayer network graph, and let  $\prec_{\mathcal{L}}$ be  an optionally provided  partial order relation over the set of layers $\mathcal{L}$. 
Given a community structure $\mathcal{C}=\{C_1, \ldots, C_k\}$ as a partitioning of the multilayer graph $G_{\mathcal{L}}$, the   \emph{multilayer modularity}  is defined as: 
\begin{equation}\label{eq:MLmodularity}
Q(\mathcal{C})      =   \frac{1}{d(V_{\mathcal{L}})}\sum_{C \in \mathcal{C}} \left[ \sum_{L \in \mathcal{L}}  \left(  d_L^{int}(C)  - \gamma(L,C) \frac{(d_L(C))^2}{d(V_{\mathcal{L}})}   +    \beta\!\!\!\sum_{L' \in \mathcal{P}(L)}  IC(C, L, L') \right) \right] 
\end{equation} 
\noindent 
where for any $C \in \mathcal{C}$ and $L \in \mathcal{L}$:
\begin{itemize}
\item   $d_L(C)$ and $d_L^{int}(C)$ are  the degree of  $C$ and the internal degree of  $C$, respectively, by considering only edges  of layer $L$; 
\item   $\gamma(L,C)$ is the value of the  resolution function;  
%$d^{ext}(V_\mathcal{L})$ is the total inter-layer degree of the multilayer network graph, 
 \item   $IC(C, L, L')$ is the value of the inter-layer coupling function for any valid layer pairings with $L$;  
 \item  $\beta \in \{0,1\}$ is a parameter to control the  exclusion /inclusion of inter-layer couplings; 
  and 
\item   $\mathcal{P}(L)$ is  the set of  valid pairings   with   $L$ defined as:
$$
\mathcal{P}(L)=
\begin{cases}
    \{L' \in \mathcal{L} \ | \ L \prec_{\mathcal{L}} L' \}, &  \text{if} \prec_{\mathcal{L}} \text{is defined}\\
    \mathcal{L} \setminus \{L\}, & \text{otherwise}
\end{cases}
$$ 
\end{itemize}
\vspace{-2mm} 
\QEDB
\end{definition}

\vspace{-2mm} 
%It should be noted that the within-layer connectivity term in    Eq.~(\ref{eq:MLmodularity}) has an analytic form that  recalls $Q_{NG}$ (Eq.~(\ref{eq:NGmodularity})) rather that of  $Q_{\textrm{ms}}$. We made this choice because of  more informativeness, in the sense  it puts better in evidence the trade-off between the intra-community connectivity to maximize and the inter-community connectivity to minimize.   

Notably, unlike  the multislice modularity in  Eq.~(\ref{muchaMod}), our proposed modularity originally introduces a resolution factor that varies with each community, and an inter-layer coupling scheme that might also depend on the layer ordering. Moreover,  Eq.~(\ref{eq:MLmodularity})  utilizes  the total degree $d(V_{\mathcal{L}})$ of the multilayer   graph  instead of the layer-specific degree (i.e., term  $2|E_{L_i}|$, for each $L_i \in \mathcal{L}$). 
 The latter difference w.r.t. the multislice modularity  is also important because, as we shall later discuss more in detail,   the total degree of the multilayer graph includes the inter-layer couplings and it might be defined in different ways depending on the scheme of inter-layer coupling.  
%Moreover, following the theoretical framework in~\cite{BrandesDGGHNW08}, we have checked that our proposed modularity varies within $[-0.5, 1]$, for different values of resolution and inter-layer coupling terms.  
 %
In the following, we elaborate on the resolution functional term, $\gamma(\cdot,\cdot)$, and the inter-layer coupling functional term,  $IC(\cdot, \cdot, \cdot)$. 
% In some cases, the network layers can follow a natural lexicographic order based on layer enumeration, or a chronological order where each layer corresponds to a time-step. In the first case, two inter-layer coupling schemes can be provided: the adjacent layer coupling scheme, considering the coupling of each layer with its adjacent, and the pairwise layer coupling scheme, considering the coupling of each layer with its subsequent ones.

\subsection{Redundancy-based resolution factor} 
The layer-specific resolution factor intuitively expresses the  relevance of a particular layer to the calculation of the expected community connectivity in that layer. 
While this   can always reflect some predetermined scheme of relevance weighting of layers, we propose a more general definition that  accounts for the strength of the contribution that   a layer takes  in determining the internal connectivity for each community.  
The key assumption underlying our approach is   that, since a high quality community should envelope   high information content among its elements, \textit{the resolution   of a layer to control the expected connectivity of    a given community  should be lowered  as the layer's  contribution to the information content of the community increases}. 

In this regard, the \textit{redundancy} measure proposed in~\cite{Berlingerio2011}  is particularly suited to  quantify  the variety of connections, such that it is higher if edges of more types (layers) connect each pair of nodes in the community.  
Let us  denote with $P_1$ the set of node pairs   connected in at least one layer in the graph, and with  $P_2$ the set of ``redundant'' pairs, i.e., the pairs of nodes connected in at least two layers. 
Given a community $C$, $P_1^C$ and $P_2^C$ denote the subset of $P_1$ and the subset of $P_2$, respectively, corresponding to nodes in   $C$. 
The redundancy of $C$, $\rho(C)$,  expresses  the number of pairs in $C$ with redundant connections, divided by  the  number of layers connecting the pairs. Formally~\cite{Berlingerio2011}:  
 %\disc{A che serve definire $\rho(C)$ dato che non lo usiamo???}
%theoretical maximum. Formally: 
%
\begin{equation}\label{eq:redundancy}
\rho(C) =   \sum_{(v,u) \in P_2^C} \frac{SL(v,u,\mathcal{L})}{|\mathcal{L}| \times |P_1^C|},   
%\frac{|\{L \in \mathcal{L} \ | \  (v,u,L) \in E_{\mathcal{L}}\}|}{|\mathcal{L}| \times |P_1^C|}  \nonumber 
\end{equation} 
with $SL(v,u,\mathcal{L})=\{L \in \mathcal{L} \ | \  (v,u,L) \in E_{\mathcal{L}}\}$.
%
%We introduce a modification in Eq. (\ref{red}) by substituting the theoretical maximum with the effective number of layers connecting the pairs.

Note that in the above formula, each of the sets $SL$ refers to the layers on which two nodes in a redundant pair are linked. Upon this concept,  we define the \textit{set of supporting layers} $sup$ for each community $C$ as:
\begin{equation}\label{eq:supportset}
sup(C,\mathcal{L})=\bigcup_{(v,u) \in P_2^C } SL(v,u,\mathcal{L}).
\end{equation} 
%
%with $SL(v,u,\mathcal{L})=\{L \in \mathcal{L} \ | \  (v,u,L) \in E_{\mathcal{L}}\}$. 
%\disc{Nell'insieme $SL$, gli $L$ sono solo quelli in cui sono presenti le redundant pairs di nodi in $C$. O questo concetto lo si ribadisce, o troviamo un modo formale per esprimerlo.}

Using the above defined $sup(C,\mathcal{L})$, we provide the following definition of \textit{redundancy-based resolution factor}.  
%express the number of times a particular layer $L$ participates in redundant pairs as:
%%
%$$
%nrp(L,C) = |\{ s=SL(v,u,\mathcal{L}) \in SL(C,\mathcal{L})  \ | \ L \in s \}|
%$$

% NOTE:  SL(C,\mathcal{L}) sosituisce p_2(\mathcal{L}, C),  mentre nrp(L,C) sostituisce   \rho(L,C) %

\vspace{2mm}
\begin{definition}[Redundancy-based resolution factor]
\em
Given a layer $L$ and a community $C$,   the  \emph{redundancy-based resolution factor} in Eq.~(\ref{eq:MLmodularity}) is defined as:
\begin{equation}\label{eq:redbasedgamma}
\gamma(L,C) = \frac{2}{1+ \log_2(1+ nrp(L,C))}
\end{equation} 
where 
$
nrp(L,C) = |\{ s\!\!=\!\!SL(v,u,\mathcal{L}) \in sup(C,\mathcal{L})  \ | \ L \in s \}|
$
expresses the number of times  layer $L$ participates in redundant pairs.
%
%\disc{Non possiamo riscrivere $nrp$ sensa usare $s$? Per esempio:}
%
%{\color{red}
%$
%nrp(L,C) = |\{ SL(v,u,\mathcal{L}) \in SL(C,\mathcal{L})  \ | \ L \in SL(v,u,\mathcal{L}) \}|
%$
%}
\QEDB
\end{definition}

\vspace{2mm}
Note that $\gamma(L,C)$ ranges in $(0, 1] \cup [2]$. In particular, it ranges   in $(0, 1]$  as long as $L$ participates in at least one redundant pair, and it decreases as $nrp(L,C)$ increases; moreover, as special case,  $\gamma(L,C)$ is equal to 2  when $nrp(L,C)~=~0$.

%%%%%%%%% RUNNING EXAMPLE GRAPH %%%%%%
\begin{figure}[t!]
\centering 
\hspace{-4.5mm}
\includegraphics[scale=0.35]{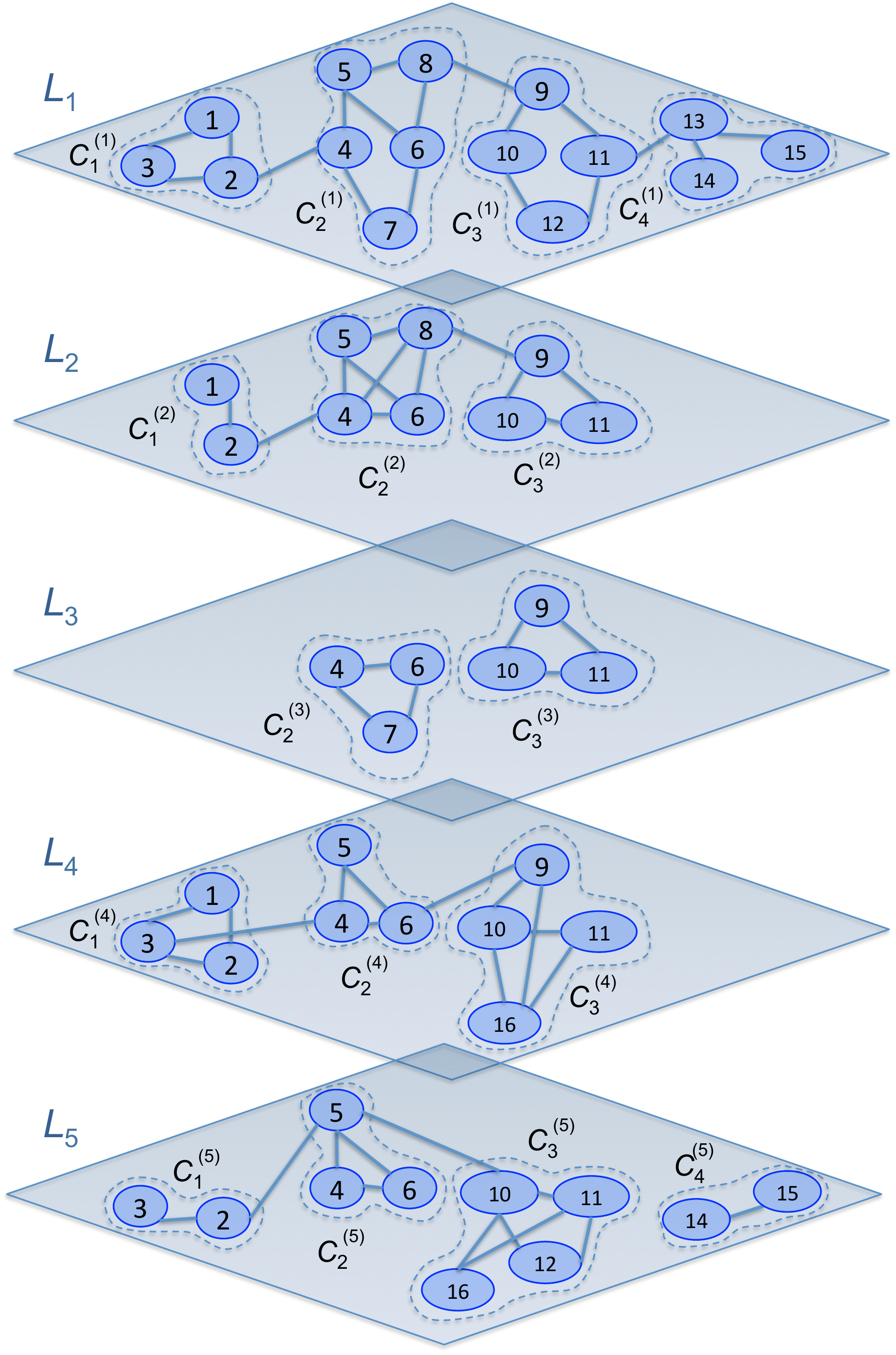}
\caption{Multilayer network for our running example}
\label{fig:example1}
\end{figure}
%%%%%%%%%%%%%%%%%%%%%%%%%%%%%%%%%%%%%%

%%%%%%%%% RUNNING EXAMPLE - step #1  (begin) %%%%%%
\begin{example}
\em 
Consider the network with  16 entities and 5 layers  shown in Fig.~\ref{fig:example1}, and let us first focus on some specific cases for the computation of the $\gamma(L,C)$ terms.    
  For instance, given community $C_4$ and  layers $L_1$ and $L_5$,  
  the corresponding values of redundancy-based resolution  are equal to 2, because $L_1, L_5$ never participate in redundant pairs for nodes of $C_4$. 
   By contrast,   layers $L_2$ and $L_5$ participate in one redundant pair for community $C_1$, which corresponds to the edge linking entities 2 and 3 for $L_5$ and the edge linking entities 1 and 2 for $L_2$; therefore,  the   values of redundancy-based resolution associated with $C_1$ on $L_2$ and $L_5$ are equal to 1. Also, the resolution for $C_1$ on $L_1$ takes a value lower than 1 since there  is  more than one redundant pair. 
   Table~\ref{tab:exampleGamma} reports on the entire set of values for the resolution factor computed on the network of Fig.~\ref{fig:example1}. 
\end{example}

\begin{table}[t!]
\caption{Redundancy-based resolution factor $\gamma(L,C)$ for each community and layer of the example network in Fig.~\ref{fig:example1}.}
\label{tab:exampleGamma}
\centering
\scalebox{0.9}{
\begin{tabular}{|c|c|c|c|c|c|}
\hline
&$L_1$&$L_2$&$L_3$&$L_4$&$L_5$\\\hline
$C_1$&0.667& 1.000& -& 0.667& 1.000 \\\hline
$C_2$&0.525& 0.558& 0.667& 0.667& 0.667 \\\hline
$C_3$& 0.602& 0.667& 0.667& 0.602& 0.558  \\\hline
$C_4$& 2.000&-&-&-&2.000 \\\hline
    \end{tabular} 
    }
\end{table}

%%%%%%%%% RUNNING EXAMPLE - step #1  (end) %%%%%%

\subsection{Projection-based inter-layer couplings}
We propose a general and versatile approach  to quantify the strength of coupling of nodes in one layer with nodes on another layer. Our key idea is \textit{to determine the fraction of nodes belonging to a community onto a layer that  appears in the projection of the community on another layer, and express the relevance of this projection w.r.t. that pair of layers}.  

Given a community $C \in \mathcal{C}$ and layers $L_i, L_j \in \mathcal{L}$, we will use symbols $C^{(i)}$ and $C^{(j)}$ to denote the \textit{projection} of $C$ onto the two layers, i.e., the set of nodes in $C$ that lay on $L_i$ and $L_j$, respectively. 
In the following, we define   two   approaches for measuring inter-layer couplings based on community projection. 

For any two layers $L_i, L_j$ and community $C$, the first approach, we call \textit{symmetric}, determines the relevance of inter-layer coupling of nodes belonging to $C$ as proportional to the fraction of nodes shared between $L_i$ and $L_j$ that belong to $C$. 

\vspace{2mm}
\begin{definition}
\em 
Given a community $C \in \mathcal{C}$ and layers $L_i, L_j \in \mathcal{L}$,  the \emph{symmetric projection-based inter-layer coupling}, denoted as $IC_s(C,L_i, L_j)$ and referring to term $IC$  in Eq.~(\ref{eq:MLmodularity}), is defined as the probability that $C$ lays on $L_i$ and $L_j$:
\begin{equation}\label{eq:ICs}
\hspace{-2mm}IC_s(C,L_i, L_j)\!=\!\Pr[C \text{~in~} L_i, C \text{~in~} L_j] = \frac{|C^{(i)} \cap C^{(j)}|}{|V_i \cap V_j|}
\end{equation}
\QEDB
\end{definition}

%---------------- dext = $\sum_{L_i,L_j \in \mathcal{L}} |V_i \cup V_j|$
The above definition assumes that the two events ``$C$ in $L_i$'' and  ``$C$ in $L_j$'' are independent to each other, and it   does not consider that  the coupling might have a different meaning depending on the \textit{relevance} a  community has on a particular layer in which it is located.  
By relevance of community, we simply mean here the fraction of nodes in a layer graph that belong to the community; therefore, the larger the community in a layer, the more relevant is w.r.t. that layer. 
However, we observe that \textit{more relevant community in a layer corresponds to less surprising projection from that layer to another}. This would imply that the inter-layer  coupling for that community is less interesting w.r.t. projections of smaller communities, and hence the strength of the coupling might be lowered.  
We capture the above intuition by the following definition of   \textit{asymmetric projection-based inter-layer coupling}.

 \vspace{2mm} 
 \begin{definition}
\em 
Given a community $C \in \mathcal{C}$ and layers $L_i, L_j \in \mathcal{L}$,  the \emph{asymmetric projection-based inter-layer coupling}, denoted as $IC_a(C,L_i, L_j)$ and referring to term $IC$  in Eq.~(\ref{eq:MLmodularity}), is defined as the probability that $C$ lays on $L_j$ given that $C$ lays on $L_i$:
 \begin{eqnarray}\label{eq:ICa}
IC_a(C,L_i, L_j) & = & \Pr[C \text{~in~} L_j | C \text{~in~} L_i]  =  \nonumber \\
& = & \frac{\Pr[C \text{~in~} L_i, C \text{~in~} L_j]}{\Pr[C \text{~in~} L_i]}
=  \nonumber \\
& = & \frac{|C^{(i)} \cap C^{(j)}|}{|V_i \cap V_j|} \times \frac{|V_i|}{|C^{(i)}|}
\end{eqnarray}
 \QEDB
\end{definition}

%\subsubsection{\bf Dealing with layer ordering}
%\label{sec:ordering}
\mpara{Dealing with layer ordering. \ }
%As previously introduced, 
Our formulation of multilayer modularity is general enough to account for an available ordering of the layers, according to a given partial order relation. 

The previously defined asymmetric inter-layer coupling is well suited to model  situations in which we might want to express the inter-layer coupling from a ``source'' layer to a ``destination'' layer. Given any two layers $L_i, L_j$, it may be the case that only comparison of  $L_i$ to $L_j$, or vice versa, is allowed. 
This is clearly motivated when    there exist layer-coupling constraints, thus only some  of the layer couplings should  be considered in the computation of $Q$.  

In practical cases, we might assume that the layers can be naturally ordered to reflect a predetermined lexicographic order, which might be set, for instance, according to a progressive enumeration of layers or to a chronological order of the time-steps corresponding 
to the layers.  
That said,   we can consider two special cases of \textit{layer ordering}:
\begin{itemize}
\item
 \textit{Adjacent layer coupling}: 
$L_i \prec_{\mathcal{L}} L_j$ iff  $j=i+1$ according to a predetermined natural order.  
\item 
\textit{Succeeding-layer coupling}:   $L_i \prec_{\mathcal{L}} L_j$ iff  $j>i$ according to a predetermined natural order. 
% * ordered set: due casi: solo coppie adiacenti ($\ell - 1$ pairs) o coppie tra layer e suoi successivi nell'ordinamento  ($(\ell^2-\ell)/2$)
\end{itemize}

\noindent 
Note that the adjacent layer coupling scheme requires $\ell - 1$ pairs to consider, while the succeeding-layer coupling scheme involves the comparison between a layer and its subsequent ones, i.e., $(\ell^2-\ell)/2$ pairs. 
%Figure~\ref{fig:esempio} illustrates an example of asymmetric inter-layer coupling over a three-layer network. 

  Moreover, it should be noted that the availability of a layer ordering enables two variants of the asymmetric projection-based inter-coupling given in Eq. (\ref{eq:ICa}).  
For any two layers $L_i, L_j \in \mathcal{L}$, such that $L_i \prec_{\mathcal{L}} L_j$ holds, we refer to as  \textbf{\emph{inner}} the direct evaluation of $IC_a(C, L_i, L_j)$, and as \textbf{\emph{outer}}  the case in which $L_i$ and $L_j$ are switched, i.e., $IC_a(C, L_j, L_i)$. 

In the inner case, the strength of coupling is higher as the projection of $C$ on the source layer (i.e., the   preceding one  in the order) is less relevant;   vice versa,  the outer case 
weights more the coupling as the projection on the destination layer  (i.e., the   subsequent one in the order) is less relevant.

%%%%%%%%% RUNNING EXAMPLE - step #2  (begin) %%%%%%
\begin{example}
\em 
Consider again the example network of Fig.~\ref{fig:example1}. The asymmetric coupling for the projection of community $C_2$ from $L_2$ to $L_3$ is $IC_a(C_2, L_2, L_3) = (2/5) \times (9/4)=9/10=0.9$ in the inner case, and  $IC_a(C_2, L_3,$ $L_2)= (2/5) \times (6/3)=4/5=0.8$ in the outer case. 
\end{example}
%%%%%%%%% RUNNING EXAMPLE - step #2  (end) %%%%%%

We hereinafter use symbols $IC_{ia}$ and $IC_{oa}$ to distinguish between the inner asymmetric and the outer asymmetric evaluation cases.

\vspace{2mm}
\textit{\underline{Time-evolving multilayer networks}.\ } 
So far we have assumed that when comparing  any two layers $L_i, L_j$, with $L_i \prec_{\mathcal{L}} L_j$,  it does not matter the number of layers between $L_i$ and $L_j$. 
Intuitively, we might want to penalize the strength of the  coupling  as more ``distant'' $L_j$ is from $L_i$. This is often the case in time-sliced networks, whereby we want to understand how community structures evolve over~time. 

In light of the above remarks, we define a refinement of the asymmetric projection-based inter-layer coupling, by introducing a multiplicative factor that smoothly decreases the value of the $IC_a$ function as the temporal distance between $L_i$ and $L_j$ increases. 

\vspace{2mm}
\begin{definition}
\em 
Given a community $C \in \mathcal{C}$ and layers $L_i, L_j \in \mathcal{L}$, such that  $L_i \prec_{\mathcal{L}} L_j$, the \emph{time-aware asymmetric projection-based inter-layer coupling}, denoted as $IC_a^t(C,L_i, L_j)$, is defined as  
 \begin{equation}\label{eq:ICat}
\hspace{-2mm}IC_a^t(C,L_i, L_j)\!=\!IC_a(C,L_i, L_j) \times \frac{2}{1+ \log_2(1+j-i)}
\end{equation}
 \QEDB
\end{definition}
 Note that the second term in the above equation is 1 for the adjacent layer coupling scheme, thus making no penalization effect when only consecutive   layers are considered.

%%%%%%%%% RUNNING EXAMPLE - step #3  (begin) %%%%%%

\begin{example} 
\em 
Referring again to the example in Fig.~\ref{fig:example1}, 
in Table \ref{tab:exampleIC} we summarize the mean and standard deviation values for the different variants of the inter-layer coupling factor. 
 One remark is  that the values for communities $C_2$ and $C_3$ are higher than those corresponding to communities $C_1$ and $C_4$. 
%This is mainly because $C_2$ and $C_3$ are present in each layer. 
This is mainly due to the representativity of  $C_2$ and $C_3$ in all layers.  The lowest values are obtained for community $C_4$, which is in fact less represented than other communities (only in layers $L_1$ and $L_5$). For instance, let us focus on this community. The mean inter-layer coupling factor $IC_{oa}^{Suc}$ for community $C_4$ is 0.11, since: $|C_4^{(5)} \cap C_4^{(1)}|=2$, $|V_5 \cap V_1|=10$ (which is exactly the size of $L_5$ without node 16), $|V_5|=11$ and $|C_4^{(5)}|=2$; this  determines an inter-layer coupling factor of 1.1, which is divided by the admissible pairings of layers, i.e., 10. On the contrary, the mean $IC_{oa}^{Adj}$ for $C_4$ is equal to zero, because the projection of this community is always empty when the adjacent coupling scheme is used. % not represented in layer $L_2$ adjacent to layer $L_1$, hence $|C_4^{(1)} \cap C_4^{(2)}|$ or $|C_4^{(2)} \cap C_4^{(1)}|$ is equal to zero. 
 
Finally, Table \ref{tab:exampleMod} reports the multilayer modularity values, including the com\-munity-specific contributions. Note that, regardless of the settings of $\gamma$ and $IC$ factors,   
%community $C_4$ has the lowest modularity values, because it is not present (i.e., its projection is empty) in most layers (i.e., $L_2$, $L_3$, and $L_4$). By contrast,  
 communities $C_2$ and $C_3$ obtain the highest  values of modularity, which is mainly determined  since   they are disconnected from the rest of the graph at layer $L_3$. %In this layer, both $C_2$ and $C_3$ have external degree which is equal to zero. 
 In general, it should be noted that the contribution given by each community is consistent w.r.t. the various settings of $\gamma$ and $IC$ factors.  
 Also, it is interesting to note that discarding the inter-layer couplings ($IC=0$, which corresponds to the 9th and 13th columns) can lead to   %(\disc{rispetto al calcolo riportato nelle colonne 2,3 e 4, suppongo. Forse bisogna dirlo})  
  values of community-modularity and global modularity that  tend  to be much higher than the corresponding cases with $IC\neq 0$.  This overestimation can also occur, though to a lesser extent, when fixing $\gamma=1$ (13th column) vs. redundancy-based $\gamma$ (9th column). 
Also, it is worth noting that using the redundancy-based resolution factor $\gamma$ with unordered layers (2nd, 3rd and 4th columns) increases the community as well as global modularity  vs. the same cases with $\gamma=1$ (10th, 11th, and 12th columns).
\end{example}

%%%%%%%%% RUNNING EXAMPLE - step #3  (end) %%%%%%

\begin{table*}[t!]
\caption{Different variants of the inter-layer coupling factor,  for each community of the example network in Fig.~\ref{fig:example1}. Values are cumulated over the admissible pairings of layers. Mean and standard deviation values are reported.}
\label{tab:exampleIC}
\centering
\scalebox{0.75}{
\begin{tabular}{|c|c|c|c|c|c|c|}
\hline
&{$IC_s$}&{$IC_{ia} \equiv IC_{oa}$}&{$IC_{ia}^{Adj}$}&{$IC_{oa}^{Adj}$}&{$IC_{ia}^{Suc}$}&{$IC_{oa}^{Suc}$}\\\hline
$C_1$&0.135 $\pm$ 0.038&	0.612 $\pm$ 0.214&	0.444 $\pm$ 0.314&	0.525 $\pm$ 0.071&	0.619 $\pm$ 0.274&	0.606 $\pm$ 0.159\\\hline
$C_2$&0.398 $\pm$ 0.096&	1.105 $\pm$ 0.284&	1.115 $\pm$ 0.157&	1.192 $\pm$ 0.320&	1.018 $\pm$ 0.245&	1.193 $\pm$ 0.305\\\hline
$C_3$&0.416 $\pm$ 0.074&	1.148 $\pm$ 0.258&	1.188 $\pm$ 0.239&	1.088 $\pm$ 0.118&	1.249 $\pm$ 0.306&	1.048 $\pm$ 0.158\\\hline
$C_4$&0.018 $\pm$ 0.000&	0.091 $\pm$ 0.000&	0.000 $\pm$ 0.000&	0.000 $\pm$ 0.000&	0.091 $\pm$ 0.000&	0.110 $\pm$ 0.000\\\hline

    \end{tabular} 
    }
\end{table*}%

\begin{table*}[t!]
\caption{Multilayer modularity for the various combinations of resolution and inter-layer coupling terms, on the example network in Fig.~\ref{fig:example1}. Community-specific values correspond to the modularity contribution given by each particular community to the overall modularity.}
\label{tab:exampleMod}
\centering
\scalebox{0.53}{
\begin{tabular}{|c||c|c|c|c|c|c|c||c||c|c|c|c|}
\hline
&$\gamma, IC_s$&$\gamma, IC_{ia}$&$\gamma, IC_{oa}$&$\gamma, IC_{ia}^{Adj}$&$\gamma, IC_{oa}^{Adj}$&$\gamma, IC_{ia}^{Suc}$&$\gamma, IC_{oa}^{Suc}$&$\gamma, IC\!=\!0$&$\gamma\!=\!1,IC_s$&$\gamma\!=\!1,IC_{ia}$&$\gamma\!=\!1,IC_{oa}$&$\gamma\!=\!1,IC\!=\!0$\\
\hline
$C_1$&0.064&	0.098&	0.098&	0.129&	0.131&	0.142&	0.141&	0.126&	0.063&	0.097&	0.097&	0.123\\\hline
$C_2$&0.164&	0.214&	0.214&	0.318&	0.320&	0.317&	0.329&	0.310&	0.162&	0.213&	0.213&	0.307\\\hline
$C_3$&0.161&	0.213&	0.213&	0.313&	0.309&	0.325&	0.312&	0.266&	0.160&	0.212&	0.212&	0.299\\\hline
$C_4$&0.022&	0.028&	0.028&	0.044&	0.044&	0.045&	0.045&	0.048&	0.022&	0.027&	0.027&	0.047\\\hline\hline
$Q$&0.411&	0.554&	0.554&	0.803&	0.804&	0.828&	0.827&	0.750&	0.408&	0.550&	0.550&	0.777\\\hline
    \end{tabular} 
    }

\end{table*}% 

\subsubsection{Relations between the resolution and inter-layer coupling factors}
 
%It is worth noting that some relation can be observed between the resolution factor $\gamma(L, C)$ and the inter-layer coupling factor $IC$.
%First, we can observe that 
Both factors take into consideration the network context, however they differ in that  
$\gamma(L, C)$ considers a ``global'' multiplex context, whereas $IC$ considers a ``local'' multiplex context. 
Intuitively, $\gamma(L, C)$  is defined for each valid   layer-community pair according to the status of the links among nodes in   community $C$ that lays on $L$ versus their status on the other layers.  By contrast,   $IC$ considers the status of the same community from one layer to  another comparable layer.  

In terms of numerical comparison,  when the size of the community structure  tends to the number of nodes of the network,  $\gamma(L, C)$ tends to increase (i.e., to the maximum value of 2) while   $IC$ tends to decrease (i.e., to zero).

 \subsection{Properties of the proposed  multilayer modularity}
We provide theoretical insights on   $Q$, focusing on the effect of increase in the size of the community structure and on the analytical derivation of the range of values of $Q$.

\subsubsection{Effect of increase in the number of communities}
\label{commSize} 
%The multilayer modularity $Q$ obtains a minimum value of 0.25 when there are no intra-community edges but only inter-community edges (e.g. this is a typical case of a bipartite graph). This value is higher than the minimum obtained by the multislice modularity $Q_{\textrm{ms}}$, which is -0.5 in the same conditions. This refers to the case of redundancy-based resolution factor $\gamma(L,C)=1$ and $\beta=0$, which indicates the absence of the inter-layer coupling factor. In the special case of $\gamma(L,C)=2$ ($\beta$ is 0), the minimum value of $Q$ is -0.5. In general, because the value of $\gamma(L,C)$ ranges between 1 and 2, the minimum value of $Q$ ranges between -0.25 and -0.5.

%Similarly to $Q_{\textrm{ms}}$, when $\gamma(L,C)=1$ and $\beta=0$, the maximum value of $Q(\mathcal{C})$ approaches to 1 in the case of a community structure composed of many communities with no inter-community edges. For $\gamma(L,C)$ values above 1, $Q(\mathcal{C})$ decreases. 
We discuss the effect of increasing $k$ (i.e., decreasing the average size of communities in $\mathcal{C}$) by distinguishing three configurations of $Q$: (i) symmetric inter-layer coupling, (ii) asymmetric inter-layer coupling, and (iii) ordered layers.

In the first case, $Q$ tends to have a monotonic decreasing trend.   
  %which is typical of the Newman and Girvan's modularity $Q_{NG}$. 
 This is easily explained by the combination of three contingencies. The first one is an average decrease in the internal degree $d_L^{int}$. The second contingency is an   increase in the redundancy-based resolution factor $\gamma$: in fact, smaller communities correspond to lower probability of observing redundant pairs within  communities over   different layers; this decreases the logarithmic term in the resolution factor,   which will progressively  tend to 2 (maximum value). The third contingency is a decrease in the inter-layer coupling factor $IC_s$, since the size of community intersection  becomes increasingly smaller as the community size decreases.
 
By contrast, when equipped with the asymmetric projection-based inter-layer coupling $IC_a$, $Q$ tends to  differ from  a monotonic decreasing trend because of the bias term $\frac{|V_i|}{|C^{(i)}|}$, which increases with communities of smaller size. 
%In this case, it is possible to identify a community number after which the divergence is visible. Also, the modularity will reach a value of 1 as soon as the number of communities equals the number of nodes in the network.

In the third case (i.e., ordered layers), $Q$ can again follow an increasing or decreasing trend. Recall that  the term $d(V_{\mathcal{L}})$    includes the contribution of the inter-layer edges, which obviously are fewer   when the  layer couplings  are order-dependent. A decrease in the number of inter-layer couplings    also makes the decrease in the actual connectivity term (i.e., $\frac{d_L^{int}(C)}{d(V_{\mathcal{L}})}$)  slower as $k$ increases, since $d(V_{\mathcal{L}})$ is smaller than in the unordered-layer  contingency. 
Consequently, the inter-layer coupling term could compensate the actual connectivity term, which will result in increasing the value of $Q$. 
Finally, considering time-aware asymmetric  inter-layer coupling, $Q$ is more likely to  follow a  decreasing  trend because of the effect due to the smoothing term $\frac{2}{1+ \log_2(1+j-i)}$, which penalizes $IC_a$ for any two  no time-consecutive layers. Consequently, since  the inter-layer coupling factor $IC_a^t$ is smaller than   $IC_a$, $Q$ could  monotonically decrease despite  the bias term $\frac{|V_i|}{|C^{(i)}|}$ in $IC_a^t$.

\subsubsection{Lower and upper bounds}
%The proposed modularity $Q(\mathcal{C})$ reaches the minimum value of 0.255 when there are no intra-community edges but only inter-community edges (e.g. this is a typical case of a bipartite graph). This value is higher than the minimum obtained by the multislice modularity $Q_{\textrm{ms}}$, which is -0.5 in the same conditions. This refers to the case of resolution parameter $\gamma(L,C)=1$ and $\beta=0$ indicating the absence of inter-layer coupling. In the extreme case of $\gamma(L,C)=2$ (while $\beta$ is 0), the minimum value of $Q(\mathcal{C})$ reaches -0.5. In general, because the value of $\gamma(L,C)$ ranges between 1 and 2, the minimum value of $Q(\mathcal{C})$ ranges between -0.255 and -0.5.
%Similarly to $Q_{\textrm{ms}}$, when $\gamma(L,C)=1$ and $\beta=0$, the maximum value of $Q(\mathcal{C})$ approaches to 1 in the case of a community structure composed of many communities with no inter-community edges. For $\gamma(L,C)$ values above 1, $Q(\mathcal{C})$ decreases. 

To determine the range of values of the basic modularity in simple graphs, the theoretical frameworks   previously studied in \cite{BrandesDGGHNW08} and \cite{FortunatoB07} define two canonical structures to support the analytical computation of the minimum and maximum value of the modularity, respectively. More specifically, the former work proved that  any \textit{bipartite} graph  with the canonical \textit{two-way clustering}   obtains the minimum value of modularity, whereas the latter work proved that the maximum modularity is reached in a graph composed of  disjoint  cliques.

 Following the lead of the above works, here  we provide  theoretical results about   the analytical derivation of the lower bound and upper bound   of our proposed multilayer modularity.   %, for different settings of the resolution and inter-layer coupling factors. 

 \begin{proposition}
 \em
 Given a multilayer network $G_{\mathcal{L}} = (V_{\mathcal{L}}, E_{\mathcal{L}}, \V, \mathcal{L})$, with $n=|\V|, \ell = |\mathcal{L}|$, and a community structure $\mathcal{C}$ for $G_{\mathcal{L}}$, 
 the {\em lower bound} of $Q$  is as follows:   
\begin{equation}
\label{eq:pippo} 
Q(\mathcal{C}) =   -\frac{n^2\ell}{(n\ell +2p)^2} + \frac{4(1+\eta)p}{n^2(n\ell + 2p)}, 
\end{equation}
with $\eta=0$ for $IC_s$ and $\eta=1$ for $IC_a$, 
and  $p= \sum_{L \in \mathcal{L}}|\mathcal{P}(L)|$ is   the total number of valid layer-pairings.
 \end{proposition}
 
 \noindent 
 \textit{Proof.\ }   Proof is reported in the {\bf \em Appendix}.

 \begin{proposition}
 \em
  Given a multilayer network $G_{\mathcal{L}} = (V_{\mathcal{L}}, E_{\mathcal{L}}, \V, \mathcal{L})$, with $n=|\V|, \ell = |\mathcal{L}|$, and a community structure $\mathcal{C}$ for $G_{\mathcal{L}}$, 
 the {\em upper bound} of $Q$  is as follows:   
\begin{equation}\label{eq:general}
\begin{split}
Q(\mathcal{C})  
 = & 2 \left[ \frac{1}{2}\frac{(\frac{n}{2}-1)\ell}{(\frac{n}{2}-1)\ell+p}-\gamma\ell \left(\frac{1}{2}\frac{(\frac{n}{2}-1)}{(\frac{n}{2}-1)\ell+p}\right)^2 + \right. \\
 + & \left.  \frac{(1+\eta)p}{{n^2}(\frac{n}{2}-1)\ell+n^2p}  \right]
\end{split}
\end{equation}
with $\eta=0$ for $IC_s$ and $\eta=1$ for $IC_a$, 
and  $p= \sum_{L \in \mathcal{L}}|\mathcal{P}(L)|$ is   the total number of valid layer-pairings.
\end{proposition}
 
 \noindent 
 \textit{Proof.\ }   Proof is reported in the {\bf \em Appendix}. 

  \vspace{2mm}
Note that,  in the special case for $\beta=0$, i.e.,  the inter-layer coupling factor is discarded, the lower bound of $Q$ is 
\begin{equation} 
Q(\mathcal{C}) =-\frac{1}{\ell}.
\end{equation}

\noindent 
Analogously, the upper bound of $Q$ is rewritten as:
 
 \begin{equation} 
Q(\mathcal{C}) =   \frac{2\ell-\gamma}{2\ell},
\end{equation}
with $\gamma=2(1+\log_2(1+\frac{\frac{n}{2}(\frac{n}{2}-1)}{2}))^{-1}$.

%==========================================
 
\section{Evaluation Methodology}
\label{sec:evaluation}

We discuss here the evaluation networks (Sect.~\ref{sec:datasets}), the multilayer community detection methods  (Sect.~\ref{sec:methods}),  and the experimental settings (Sect.~\ref{sec:setting}).

\subsection{Datasets}
\label{sec:datasets}

Our selection of network datasets was motivated to fulfill the \textit{reproducibility} requirement: in fact, all of our evaluation datasets, including both real-world networks and synthetic generators, are publicly available. Moreover, we also took the opportunity of diversifying the choice of real-world networks by considering various domains that are profitably modeled as multilayer networks.

 \subsubsection{Real-world network datasets}
We considered 10 real-world multilayer network datasets. 
%, namely: 
%\textit{AUCS}, \textit{DBLP}, \textit{EU-Air transportation}, \textit{\fftwyt}, \textit{Higgs-Twitter},  \textit{London transportation}, and \textit{7thGraders}.  
%
%
\textit{AUCS} ~\cite{KimL15,Rossi2015}  %~\cite{Rossi2015} 
describes relationships among university employees:   work together, lunch together, off-line friendship, friendship  on Facebook, and coauthorship. 
%
%
%\textit{DBLP}~\cite{KimL15}     %~ \cite{Ley2002} 
%represents  co-authorships over different time slices, which correspond to the publication years in the period 1971-2014. 
%
%
EU-Air transport  network~\cite{KimL15}  %~\cite{Cardillo2013} 
(\textit{EU-Air}, for short) 
represents European airport connections considering  different  airlines. 
FAO Trade network (\textit{FAO-Trade}) \cite{Domenico2015} represents different types of trade relationships among countries, obtained from FAO (Food and Agriculture Organization of the United Nations).  
\textit{\fftwyt} (stands for FriendFeed, Twitter, and YouTube)~\cite{Magnanibook}  was built by exploiting the feature of FriendFeed as social media aggregator to align registered users who were also members of Twitter and YouTube. 
Flickr refers to the  dataset studied in~\cite{cha2009www}. 
% which contains time information about the  favorite markings assigned to   uploaded photos. 
We used the corresponding    timestamped interaction network whose links express ``who puts a favorite-marking to a photo of whom''.   
We extracted the layers on a month-basis and aggregated every six (or more)  months. % to achieve a quite balanced multilayer network.
\textit{\ghsotw} (stands for GitHub, StackOverflow and Twitter)~\cite{kdweb2015} is another cross-platform network   % originally built to characterize specialist expertise of users engaged in Web-mediated professional activities, 
where  edges express followships on Twitter and GitHub, and  ``who answers to whom'' relations on   StackOverflow.
\textit{Higgs-Twitter}~\cite{KimL15}  %~\cite{DeDomenico2013} 
represents  friendship,   reply,   mention, and retweet relations among Twitter users. 
London   transport network~\cite{ZhangWLY16}  % ~\cite{DeDomenico2014} 
 (\textit{London}, for short) models three types of connections of train stations in London: underground lines, overground, and DLR.   
ObamaInIsrael2013~\cite{Omodei2015}
(\textit{Obama}, for short) models retweet, mention, and reply relations of users of  Twitter during   Obama's visit to Israel in 2013.
7thGraders ~\cite{ZhangWLY16}  %~\cite{Vickers81} 
(\textit{VC-Graders}, for short) represents  students involved in    friendship,  work together, and affinity relations in the class. 
  Table~\ref{tab:datasets} reports for each dataset, the size of set $\V$, the number of edges in all layers, %the description of layers along with the percentage of nodes in each particular layer, 
 and the average coverage of node set (i.e.,  $ 1/|\mathcal{L}| \sum_{L_i \in \mathcal{L}} (|V_i|/|\mathcal{V}|)$). %, and the average coverage of edge set (i.e., $1/|\mathcal{L}|  \sum_{L_i \in \mathcal{L}} (|E_i|/ \sum_{L_i} |E_i|)$). 
The table also shows basic, monoplex structural statistics (degree, average path length, and clustering coefficient)  for the layer graphs of each dataset.

\begin{table*}[t!]
\caption{Main characteristics of our evaluation network datasets. Mean and standard deviation over the layers are reported for degree, average path  length, and clustering coefficient  statistics.}
\centering 
\scalebox{0.68}{
\begin{tabular}{|l||c|c|c|c|c|c|c|}
\hline
&\#entities &\#edges &\#layers & node set &   degree & avg. path & clustering \\
& $(|\V|)$ & & $(\ell)$ & coverage &    & length & coefficient \\
\hline \hline
\textit{AUCS} & 61 & 620 & 5 &0.73     & 10.43  $\pm$ 4.91 & 2.43 $\pm$ 0.73 &  0.43 $\pm$   0.097 \\
\hline
%DBLP & 1\,314\,050 & 7\,647\,677 & 44 & 0.06 & 0.02 & 7.46 $\pm$ 3.06 & 8.59  $\pm$ 1.39 & 0.69 $\pm$ 0.13 \\
%\hline 
\textit{EU-Air} & 417 & 3\,588 & 37 & 0.13   & 6.26 $\pm$ 2.90 & 2.25 $\pm$ 0.34 & 0.07 $\pm$ 0.08 \\
\hline
 \textit{FAO-Trade} &  214 & 318\,346  & 364  & 1.00   &  7.35$\pm$6.17  &  2.43$\pm$0.39  &  0.31$\pm$0.11  \\
\hline
\textit{\fftwyt} & 6\,407 & 74\,836 & 3 & 0.58   &  9.97 $\pm$ 7.27 & 4.18 $\pm$ 1.27  & 0.13  $\pm$  0.09 \\
\hline
\textit{Flickr} & 789\,019 &17\,071\,325  & 5 &0.33   &  23.15 $\pm$   5.61 &  4.50 $\pm$    0.60 &  0.04 $\pm$   0.01  \\
\hline 
\textit{\ghsotw} &55\,140  &2\,944\,592   & 3 &0.68      &  41.29 $\pm$ 45.09 &   3.66 $\pm$   0.62 &  0.02 $\pm$   0.01 \\
\hline
\textit{Higgs-Twitter} & 456\,631 & 16\,070\,185 & 4 &0.67    & 18.28 $\pm$ 31.20 & 9.94 $\pm$ 9.30 & 0.003 $\pm$ 0.004\\
\hline
\textit{London} & 369 & 441 & 3 &0.36    & 2.12 $\pm$ 0.16 & 11.89 $\pm$ 3.18 & 0.036 $\pm$ 0.032 \\
\hline
\textit{Obama} &2\,281\,259  &4\,061\,960  & 3 &0.50   &  4.27 $\pm$ 
 1.08 &  13.22 $\pm$   4.49 &   0.001 $\pm$    0.0005  \\
\hline
\textit{VC-Graders} & 29 & 518 & 3 &1.00  & 17.01 $\pm$ 6.85 & 1.66 $\pm$ 0.22 & 0.61 $\pm$ 0.89 \\
\hline
\end{tabular}
}
\label{tab:datasets}
\end{table*}

 \subsubsection{Synthetic network datasets}
 \label{sec:snd}
Besides the real-world network data, we generated four synthetic multilayer network datasets. Our goal was the evaluation of the multilayer modularity $Q$ on different network models.
 Two out of the four networks are composed of 2 layers and 256 entities. In one   network, hereinafter referred to as \textit{ER-ER},  the two layers are Erd\"os-R\'enyi (ER) random graphs. In the second network, dubbed \textit{LFR-ER}, the first layer is generated by the Lancichinetti-Fortunato-Radicchi (LFR) benchmark, while the second layer is an Erd\"os-R\'enyi random graph. The other two networks are composed of 4 layers and 128 nodes. Both networks are characterized by two Erd\"os-R\'enyi  layers  and two layers built as Girvan-Newman (GN) graphs, but they differ in the layer ordering:  \textit{GN-ER-GN-ER} in the first network, and \textit{GN-ER-ER-GN} in the second network.  

Moreover, mainly for purposes of efficiency evaluation, 
%in order to evaluate the computation time of the multilayer modularity $Q$ in the different configurations,  
 we  generated a set of synthetic multilayer networks using the Lancichinetti-Fortunato-Radicchi (LFR) benchmark. In particular, single-layer network datasets were provided by the LFR benchmark using a variable number of nodes with steps of 128 until 1024. Also, the maximum and average node degrees were set to 16, and the mixing coefficient $\mu$ was set to 0.1.   Each network dataset was characterized by four communities.  
From each of such networks, a multilayer network   was created by replicating the LFR single-layer     from 2 to 10.

%======================================
\subsection{Community detection methods}
\label{sec:methods}
 
We resorted to   state-of-the-art methods for  community  detection in multilayer networks, which belong to the two  major  approaches, namely  \textit{aggregation} and \textit{direct} methods. The former  detect a community structure separately for each network layer, after that  an aggregation mechanism is used to obtain the final community structure, while the latter directly work on the multilayer graph by    optimizing a multilayer quality-assessment criterion. 
(Note that while it is possible to flatten the multilayer graph in order to apply on it any conventional  community detection algorithm, this approach can be too simplistic, since, e.g.,  it would not permit to investigate about the temporal evolution of communities.)

As   exemplary methods of the   aggregation approach, we used  \textit{Principal Modularity Maximization} (PMM)~\cite{TangWL09} and \textit{Enhanced Modularity-driven Ensemble-based Multilayer Community Detection} (\greedystar)~\cite{Tagar18}. 
 PMM aims to find a concise representation of features from the various layers (dimensions) through   
 structural feature extraction and cross-dimension integration.  
Features from each dimension are first extracted via modularity maximization, then concatenated and subjected to PCA to select the top eigenvectors, which represent possible community partitions. Using these eigenvectors, a low-dimensional embedding is computed  to   capture  the principal patterns across all the dimensions of the network, finally a   $k$-means on this embedding is carried out to discover a community structure.  
% find out the discrete community assignment.   
 %
 \greedystar is a parameter-free extension of  the \greedy method proposed in~\cite{Tagar17}. Given an ensemble of community structures available for a multilayer network, \greedy optimizes a consensus objective function  to discover a consensus solution with maximum modularity, subject   to the constraint of being searched over a hypothetical space of consensus community structures that are valid w.r.t. the input ensemble and topologically bounded by two baseline solutions.  To detect the initial cluster memberships of nodes,  \greedy utilizes a   consensus or co-association matrix, which stores the fraction of clusterings in which any two nodes are assigned to the same cluster. To filter out noisy, irrelevant co-association, a user-specified threshold must be specified.  
Besides introducing flexibility in community assignments of nodes during the modularity optimization,  \greedystar overcomes the limitation of setting such a parameter of minimum co-association,   by providing a parameter-free identification of consensus clusters   based on generative models for graph pruning. 
%which enable edge-removal decisions with no user-specified parameters. The latter is solved by a three-steps process based on intra-community connectivity refinement, community partitioning, and relocation of nodes from a community to a neighboring one.}
 
As for the direct methods, we resorted to \textit{Generalized Louvain} (GL)~\cite{Mucha10} and \textit{Locally Adaptive Random Transitions} (LART)~\cite{KunchevaM15}. 
GL extends the classic Louvain method using  multislice  modularity, so that each node-layer tuple is assigned separately to a community.  Majority voting is adopted to decide the final assignment of an entity node to the  community that contains the majority of its layer-specific instances.  
LART is a   random-walk based method. It first runs a different random walk for each layer, then a dissimilarity measure between nodes is obtained leveraging the per-layer transition probabilities. A hierarchical clustering method is used   to produce a hierarchy of  communities which is eventually cut at the level corresponding to the best value of multislice modularity.

It should be emphasized that  we selected the above methods because, while having  different characteristics, they all use modularity either as optimization criterion (GL, PMM   and \greedystar) or as evaluation criterion to produce the final community structure (LART).  
%
%GL extends the classic Louvain method using a multislice modularity measure,  PMM relies on a lower-dimensional embedding, and LART is a random-walk based approach. 

\begin{table}[t]
\caption{Number of communities found by GL, LART, PMM and \greedystar with MLF model-filter on the  real-world network datasets}
\centering
%\color{blue}
\scalebox{0.9}{
\begin{tabular}{|c||c|c|c|c|}\hline  
&GL&LART&PMM&\greedystar\\
\hline\hline
\textit{AUCS}&5&27&2&13\\\hline
\textit{EU-Air}&10&381&5&39\\\hline
\textit{FAO-Trade}&12&-&10&11\\\hline
\textit{\fftwyt}&749&-&10&115\\\hline
\textit{\ghsotw}&87&-&10& 392 \\\hline
\textit{Flickr}&12290&-&10& 7660 \\\hline
\textit{Higgs-Twitter}&15218&-&10&121\\\hline
\textit{London}&21&339&30&46\\\hline
\textit{Obama}&297062&-&10&328367\\\hline
\textit{VC-Graders}&3&6&2&16\\\hline
    \end{tabular} 
    }
\label{tab:ncomm}
\end{table}

Note also that  PMM requires the desired number of communities ($k$) as input. 
Due to different size of our evaluation datasets, we devised several configurations of variation of parameter $k$ in PMM, by reasonably adapting each of the configuration range and increment step   to the network size.  Concerning \greedystar, we used  the \textit{marginal likelihood filter} (MLF) to perform parameter-free detection of the number of communities~\cite{Tagar18}.

It should be noted that the selected methods actually discover different community structures, thus supporting our choice in terms of diversity of evaluation scenarios for  the two competing modularity measures under study. 
 Table \ref{tab:ncomm} reports the number of communities of the solutions found by the various methods on the  real-world network datasets. (The number of communities $k$ in PMM is selected according to the solution with highest modularity value.)  
We found  that GL tends to discover a high number of communities for larger networks (i.e.,   \textit{Flickr}, \textit{Higgs-Twitter},  \textit{\fftwyt},    and \textit{Obama}), and 
the size distribution   of these communities (results not shown)   is highly right-skewed  %(i.e., tail stretching toward the right) 
on the larger networks, while it is moderately left-skewed on the remaining datasets.  A similar result can be observed in \greedystar for the different networks, although in \textit{Higgs-Twitter} and \textit{\fftwyt} (resp. \textit{\ghsotw})  the number of communities is much lower (resp. higher) than in GL. 
By contrast,  the best performances of PMM  usually correspond to a low and quite stable number of communities.  Also,   LART generally tends to produce much more communities than the other methods, on the networks for which it is able to discover communities. 

As a final general remark, we used the original implementations of the selected methods, based on the source code made available by the respective authors. We emphasize that it is beyond the goals of this work to make any performance improvement in the community detection methods under study, which hence are considered here with no intent of comparative evaluation and with all their limitations. (This justifies, in particular, the inability of LART in terminating the task for some network datasets.)

\subsection{Experimental settings}
\label{sec:setting}

We carried out GL, PMM, LART  and   \greedystar  methods on each of the network datasets and measured, for each community structure solution, our proposed multilayer modularity ($Q$) as well as the  Mucha et al.'s multislice modularity ($Q_{\textrm{ms}}$).  

 We   evaluated $Q$ using the redundancy-based resolution factor $\gamma(L,C)$ with either  the symmetric   ($IC_s$) or the asymmetric ($IC_a$) projection-based inter-layer coupling.   
 We  also considered   cases corresponding to   ordered   layers, using either the adjacent-layer scheme or the succeeding-layer scheme, and for both schemes considering inner ($IC_{ia}$) as well as outer   ($IC_{oa}$) asymmetric coupling.  
We further evaluated the case of temporal ordering, using   the time-aware asymmetric projection-based inter-layer coupling. 
Yet, we considered the particular setting  of uniform resolution (i.e., $\gamma(L,C)=1$, for each layer $L$ and community $C$).

As for $Q_{\textrm{ms}}$, we   devised two settings: the first by varying  
$\omega$  within $[0..2]$  while fixing $\gamma=1$,  
%$\gamma$ within $[0..2]$ while fixing $\omega=0$ (i.e., no inter-layer couplings), 
 the second by varying $\gamma$ and $\omega=1-\gamma$~\cite{Mucha10}.

%==========================================

\section{Results}
\label{sec:results}

We organize our main experimental results into two parts, depending on whether layer ordering  was considered in the evaluation networks. 
Experiments were  carried out on an Intel Core i7-3960X CPU @3.30GHz, 64GB RAM machine.

\subsection{Evaluation with unordered layers}

 \begin{table*}[t!]
\caption{Multilayer modularity $Q$ and multislice modularity $Q_{\textnormal{ms}}$ on GL community structures of the four synthetic networks.}
\label{tab:Synth}
\centering
\scalebox{0.8}{
\begin{tabular}{|l||c|c||c|c|c|c|c|}
\hline
  &\#comm.&$Q_{\textrm{ms}}$& $\gamma=1$,  &  $\gamma=1$, &$\gamma$,  &$\gamma$, &$\gamma$, \\
 &  by GL  & & $IC_s$ & $IC_{ia} \equiv IC_{oa}$ & $\beta=0$ & $IC_s$ & $IC_{ia} \equiv IC_{oa}$ \\
\hline\hline
   ER-ER&10&0.249  &0.192    & 0.196         & 0.290  &0.258&0.262\\
   LFR-ER&16&0.486&0.404&0.411&0.486&0.434&0.441\\
   GN-ER-GN-ER&4&0.429&0.432&0.436&0.552&0.471&0.475\\
   GN-ER-ER-GN&4&0.429&0.432&0.436&0.552&0.471&0.475\\\hline
    \end{tabular}
}
\end{table*}%

\subsubsection{Synthetic network datasets}
 
Table \ref{tab:Synth} reports the multilayer modularity $Q$, multislice modularity $Q_{\textrm{ms}}$ and number of communities obtained by the GL solution on the four synthetic networks.

One first  remark is that using the redundancy-based resolution factor $\gamma$ always leads to higher $Q$ w.r.t. the cases corresponding to $\gamma$ fixed to 1.  In particular,  we observe gains up to 0.1 on ER-ER, 0.07 on LFR-ER, and 0.12 on GN-ER-GN-ER and GN-ER-ER-GN. %It is also worth noting that the redundancy-based resolution factor $\gamma$ increases $Q$ w.r.t. $\gamma$, $IC_s$ and $\gamma$, $IC_a$. 

Another  remark is that the fully combination of resolution and inter-layer coupling factors (i.e., rightmost two columns) tends to lower the value of $Q$ w.r.t. the cases corresponding to varying $\gamma$ with $\beta=0$ (i.e., third last column); moreover, the asymmetric inter-layer coupling results in a higher $Q$ w.r.t. the symmetric setting of $IC$.  This would hint that when the normalization term in the $Q$ equation accounts for  the inter-layer couplings, this results in lowering the value of  $Q$, which is turn smoother when the asymmetric setting is used.

Comparing $Q$ and $Q_{\textrm{ms}}$, it should be noted that the two measures behave consistently on ER-ER vs. LFR-ER, i.e., the presence of a layer with a (LFR) modular structure actually leads to an increase in both modularity measures w.r.t. ER-ER. 
By contrast,  $Q$ tends increase faster  than $Q_{\textrm{ms}}$    on the two GN-ER networks: this can be explained since a higher number of layers  (as occurs for the two GN-ER networks than for the ER-ER and LFR-ER networks) has a higher effect on  the inter-layer coupling factor $IC$, which is not present in  $Q_{\textrm{ms}}$.

\vspace{2mm}
\subsubsection{Real-world network datasets}

\begin{figure}[t!]
\centering
\begin{tabular}{ccc}
\hspace{-5mm}
\includegraphics[scale=0.075]{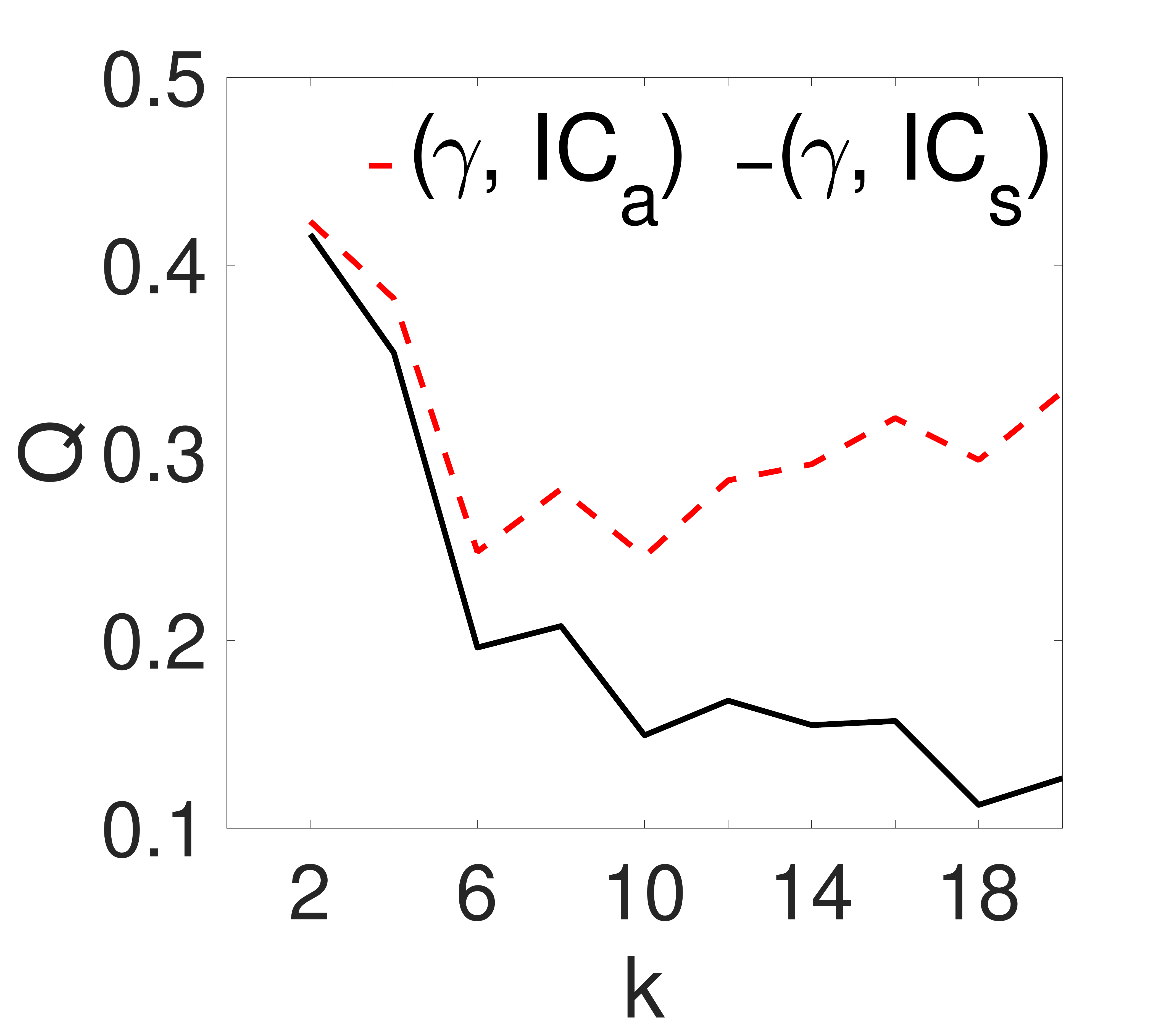} &  
\hspace{-8mm}
\includegraphics[scale=0.075]{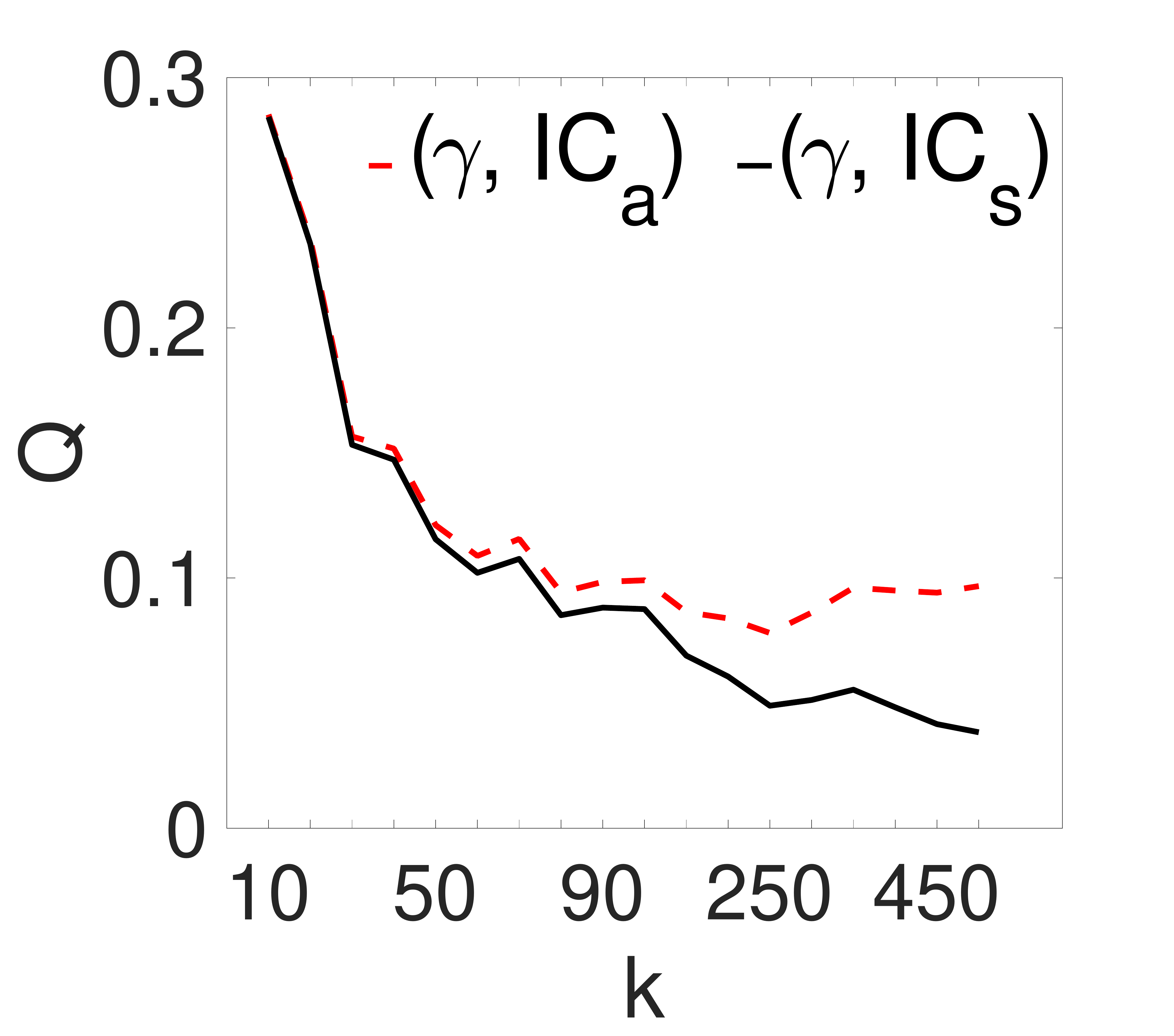} &
\hspace{-8mm}
\includegraphics[scale=0.075]{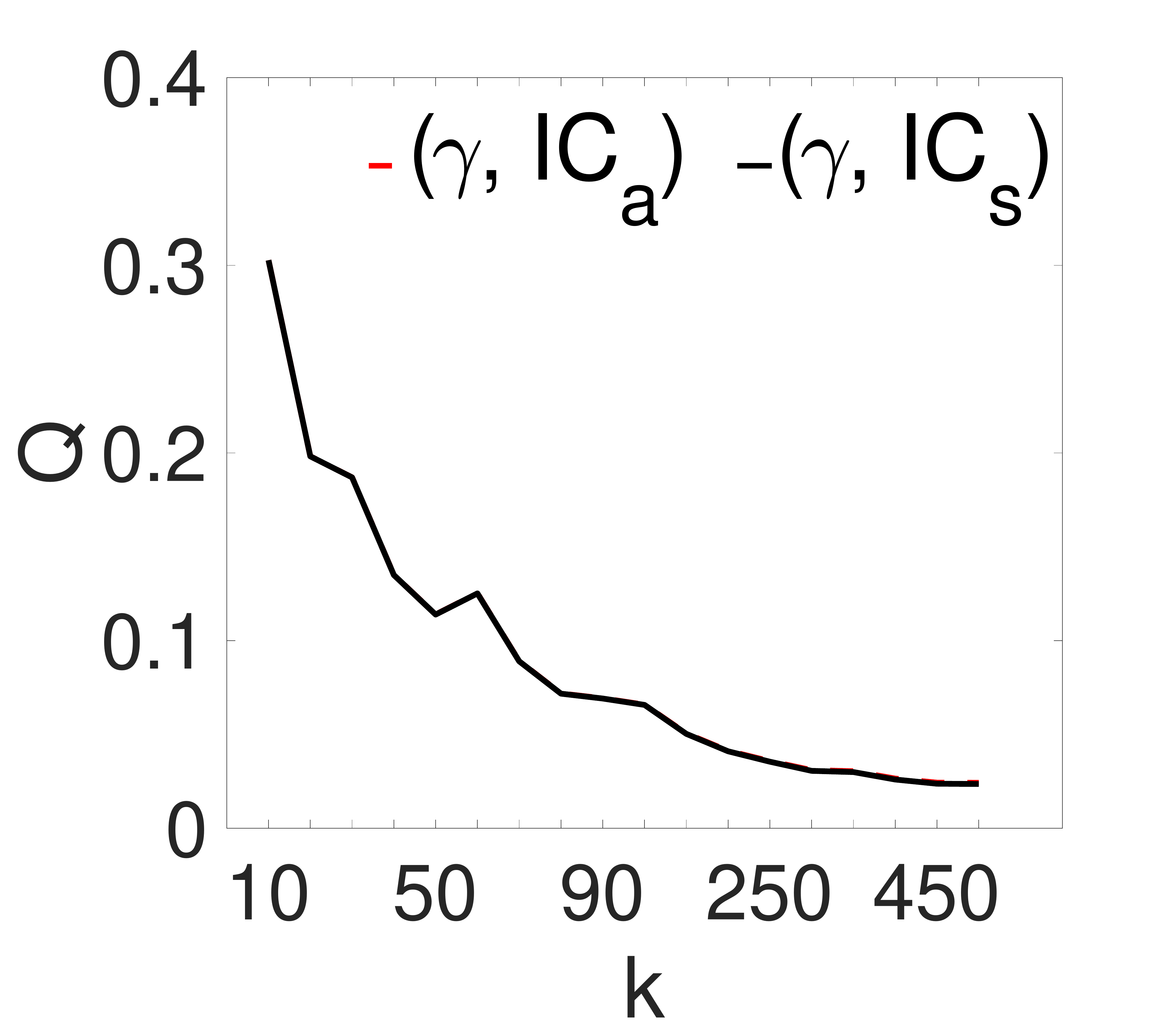} \\ 
\hspace{-7mm} (a) \textit{AUCS} & \hspace{-8mm}
 (b) \textit{\fftwyt}& \hspace{-9mm}  (c) \textit{Flickr}  \\
\hspace{-5mm}
\includegraphics[scale=0.075]{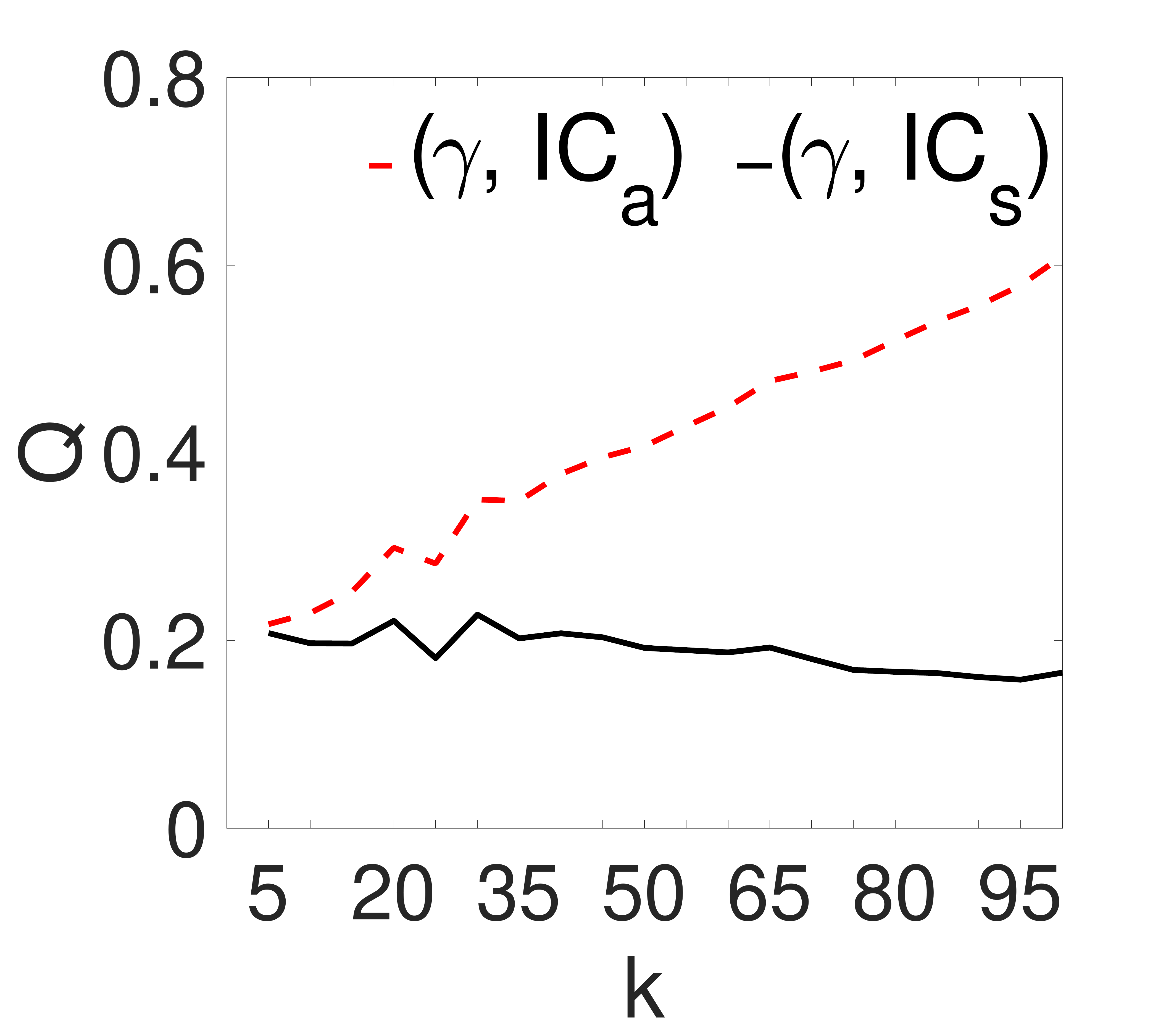} & 
\hspace{-8mm}
\includegraphics[scale=0.075]{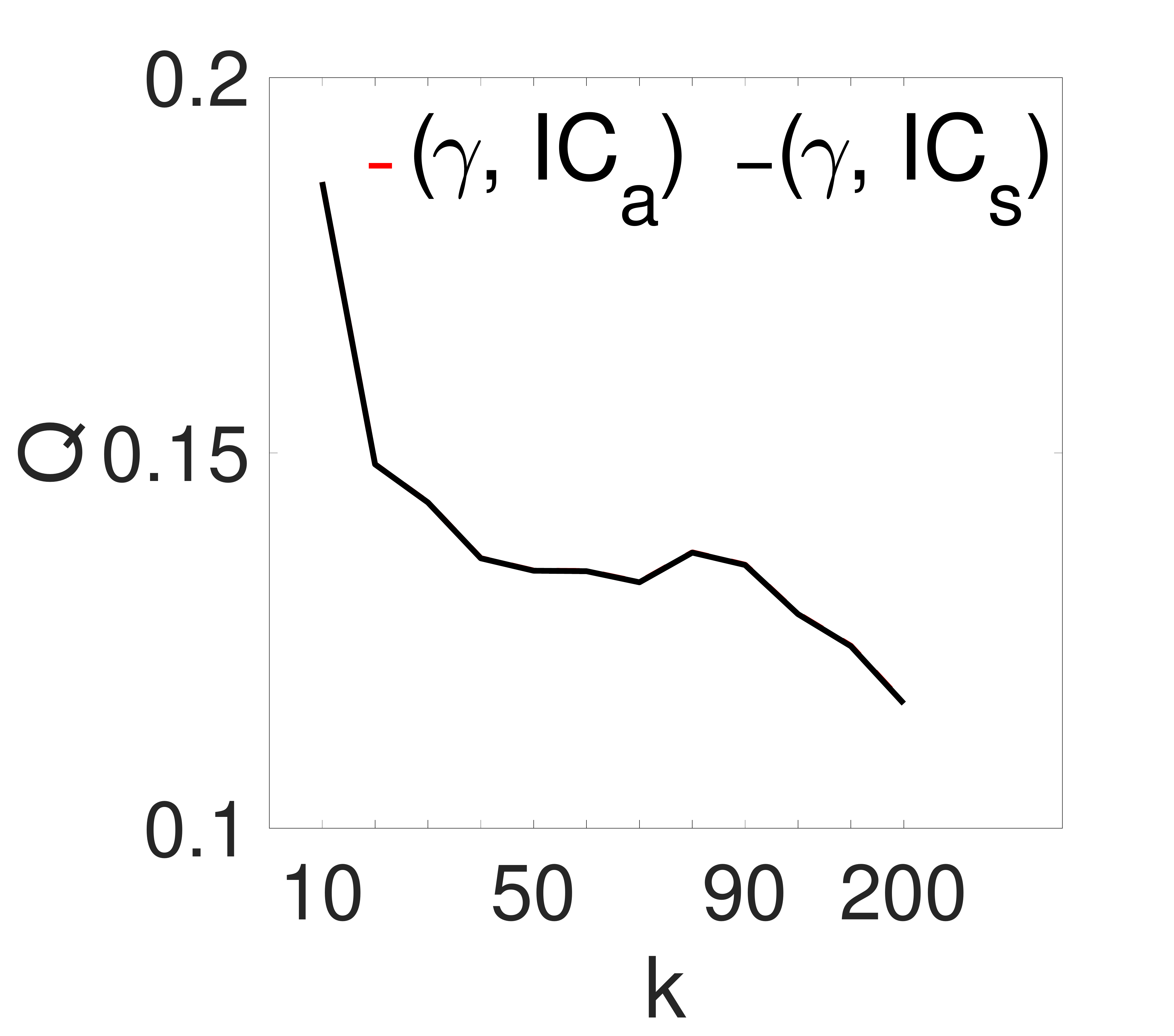} & 
\hspace{-8mm}
\includegraphics[scale=0.075]{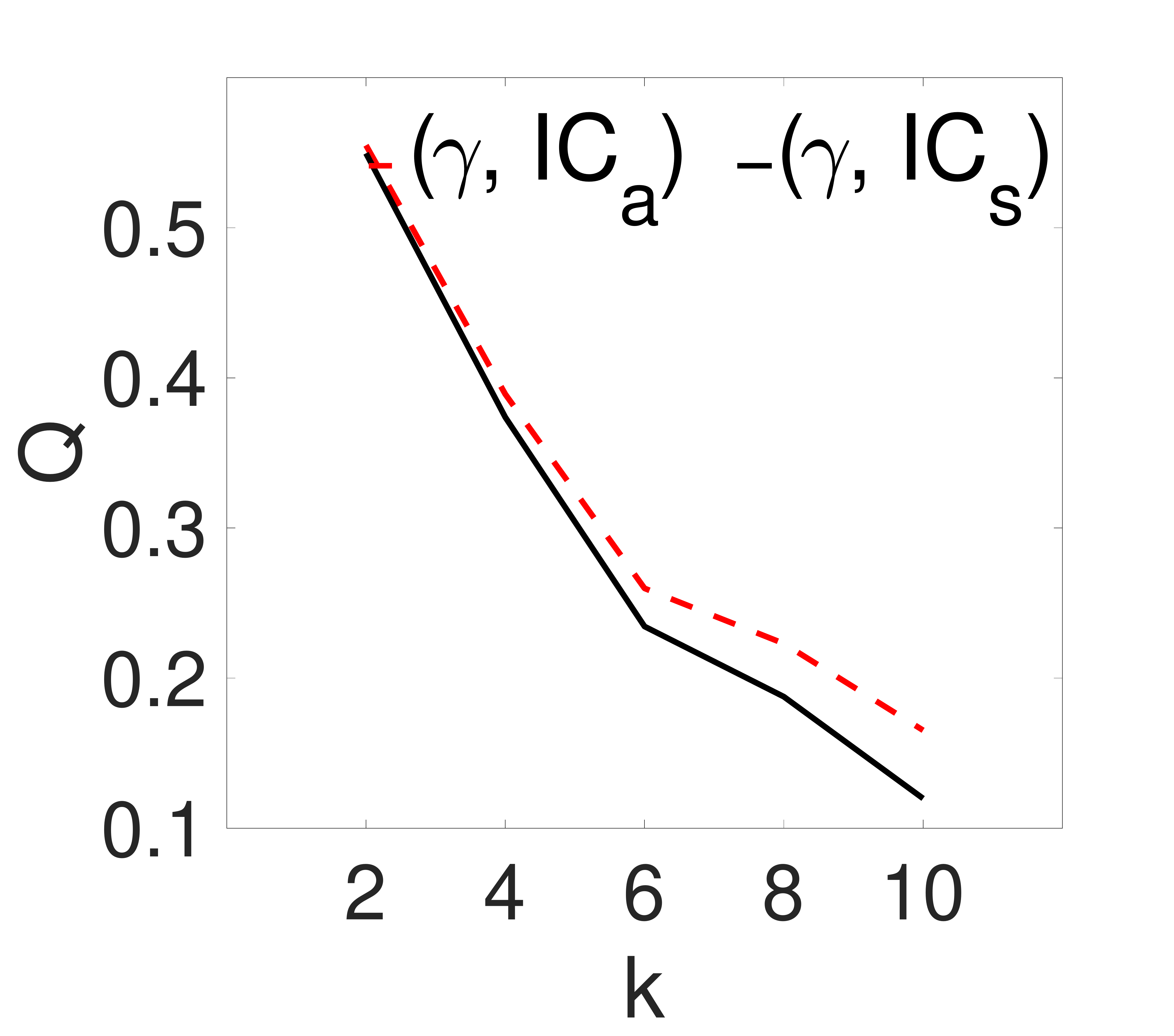} \\
\hspace{-7mm} (d) \textit{London} & \hspace{-9mm}
 (e) \textit{Obama} & \hspace{-9mm} (f) \textit{VC-Graders} \\
\end{tabular}
\caption{Multilayer modularity $Q$ on PMM community structure solutions}
\label{fig:gammaICasbeta0}
\end{figure}

 Tables~\ref{tab:GL}--\ref{tab:EMCD}  and 
Fig.~\ref{fig:gammaICasbeta0} report $Q$  measurements on the community structure solutions obtained by the various  community detection methods. 

Concerning GL (Table~\ref{tab:GL}), 
 we observe that with the exception of \textit{\ghsotw} on which effects on $Q$ are equivalent,  
 using $IC_a$ leads to higher  $Q$  than  $IC_s$.  
On average over all networks, using  $IC_a$ yields an increment of  13.4\% and 14.6\% (with $\gamma$ fixed to 1) w.r.t. the value of $Q$ corresponding to $IC_s$.  
 This   higher performance of $Q$ due to $IC_a$ supports our initial hypothesis on the opportunity of   asymmetric inter-layer coupling.     
It is also interesting to note that, when  fixing $\gamma$ to 1, $Q$ decreases w.r.t. the setting  with  redundancy-based resolution $\gamma(L,C)$  --- 
decrement of 11\% and 12\%   using $IC_a$ and $IC_s$, respectively.   

Table~\ref{tab:LART} shows results obtained from LART solutions. 
%\disc{qui dobbiamo essere + incisivi: ossia dire che stiamo utilizzando il source originale, dire che va oltre gli obiettivi usare implementazioni performanti dei metodi, che non facciamo valutazioni comparative....che meglio e' avere rius di un metodo che non averne proprio....} 
 (Due to memory-resource and efficiency issues shown by the currently available implementation of LART, we are able to report  results only on some networks). 
   We observe that  the relative performance difference  between $IC_s$ and $IC_a$ settings is consistent with results found in the GL evaluation; this difference is even extreme (0.98 or 0.99) on  \textit{EU-Air} and \textit{London}, which is likely due also to the different sizes of community structures detected by the two methods (cf. Sect.~\ref{sec:methods}).  
%%On these two datasets, we also observe a drastic reduction of $Q$ also when the inter-layer couplings are not considered ($\beta=0$), while in other datasets the value of $Q$ is comparable to those corresponding to $IC_s$ e $IC_a$.  
 
 Table~\ref{tab:EMCD} shows results obtained by \greedystar solutions. Also in this case, %with the only exception of \textit{Higgs-Twitter} on which effects on $Q$ are equivalent,  
 using $IC_a$ generally leads to better   $Q$  than  $IC_s$, regardless of the setting of $\gamma$. In particular, the observed increase is higher  in \textit{VC-Graders} and \textit{London}  (0.2),  followed by \textit{AUCS} (0.18) and \textit{Obama} (0.13). Moreover, when  fixing $\gamma$ to 1, in most cases $Q$ decreases (0.01-0.06) w.r.t. the setting  with  redundancy-based resolution $\gamma(L,C)$.

 \begin{table}[t!]
\caption{Multilayer modularity $Q$ on GL community structures}\label{tab:GL}
\centering
\scalebox{0.9}{
\begin{tabular}{|l||c|c||c|c|}
\hline
& \multicolumn{2}{c||}{$\gamma(L,C)$} & \multicolumn{2}{c|}{$\gamma=1$}
\\
\cline{2-5}
    & $IC_a$&$IC_s$& $IC_a$& $IC_s$\\
  \hline\hline
\textit{AUCS}&    0.41 &   0.37 &      0.39  &  0.35\\\hline
\textit{EU-Air}&0.04 &   0.03 &       0.04&    0.03\\\hline
 \textit{FAO-Trade}& 0.11& 0.03& 0.11& 0.03\\\hline
\textit{\fftwyt}&    0.50  &  0.42    &  0.42  &  0.34\\\hline
\textit{Flickr}&    0.32  &  0.31 &      0.28 &   0.27\\\hline
\textit{\ghsotw}&    0.40 &   0.40   &   0.35 &   0.35\\\hline
\textit{Higgs-Twitter}&    0.15&    0.13 &      0.14 &   0.12\\\hline
\textit{London}&    0.35  &  0.26&      0.34&    0.26\\\hline
\textit{Obama}&    0.43 &   0.32 &       0.43 &   0.32\\\hline
\textit{VC-Graders}&    0.54 &   0.53&        0.44 &   0.43\\\hline
    \end{tabular}
}
\end{table}%

  \begin{table}[t!]
\caption{Multilayer modularity $Q$ on LART community structures}
\label{tab:LART}
\centering
\scalebox{0.9}{
\begin{tabular}{|l||c|c||c|c|}
\hline
& \multicolumn{2}{c||}{$\gamma(L,C)$} & \multicolumn{2}{c|}{$\gamma=1$}
\\
\cline{2-5}
    & $IC_a$&$IC_s$& $IC_a$& $IC_s$\\
  \hline\hline
   \textit{AUCS}&  0.47&    0.19&         0.43 &   0.15\\\hline
\textit{EU-Air}&1.00 &   0.02&             1.00 &   0.02\\\hline
\textit{London}&    1.00&    0.01  &           1.00 &   0.01\\\hline
\textit{VC-Graders}&   0.30 &   0.28 &       0.22  &  0.20\\\hline
    \end{tabular}
}
\end{table}%

 \begin{table}[t!]
 %\color{blue}
\caption{Multilayer modularity $Q$ on \greedystar community structures}\label{tab:EMCD}
\centering
\scalebox{0.9}{
\begin{tabular}{|l||c|c||c|c|}
\hline
& \multicolumn{2}{c||}{$\gamma(L,C)$} & \multicolumn{2}{c|}{$\gamma=1$}
\\
\cline{2-5}
    & $IC_a$&$IC_s$& $IC_a$& $IC_s$\\
  \hline\hline
\textit{AUCS}&0.51     &0.33    & 0.50       & 0.32 \\\hline
\textit{EU-Air}&0.20 &0.14    & 0.20      &0.14    \\\hline
\textit{FAO-Trade}&0.02&0.03&0.02&0.03\\\hline
\textit{\fftwyt}& 0.47     & 0.41     &0.47    &0.41  \\\hline
\textit{Flickr}& 0.37     &0.35   & 0.31      &0.29   \\\hline
\textit{\ghsotw}& 0.64   & 0.63    & 0.61   &0.60   \\\hline
\textit{Higgs-Twitter}&0.58    &0.58    & 0.52       & 0.52\\\hline
\textit{London}& 0.46     &0.26  &0.46      &0.25    \\\hline
\textit{Obama}&0.42     &0.29    & 0.42       &0.29   \\\hline
\textit{VC-Graders}& 0.52    & 0.32  & 0.50       & 0.30  \\\hline
    \end{tabular}
}
\end{table}%

 Figure~\ref{fig:gammaICasbeta0} shows how $Q$ varies in function of the number ($k$) of clusters given as input to   PMM. 
 One major remark is that  $Q$ tends to decrease as $k$ increases. This holds consistently   for the configuration of $Q$ with symmetric inter-layer coupling; in fact, as discussed in Sect.~\ref{commSize}, the decrease of $Q$ for increasing $k$  depends on a combination of  decrease of the internal degree $d_L^{int}$, decrease of the symmetric inter-layer coupling factor $IC_s$, and     increase of the redundancy-based resolution factor $\gamma(L, C)$.
 Moreover, values of $Q$ corresponding to $IC_a$ tend to be close to the ones obtained for $IC_s$ on the large networks, while on the smaller ones, $IC_a$ trends  are above $IC_s$, by diverging for high $k$ in some cases; in particular,   in \textit{London} %and \textit{EU-Air},  
modularity for $IC_a$ follows a rapidly, roughly linear  increasing trend with $k$; even more evident is the divergence of the    $IC_s$ and $IC_a$ trends for \textit{AUCS}.  Again, as we previously discussed in   Sect.~\ref{commSize}, this is due to the bias term $\frac{|V_i|}{|C^{(i)}|}$ of $IC_a$, which increases with communities of smaller size. 
% In this case, it is possible to identify a community number from which the divergence can be detected, i.e. 6 communities for \textit{AUCS}, 40 communities for \textit{\fftwyt}, 5 communities for \textit{London}, and 3 communities for \textit{VC-Graders}. 
 %
Note that, from an inspection of  the behavior of $Q$ for higher regimes of $k$, we also found that  $Q$ values eventually tend to stabilize below 1.    
 As concerns  the   setting with $\gamma$ fixed to 1 (results not shown), while the trends of $Q$ for $IC_a$ and for $IC_s$ do not change significantly, the values are typically lower than those obtained with redundancy-based resolution, which is again consistent with results observed for GL, LART and \greedystar evaluations.

 \begin{figure*}[t!]
\centering 
\begin{tabular}{ccc}  
\includegraphics[height=3cm, width=5.8cm]{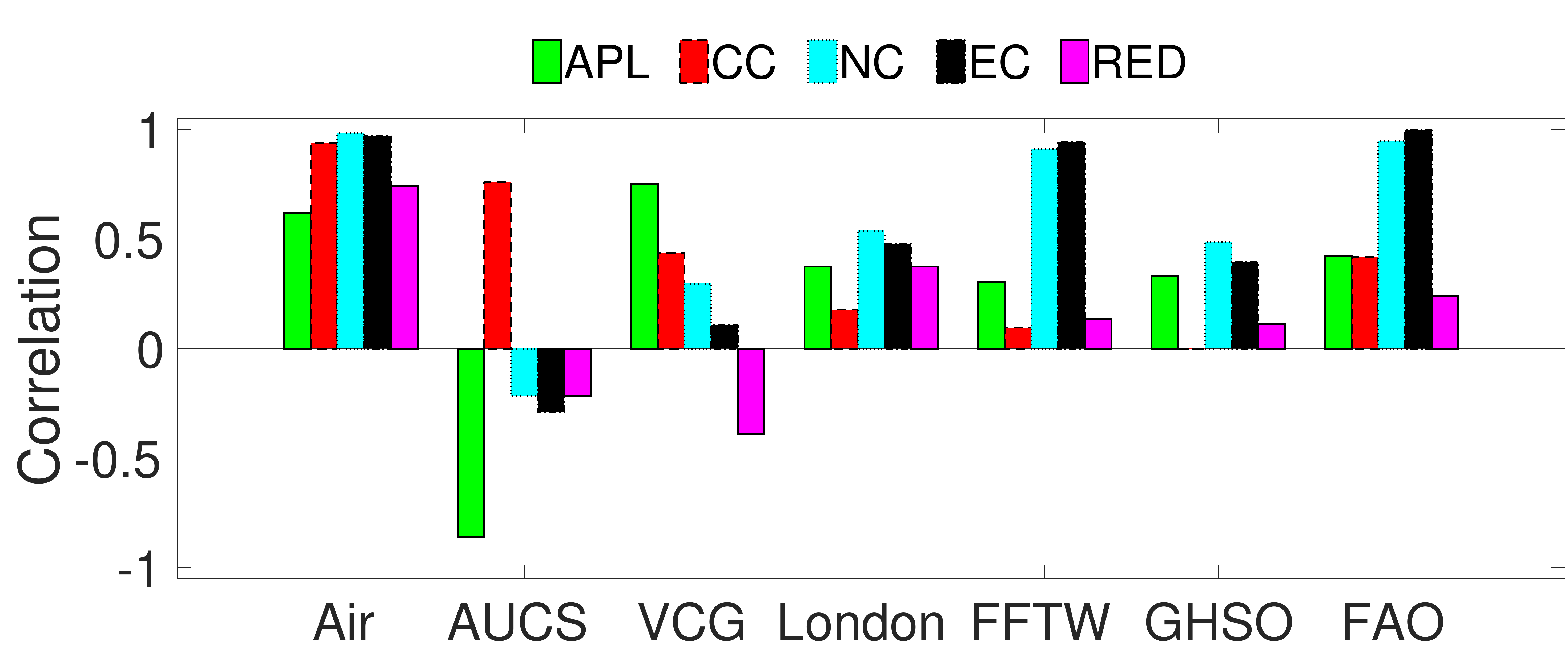}  \\  
(a) \\
\includegraphics[height=3cm, width=5.8cm]{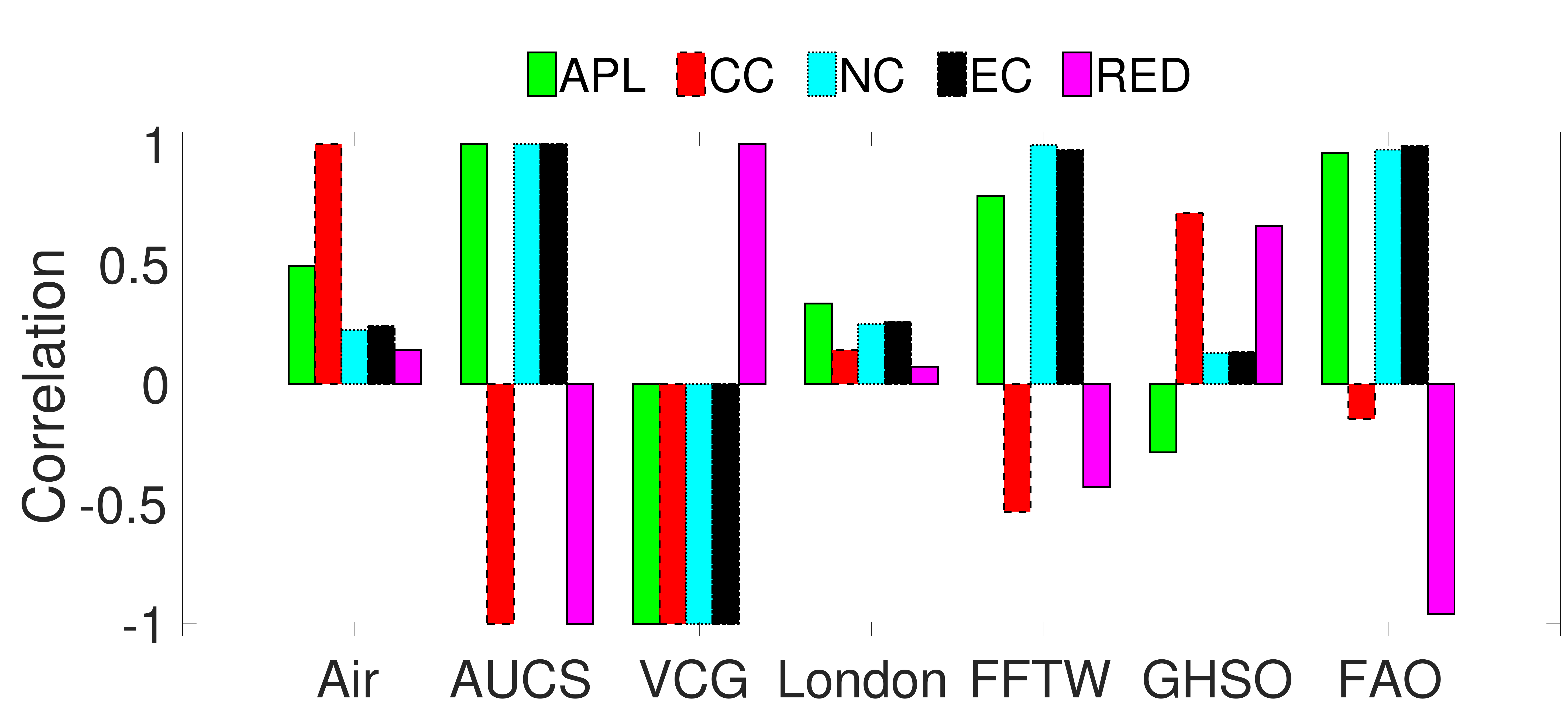} \\
(b) \\
\includegraphics[height=2.8cm, width=5.8cm]{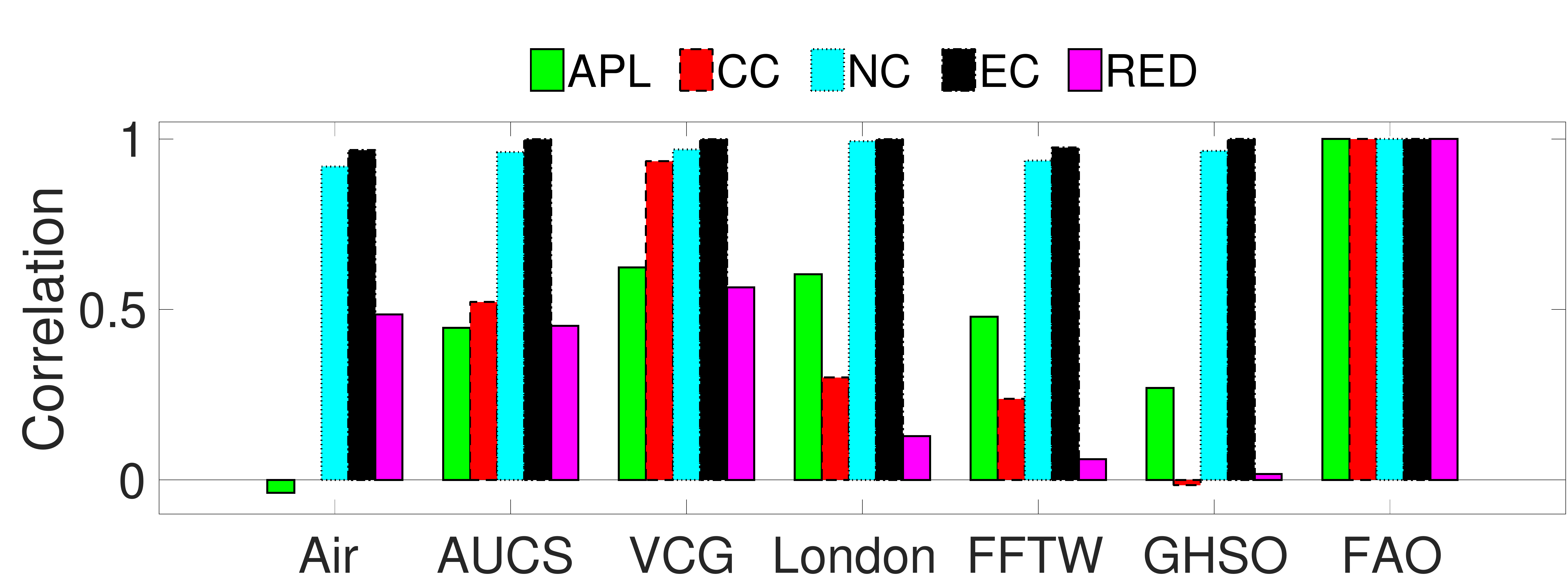}\\
 (c)
\end{tabular}
\caption{Pearson correlation coefficient between average path length (APL), clustering coefficient (CC), node-set coverage (NC), edge-set coverage (EC), and redundancy (RED) and the multilayer modularity $Q$ with $\gamma(L,C)$ and $IC_s$ computed on the solution found by (a) GL, (b) PMM, and (c) \greedystar for selected networks. Each statistic is computed at community-level}
\label{statGL}
\end{figure*}

 \begin{figure*}[t!]
  \centering
\begin{tabular}{c} 
% \subfigure[]{
% \includegraphics[height=4cm, width=4cm, keepaspectratio]{./../img/lart_u.eps}
% }
% \subfigure[]{
% \includegraphics[height=4cm, width=4cm, keepaspectratio]{./../img/lart_d.eps}
% }
\includegraphics[height=6cm, width=5.8cm, keepaspectratio]{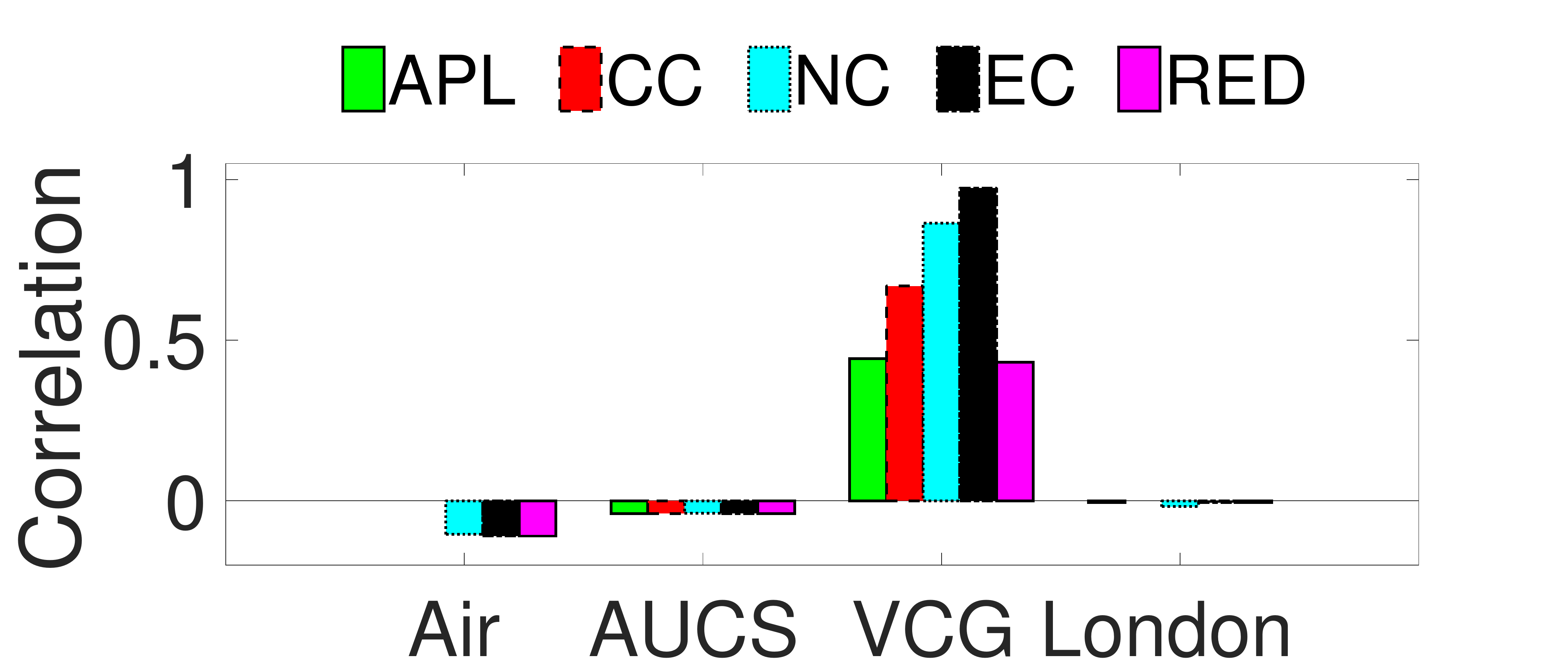} \\
(a) \\
\includegraphics[height=6cm, width=6cm, keepaspectratio]{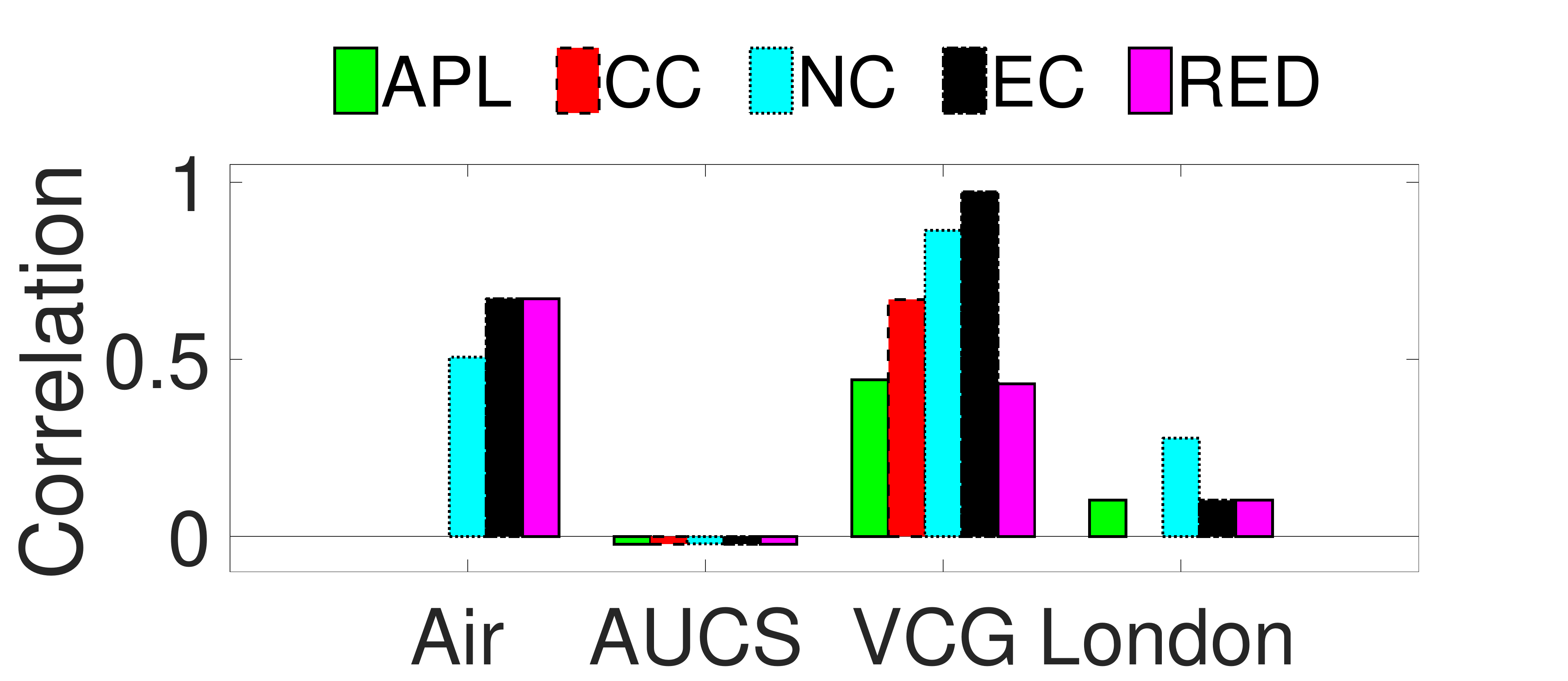} \\
(b) \\
\includegraphics[height=6cm, width=6cm, keepaspectratio]{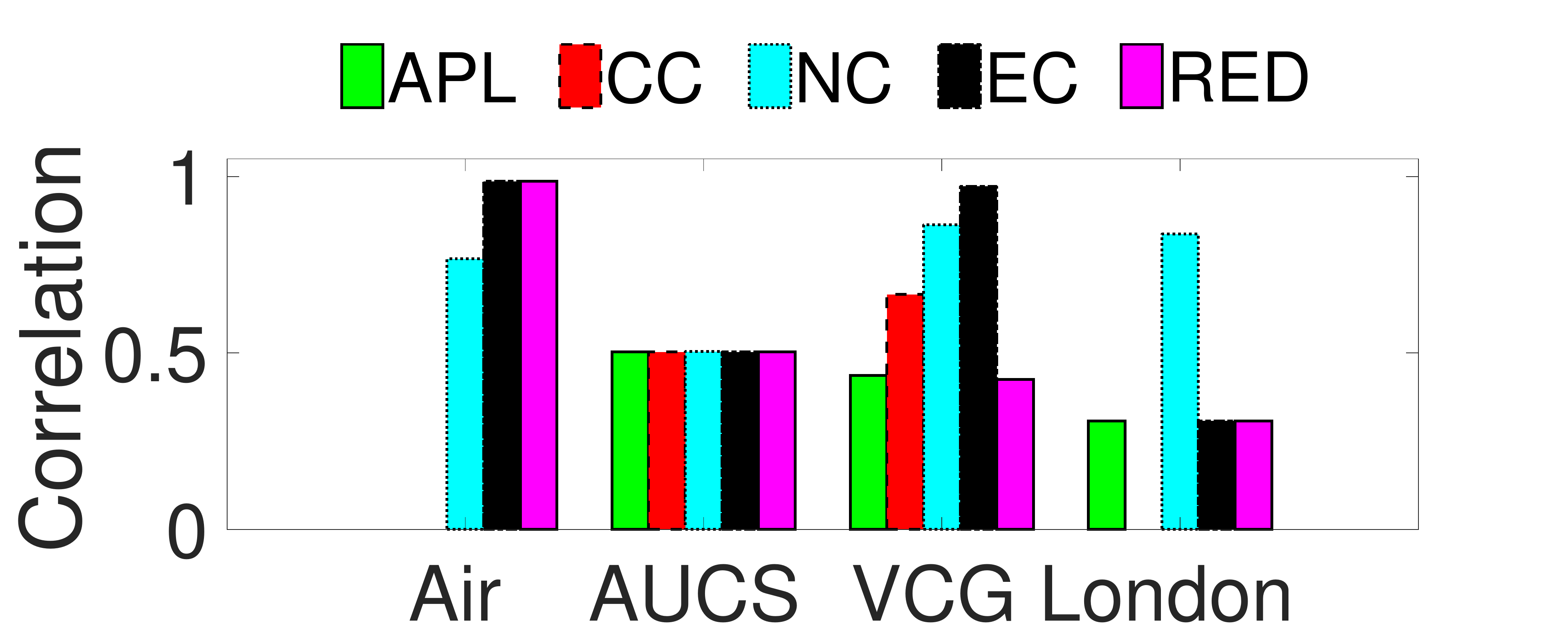} \\
(c) 
\end{tabular} 
\caption{Pearson correlation coefficient between average path length (APL), clustering coefficient (CC), node-set coverage (NC), edge-set coverage (EC), and redundancy (RED) and the multilayer modularity $Q$ with: %(a) $IC_s$, (b) $IC_a$, 
(a) $\gamma(L,C)$, (b) $\gamma(L,C)$ and $IC_s$, and (c) $\gamma(L,C)$ and $IC_a$ computed on the solution found by LART for the different real-world network datasets. Each statistics is computed at community-level %\disc{ricordiamoci di dire nel testo che le contingenze ex (a) e (b) sono state omesse in quanto similissime alle attuali (b) e (c)}
}
\label{stat1GL}
\end{figure*}

\textbf{Correlation analysis.\ } 
We investigated whether any correlation may exist at \textit{community-level} between the value of $Q$   and selected   statistics based on   structural characteristics of the input network. 
For this purpose, we focused on the average path length, clustering coefficient, redundancy, and node- and edge-set coverage,  for each community in an evaluation network; note that the latter two statistics are computed as, given a community $C$, the fraction of nodes (resp. edges) in a layer $L_i$ that belong to  $C$, averaged over all layers in the network.  
%%%%%%%%%%%%%%%
%%%%%  EC(C) = \frac{1}{|\L|} \sum_{L_i} \frac{|E_i(C)|}{|E_i|}
%%%%%%%%%%%%%%%

Figure~\ref{statGL} shows the correlation between each of the above structural characteristics and the values of  $Q$, with redundancy-based resolution factor $\gamma(L,C)$ and $IC_s$, on the solution found on selected networks by GL, \greedystar and PMM; for the latter,   $k$ was chosen as that corresponding to the best modularity performance. 
 Note also that  the correlation results obtained by $Q$ with $\gamma=1$ and $IC_s$, $\gamma=1$ and $IC_a$, $\gamma(L,C)$ only, and combination of  $\gamma(L,C)$ and $IC_a$, do not show significant differences, hence their presentation is discarded.   
Looking at the three plots in the figure, we observe a mid-high positive correlation of $Q$ with the topological measures  in most cases.  More in detail, in Fig. \ref{statGL} (a) the modularity of the solution found by GL on \textit{EU-Air} shows an average correlation of 0.85 with the other measures. Also, an average correlation of 0.95 and 0.96 is obtained between $Q$  and respectively node-set  and edge-set coverage on \textit{\fftwyt}. For \textit{AUCS}, $Q$ has a positive correlation of 0.76 with   clustering coefficient and a negative correlation with the other measures. For \textit{VC-Graders}, $Q$ shows a positive correlation with all measures except   with   redundancy. For \textit{London} and \textit{\ghsotw}, the correlation is up to 0.5.   For \textit{FAO-Trade}, $Q$ shows a higher correlation up to 1 with node-set and edge-set coverage, and a lower correlation up to 0.5 with average path length, clustering coefficient and redundancy.
Considering  Fig. \ref{statGL} (b), the multilayer modularity $Q$ of the solution found by PMM shows an average correlation of 0.99 with   clustering coefficient for \textit{EU-Air}, of 1 and 0.92 with   average path length, node-set and edge-set coverage for  \textit{AUCS} and \textit{\fftwyt}, respectively, and of 1 with   redundancy for \textit{VC-Graders}.
 On the contrary, a correlation of -1 is obtained between $Q$ and the clustering coefficient and redundancy for \textit{AUCS}, and between $Q$ and all measures except the redundancy for \textit{VC-Graders}.  For \textit{FAO-Trade}, $Q$ shows a positive correlation up to 1 with average path length, node-set and edge-set coverage, and a negative correlation up to -1 with the redundancy. A weakly negative correlation is shown between $Q$ and clustering coefficient. In the other cases, the correlation ranges between -0.5 and 0.5.
  Finally, considering Fig. \ref{statGL} (c), the multilayer modularity $Q$ of the solution found by \greedystar shows a very high   correlation   with node-set and edge-set coverage in all networks. Also, $Q$ shows a correlation with the average path length which is up to 0.5 in all networks, with the only exception of \textit{EU-Air} and \textit{GH-SO-TW}. For redundancy and clustering coefficient,  $Q$ obtains a high correlation   with clustering coefficient and redundancy in \textit{FAO-Trade} and with clustering coefficient in \textit{VC-Graders}.

    Figure \ref{stat1GL} shows the correlation between various settings of  $Q$ and the previously analyzed set of statistics for solutions obtained by LART. Looking at the plots,  $Q$ obtains the highest correlation with the edge-set coverage, followed by the node-set coverage, clustering coefficient and redundancy.  
    Overall, results by LART confirm the trends observed for GL and PMM, with even higher tendency to positive correlation in general. Remarkably, this particularly  holds when $Q$ involves the inter-layer coupling terms, with $IC_a$ leading to higher correlation than $IC_s$.

\subsection{Evaluation with ordered layers} 
In this section we focus on   evaluation scenarios that  correspond to the specification of an ordering of the set of layers. 
 We will present results on  
the real-world networks \textit{EU-Air} and \textit{Flickr}. The former was chosen because of its highest dimensionality (i.e., number of layers) over all datasets, the latter is a time-evolving multilayer network and   was chosen for evaluating the time-aware asymmetric inter-layer coupling.

\begin{table}[t!]
\caption{Multilayer modularity $Q$, with layer ordering,  from GL,  LART and \greedystar  community structures,  on \textit{EU-Air}}
\centering
\scalebox{0.75}{
\begin{tabular}{|l||c|c|c|c||c|c|c|c|}
\hline
&\multicolumn{4}{c||}{$\gamma(L,C)$}&\multicolumn{4}{c|}{  $\gamma=1$}\\\hline
&$IC^{\textrm{Suc}}_{ia}$&$IC^{\textrm{Suc}}_{oa}$&$IC^{\textrm{Adj}}_{ia}$&$IC^{\textrm{Adj}}_{oa}$&$IC^{\textrm{Suc}}_{ia}$&$IC^{\textrm{Suc}}_{oa}$&$IC^{\textrm{Adj}}_{ia}$&$IC^{\textrm{Adj}}_{oa}$\\\hline\hline
GL  & 0.786 &0.734 &0.512  & 0.511 & 0.783&0.729&0.504&0.503 \\\hline
  LART\!\!\!& 0.981 &0.972&0.665&0.656 &0.981&0.972&0.664 &0.656 \\\hline
   \greedystar\!\!\!&  0.997  & 0.998&  0.970& 0.969 & 0.997&0.999& 0.974 & 0.973\\\hline
    \end{tabular} 
    }
\label{tab:GL-LART-Air}
\end{table}%

  \begin{figure}[t!]
\centering
\begin{tabular}{ccc}
\hspace{-5mm}
\includegraphics[scale=0.07]{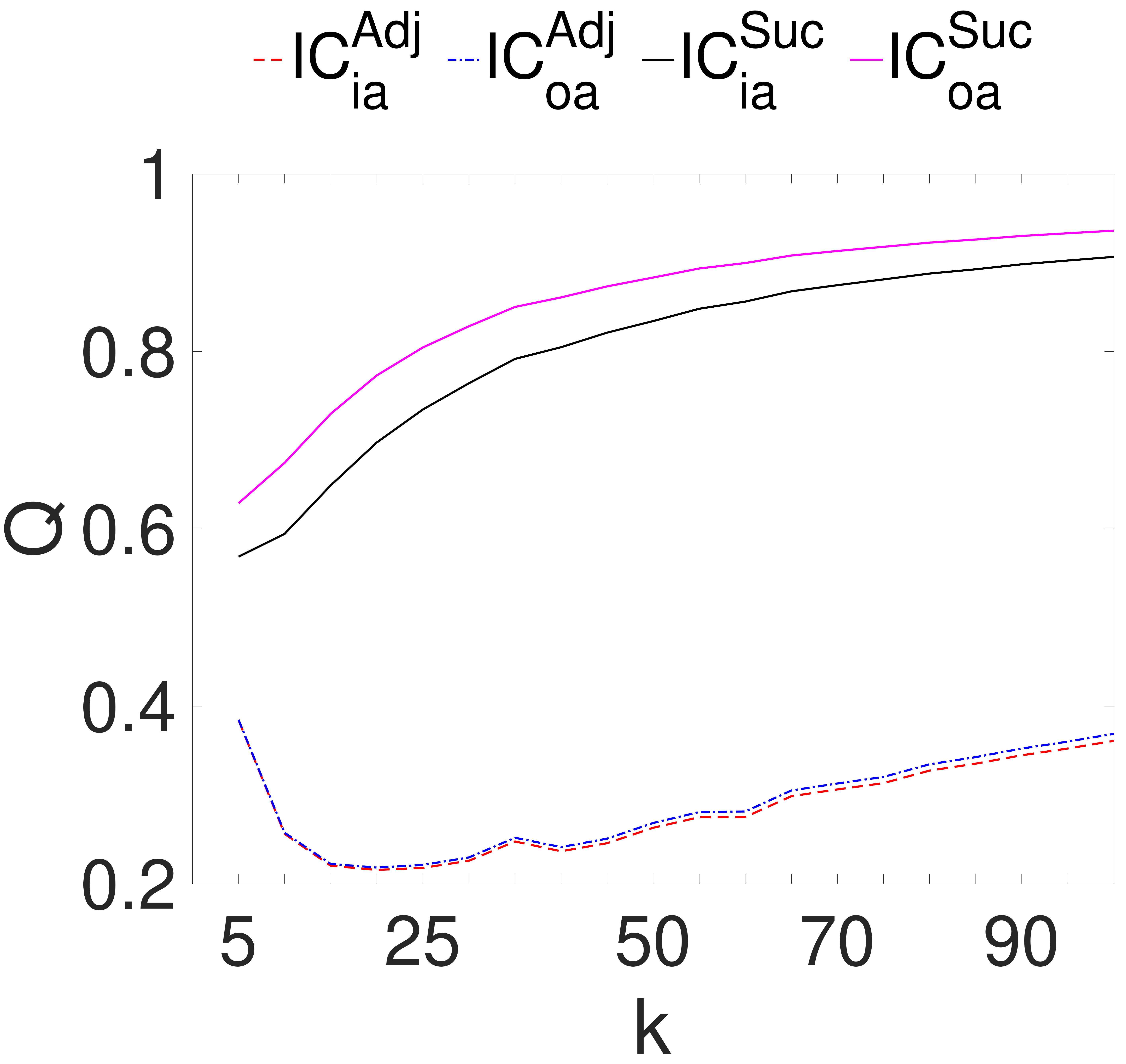}  &
\hspace{-3mm}
\includegraphics[scale=0.07]{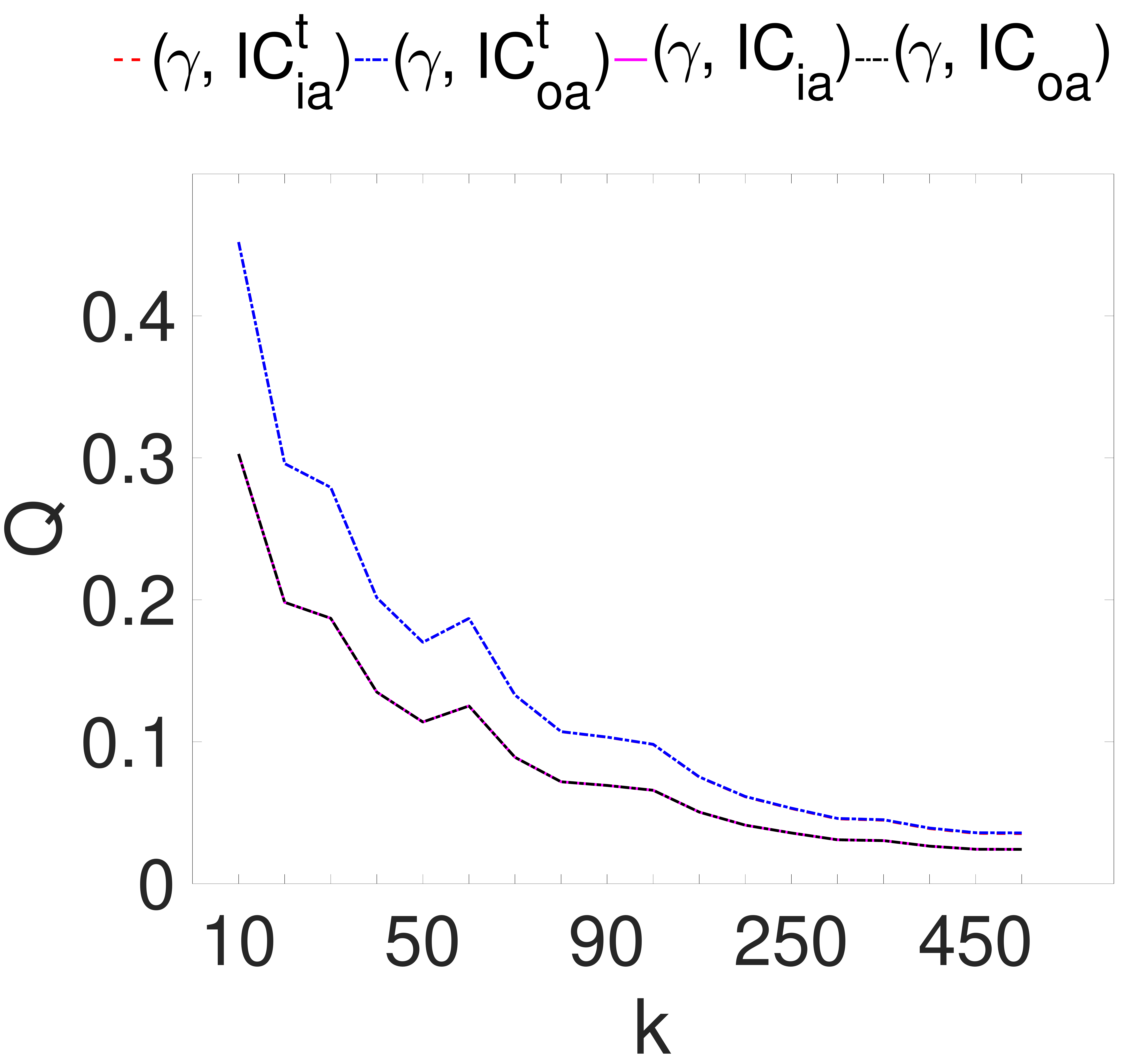} \\
 (a) \textit{EU-Air} (ascend.) &    (b) \textit{Flickr}
\end{tabular}
\caption{Multilayer modularity $Q$ of PMM solutions   with layer ordering.}% on (a) \textit{EU-Air} with ascending order and (b)  \textit{Flickr}}
\label{fig:orderedPMM}
\end{figure}

 \begin{figure*}[t!]
\centering
\begin{tabular}{cc}
\includegraphics[height=2.5cm, width=4cm]{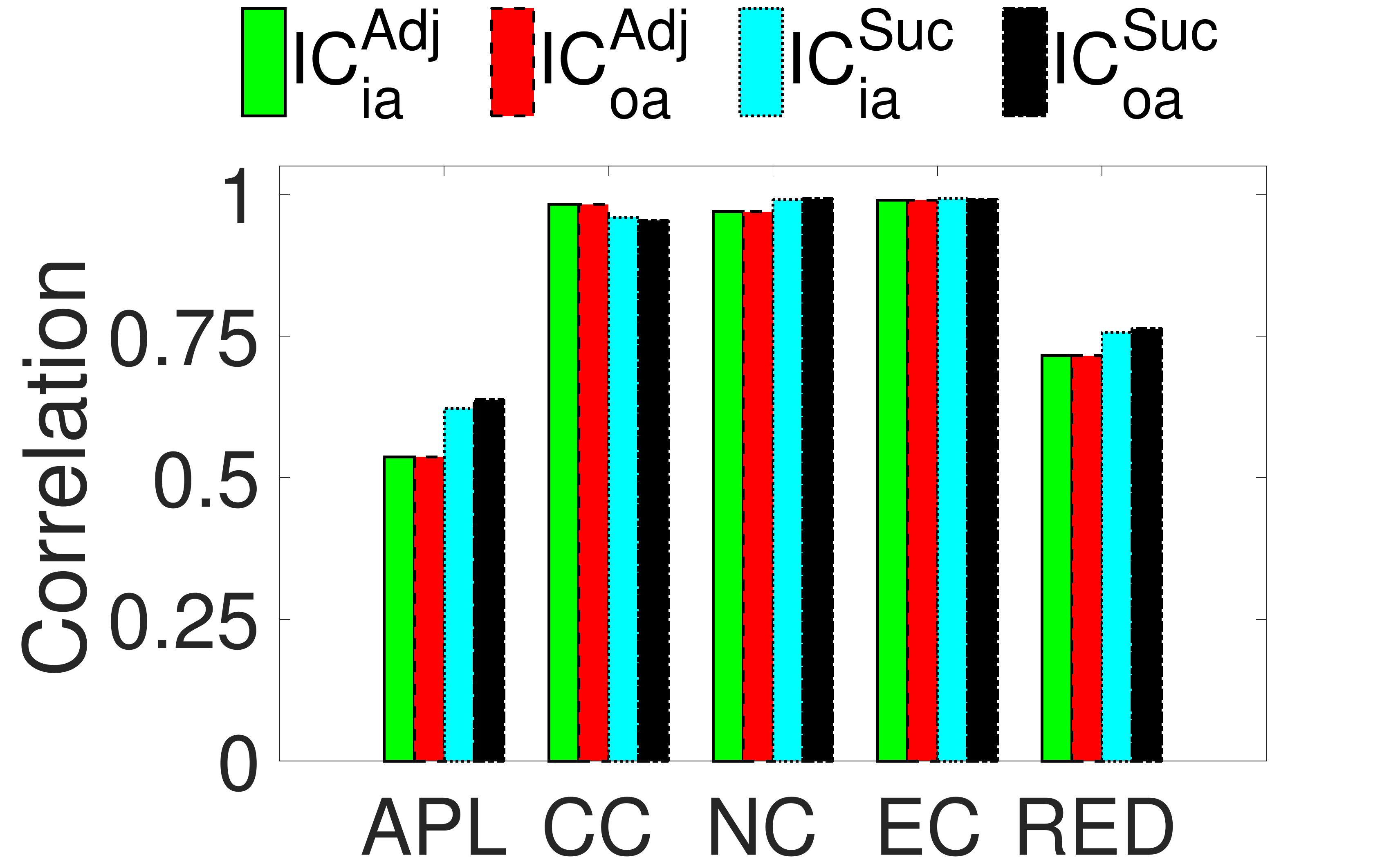} & 
\includegraphics[height=2.5cm, width=4cm]{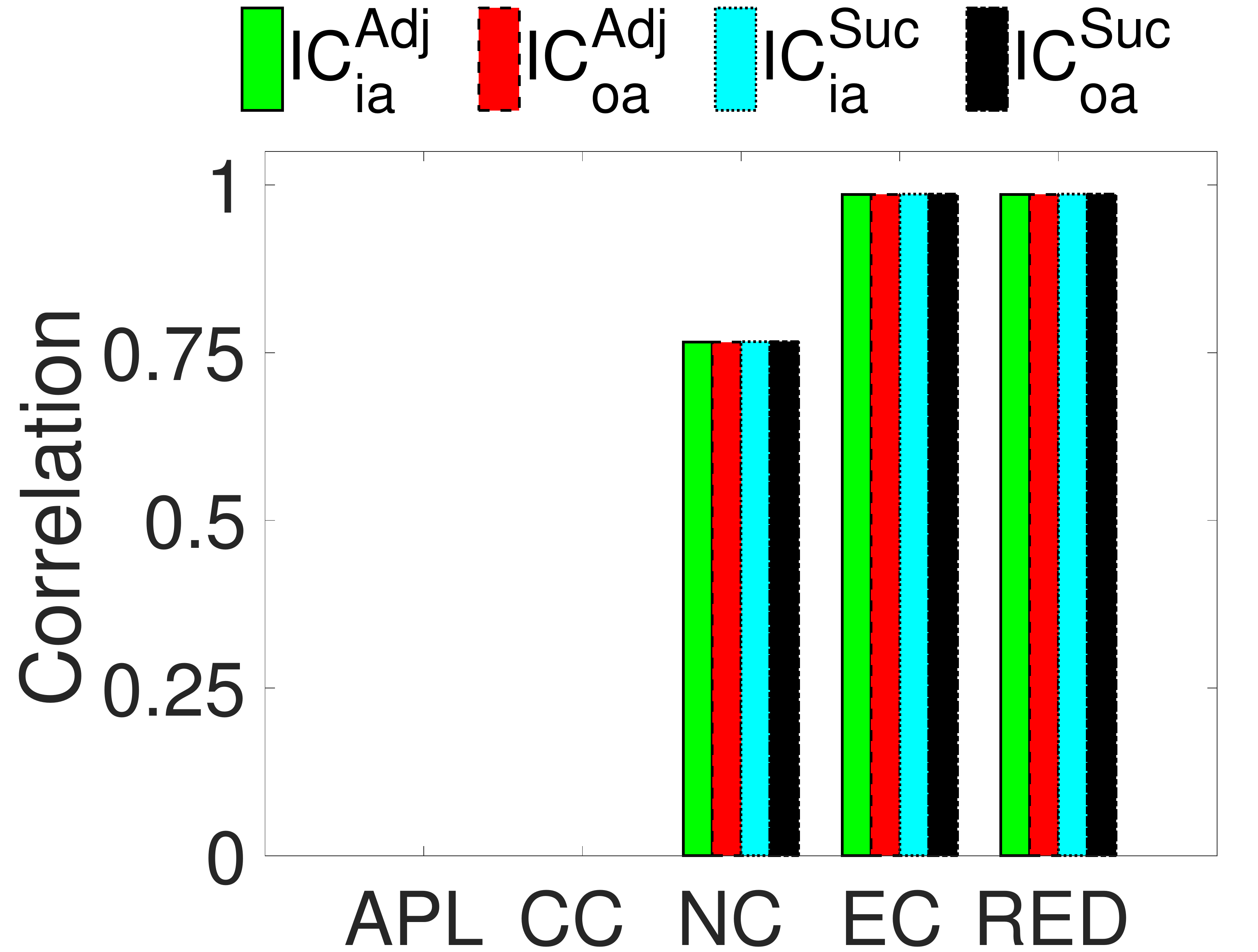} \\
 (a) & (b) \\
\includegraphics[height=2.5cm, width=4cm]{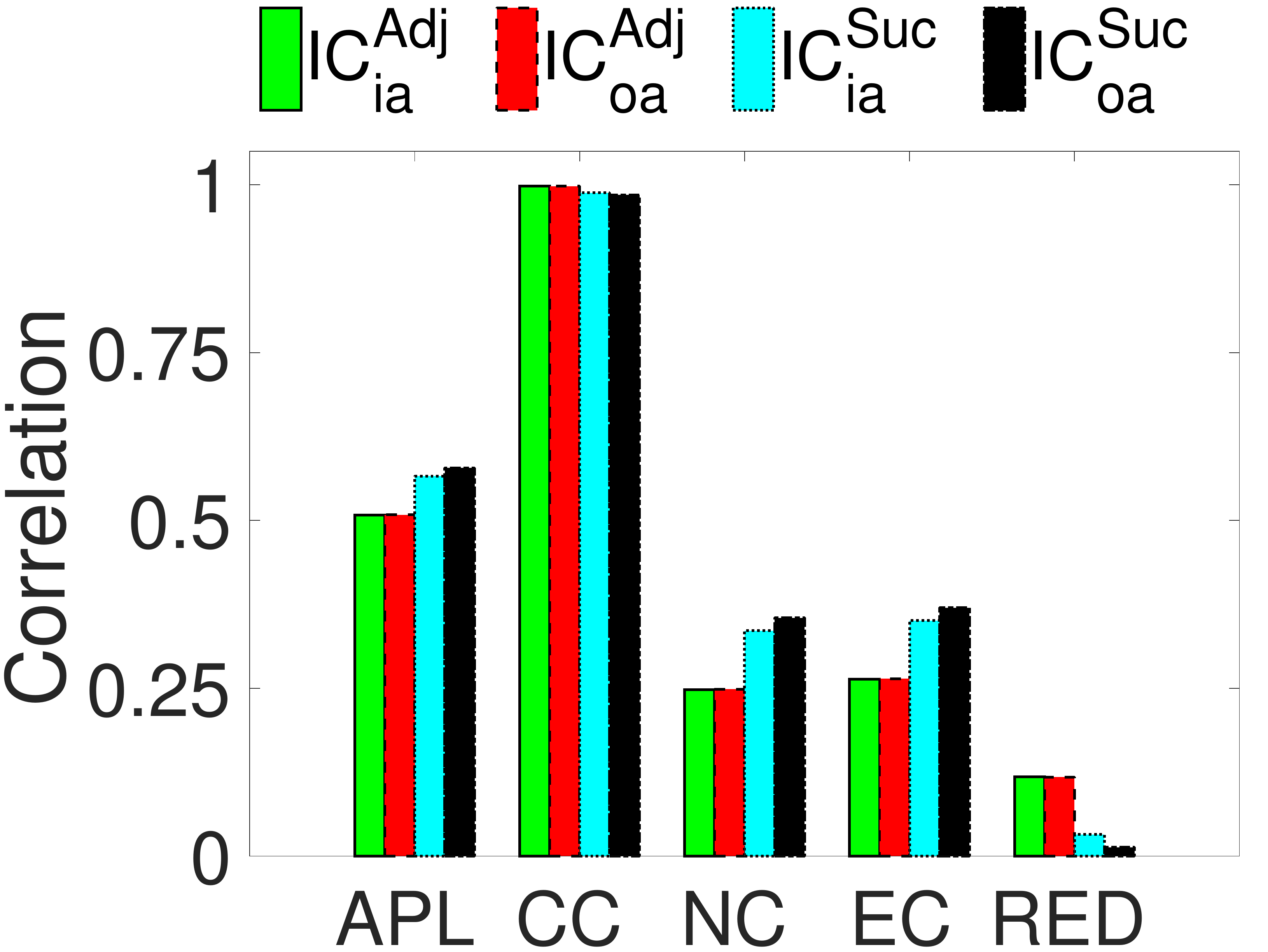} &
\includegraphics[height=2.5cm, width=4cm]{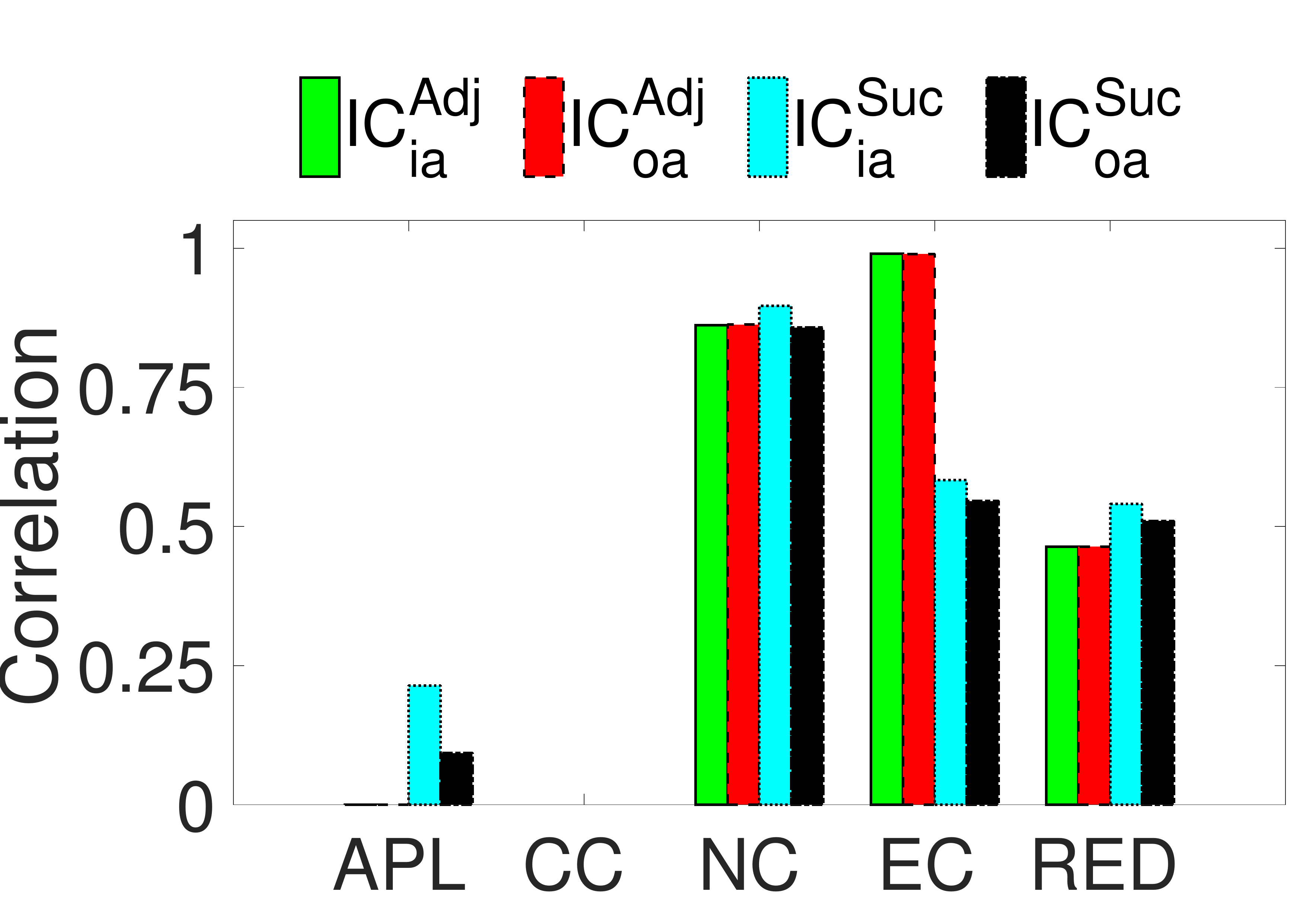} \\
   (c) & (d) \\
\end{tabular} 
\caption{Pearson correlation coefficient between average path length (APL), clustering coefficient (CC), node coverage (NC), edge coverage (EC), redundancy (RED), and the multilayer modularity $Q$ with $\gamma(L,C)$ and $IC_{ia}^{Adj}$, $IC_{oa}^{Adj}$, $IC_{ia}^{Suc}$, $IC_{oa}^{Suc}$, and ascendent layer ordering,  computed on the solution found by 
(a) GL,  (b) LART, (c) PMM, and (d) \greedystar 
on the \textit{EU-Air} network. Each statistics is computed at community-level}
\label{stat2GL}
\end{figure*}

Table~\ref{tab:GL-LART-Air} summarizes results by   GL, LART and \greedystar  on \textit{EU-Air}, corresponding to adjacent and succeeding-layer coupling.  
We observe that, regardless of the setting of the resolution factor, values of $Q$ with  succeeding-layer coupling are higher than the corresponding ones for the adjacent layer coupling scheme. This   suggests that  the impact on the inter-layer coupling term is higher when all ordered pairs of layers are taken into account, than when only adjacent pairs are considered. In this regard,  recall that the total degree of the multilayer graph, which normalizes the inter-layer coupling term as well, is properly computed according to the actual number of inter-layer couplings considered,  depending on whether   adjacent or   succeeding-layer scheme was selected. 

 The above result is also confirmed by PMM, as shown in Fig.~\ref{fig:orderedPMM}, where the plots for the succeeding-layer scheme superiorly bound those for the adjacent scheme, over the various $k$. %\disc{Come si spiega la diversita' rispetto alle curve del caso ordinato?}
 Note also that, while results on \textit{EU-Air} are shown only for  the ascendent layer ordering, by inverting this order we will have    a switch between results corresponding to the inner asymmetric case with results corresponding to the outer asymmetric case.  
% in Fig.~\ref{fig:orderedPMM}(b) correspond to a descending natural ordering of the layers, by inverting this order we will have clearly  a switch between results corresponding to the inner asymmetric case with results corresponding to the outer asymmetric case (see Fig.~\ref{fig:orderedPMM}(a)). 
   Moreover, Fig.~\ref{fig:orderedPMM}(b) compares  the effect of  asymmetric inter-layer coupling on \textit{Flickr} with and without time-awareness, for PMM solutions. Here we observe that  both $IC^t_{ia}$ and $IC^t_{oa}$ plots are above those corresponding to $IC_{ia}$ and $IC_{oa}$.   This indicates that considering a smoothing   term for the temporal distance between layers
   (Eq.~(\ref{eq:ICat}))   leads to an increase in modularity. 
   This general result is also confirmed by GL, LART and \greedystar (results not shown); for instance, GL achieved on \textit{Flickr}  modularity 
 0.462 for $IC^t_{ia}$,  0.468 for $IC^t_{oa}$, and 0.460 for $IC^t_{s}$, which compared to results shown in Table~\ref{tab:GL}  represent increments in $Q$ of    43\%. Similarly, \greedystar obtained on \textit{Flickr}  modularity 
 0.975 for $IC^t_{ia}$, $IC^t_{oa}$, and $IC^t_{s}$, which is higher than the corresponding values reported in Table \ref{tab:EMCD} for $IC_a$ and $IC_s$, respectively.

\textbf{Correlation analysis.\ } 
Analogously to correlation analysis performed for the unordered case,  
 we compare different settings of $Q$ with selected  statistics on topological properties.  Figures \ref{stat2GL} %--\ref{statEMCD} 
  show results on   \textit{EU-Air}   %with ascendent and descendent layer ordering, 
   obtained by GL, LART, PMM and \greedystar, respectively.  Again, for PMM, $k$ was set to the number of communities corresponding to the best modularity performance achieved by the method.  
 %redundancy-based resolution factor $\gamma(L,C)$ and $IC_{ia}^{Adj}$, $IC_{oa}^{Adj}$, $IC_{ia}^{Suc}$, $IC_{oa}^{Suc}$ computed on the solution found by GL (see Fig. \ref{stat2GL}), LART (see Fig. \ref{statLART}) and PMM (see Fig. \ref{statPMM}), and the average path length, clustering coefficient, node and edge coverage, and redundancy, on the \textit{EU-Air} network dataset with ascendent and descendent layer ordering. The $k$ value corresponding to the highest multilayer modularity is chosen for PMM in order to evaluate the correlation. Because the correlation obtained by $Q$ with $\gamma=1$ shows minimum variations, it is not visualised. The corresponding values of correlation are reported in 
%   ........ Table \ref{corrDescend}.   

As a general remark  $Q$ is always non-negatively  correlated with all topological measures. More specifically,  the correlation is highly positive with all measures, when GL is used,  and with all measures but average path length and clustering coefficient, when LART and \greedystar are used; for PMM, correlation is very high with clustering coefficient, and mid-low with the other measures. 
 % 
% The multilayer modularity $Q$ obtains a positive correlation with all other measures with the only exception of LART where $Q$ shows no correlation with average path length and clustering coefficient. 
 When equipped with   succeeding-layer coupling, % $Q$ is higher correlated  
 correlation is higher than in the   adjacent-layer setting with average path length (up to +0.14), node-set coverage (up to +0.02) and redundancy (up to +0.05) for the solution found by GL and \greedystar, and with average path length (up to +0.07), node-set coverage (up to +0.11) and edge-set coverage (up to +0.11) for the solution found by PMM. 
 %For the solution found by GL, the succeeding-layer coupling is more correlated than the adjacent layer coupling with the average path length up to 0.14, with the node coverage up to 0.02, with the edge coverage up to 0.003, and with the redundancy up to 0.05 (see Table \ref{corrDescend}). For the solution found by PMM, the succeeding-layer coupling is more correlated than the adjacent layer coupling with the average path length up to 0.07, with the node coverage up to 0.11, and with the edge coverage up to 0.11 (see Table \ref{corrDescend}).  
  We also found that  the layer ordering does not provide meaningful variations on the correlation values --- plots regarding descendent layer ordering are reported in   the {\bf \em Appendix}. %Even more, on LART,    succeeding-layer and adjacent-layer settings have the same effect on the correlation of $Q$ with the topological measures (i.e., 0.77  with   node-set coverage, 0.99 with  edge-set coverage and redundancy, 0 with the other two measures).

 %node-set coverage, edge-set coverage and redundancy independently from the layer ordering. In particular, $Q$ shows a correlation of 0.77 with the node coverage, and 0.99 with the edge coverage and redundancy.

\begin{figure}[t!]
\centering
\subfigure[\fftwyt]{
\includegraphics[scale=0.08]{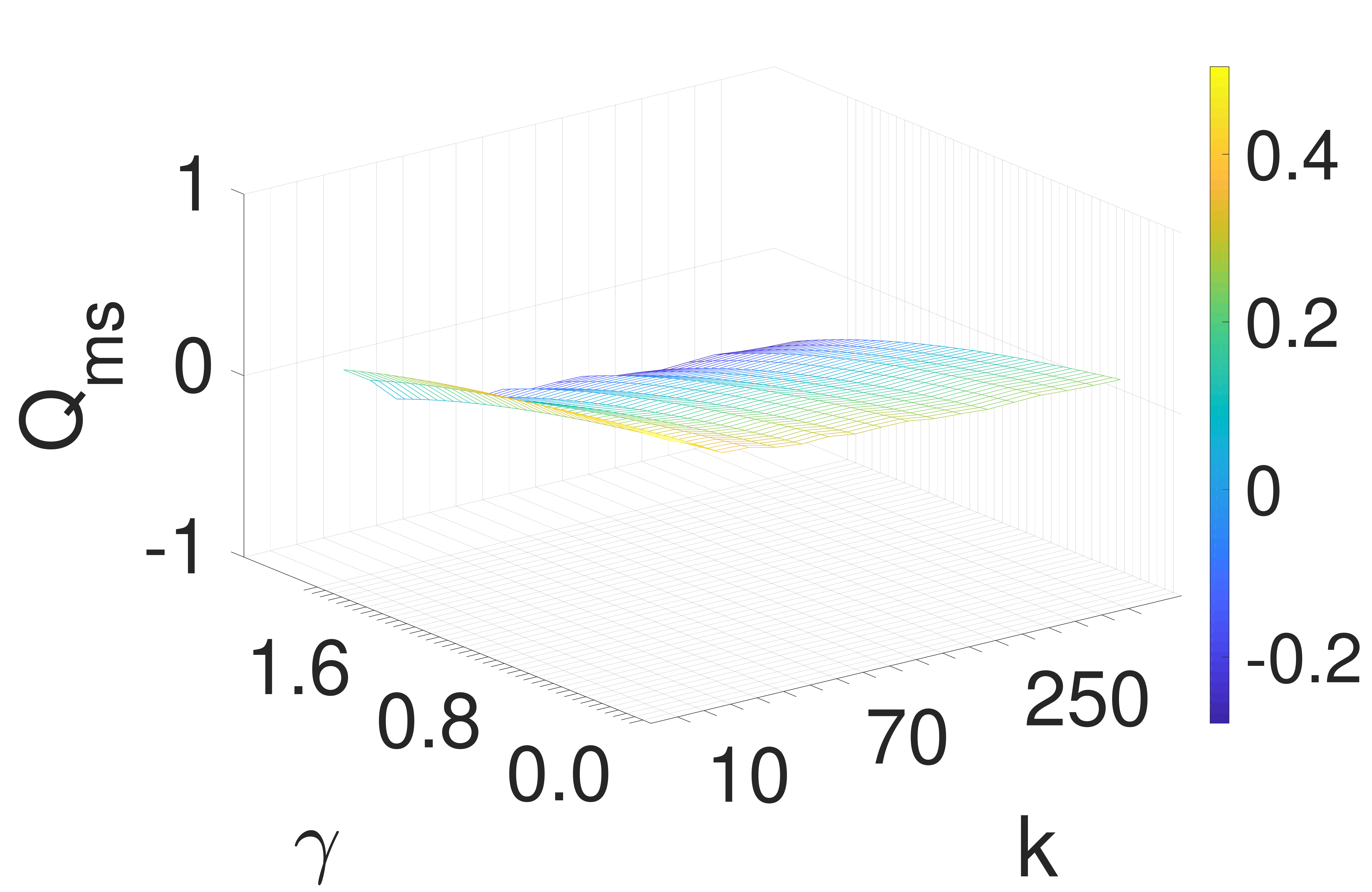} %
}\hspace{-0.4cm}
\subfigure[Flickr]{
\includegraphics[scale=0.08]{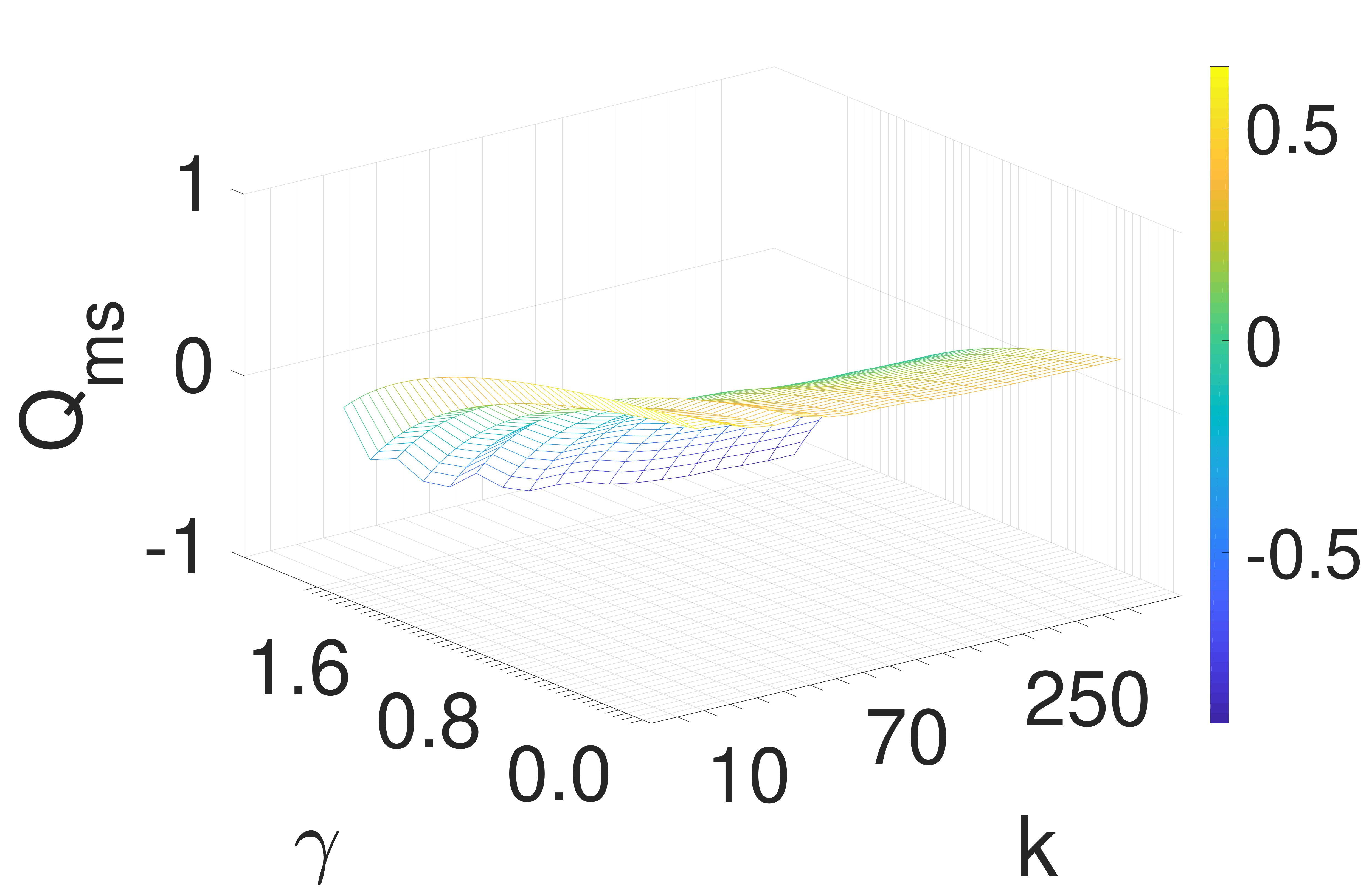} %
}
\subfigure[VC-Graders]{
\includegraphics[scale=0.08]{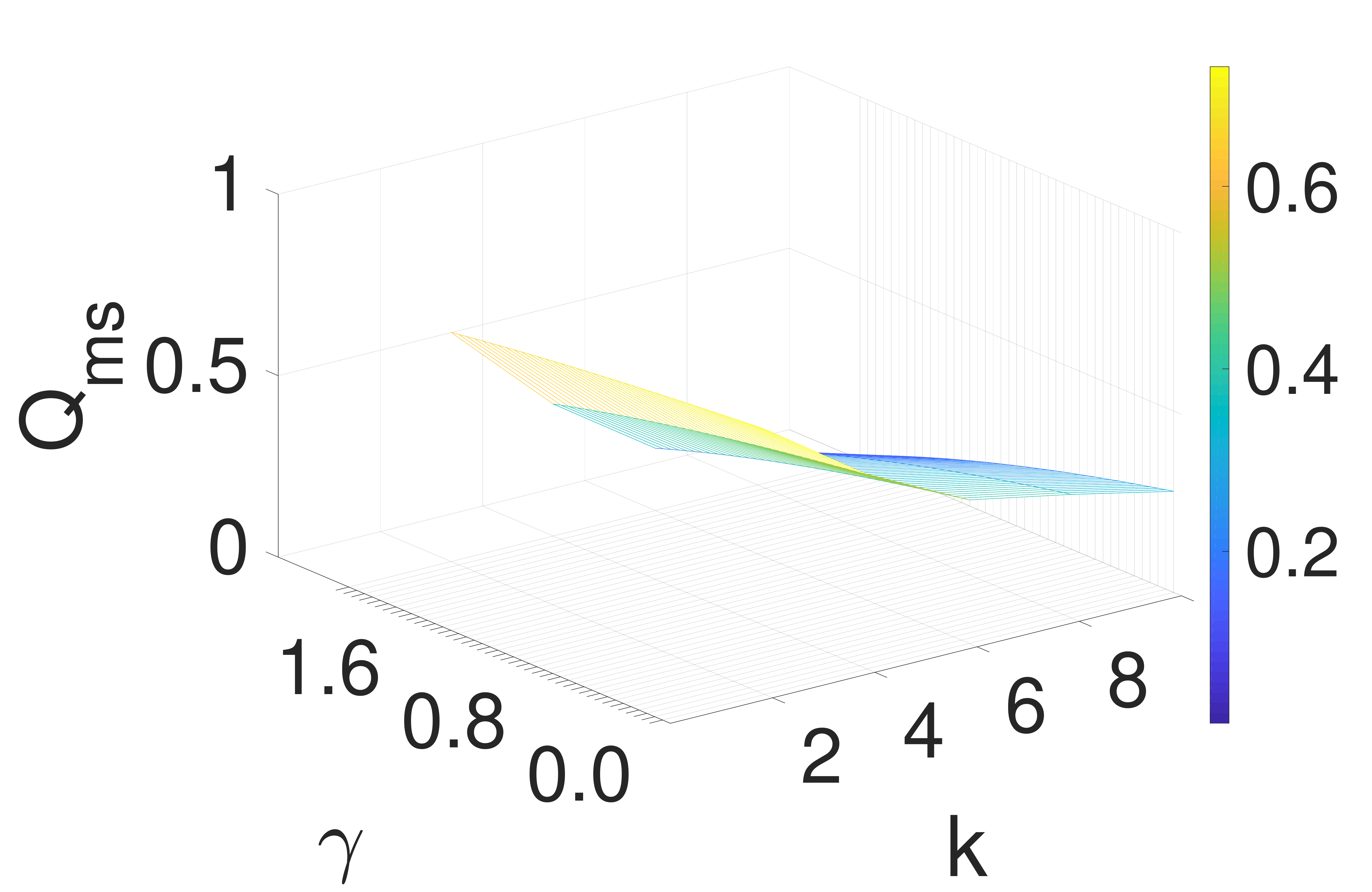} %
}\hspace{-0.4cm}
\subfigure[FAO-Trade]{
\includegraphics[scale=0.08]{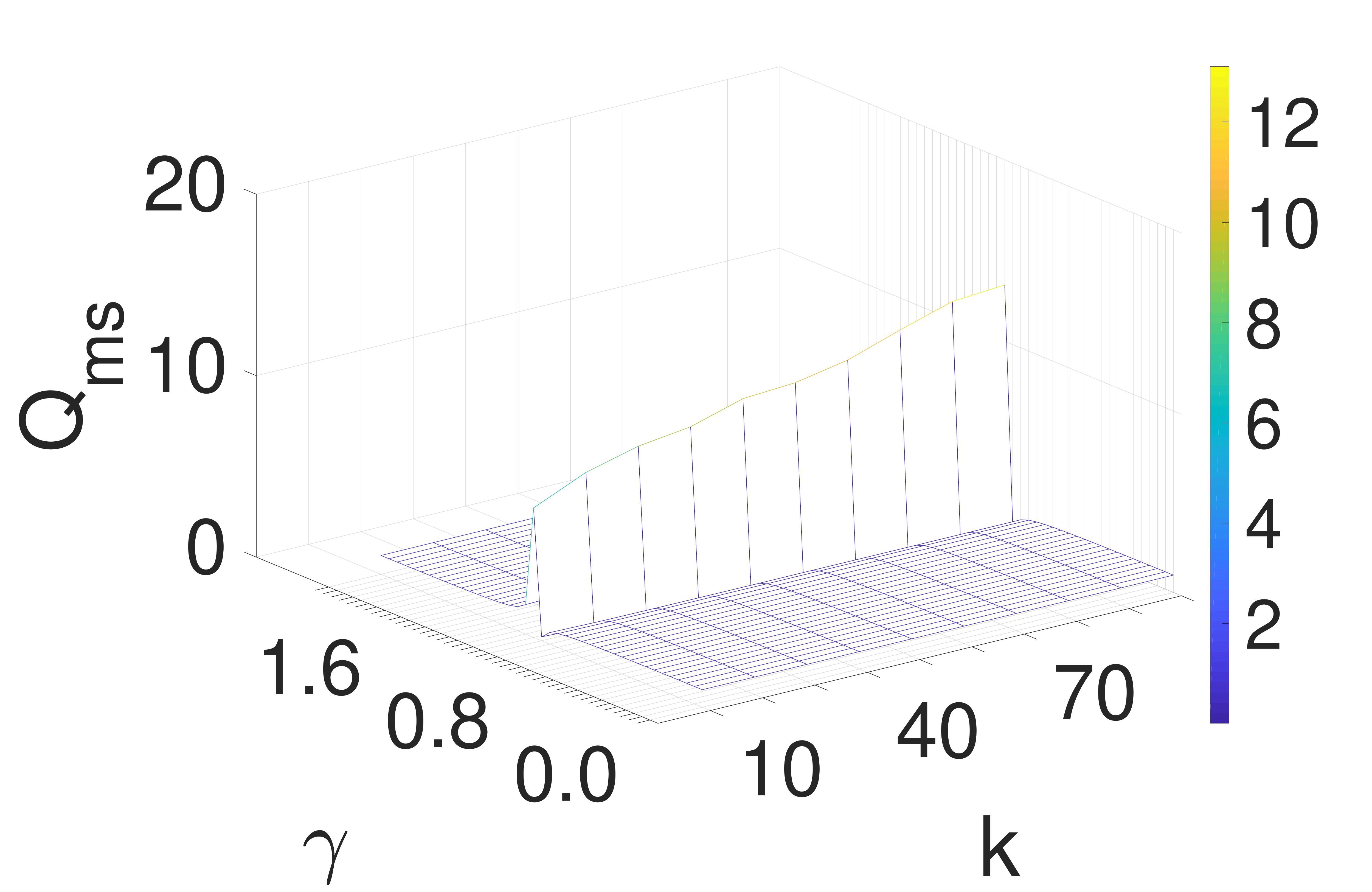} %
}\hspace{-0.4cm}
\subfigure[AUCS]{
\includegraphics[scale=0.08]{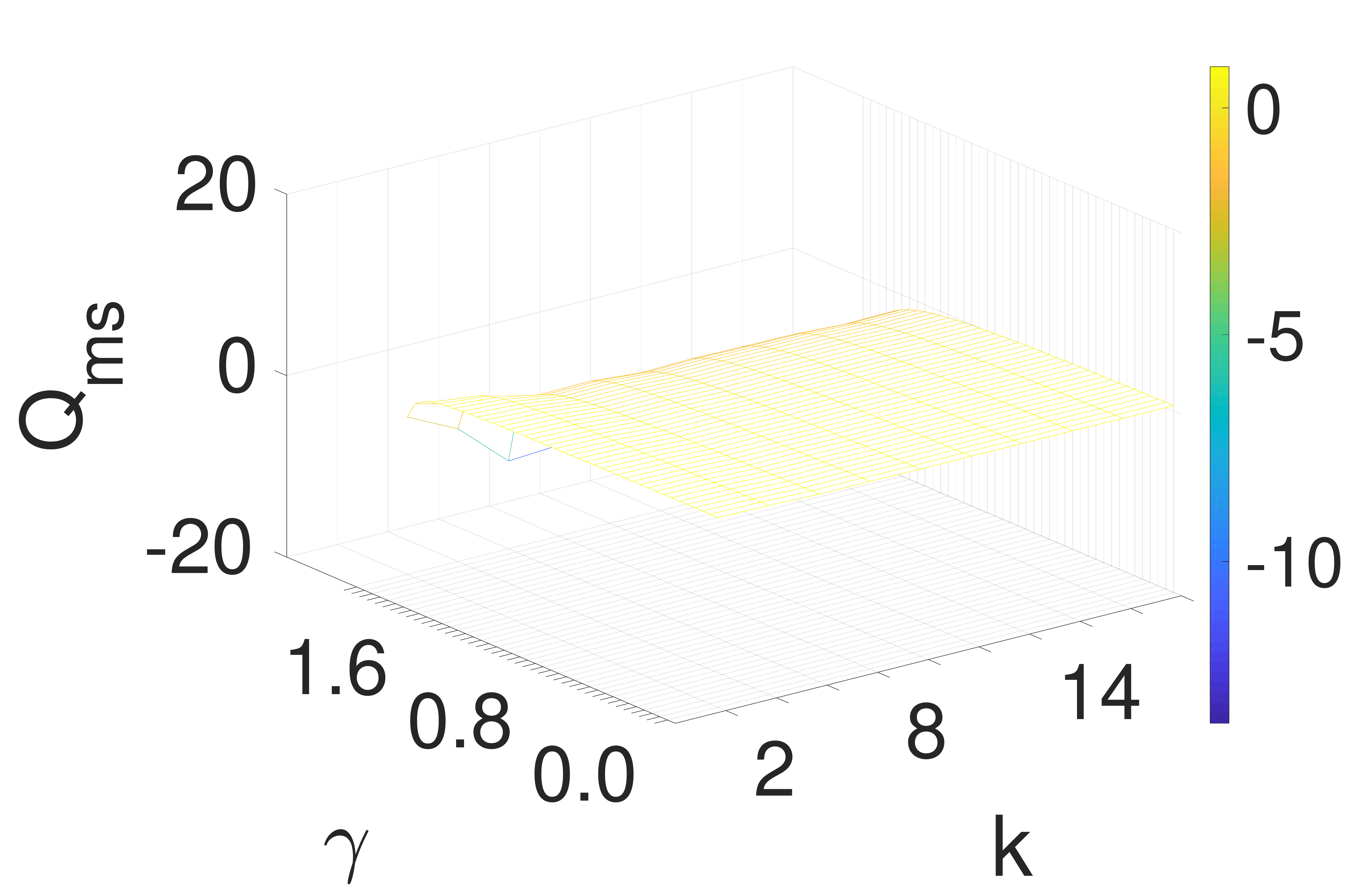}
}
\subfigure[Higgs-Twitter]{
\includegraphics[scale=0.08]{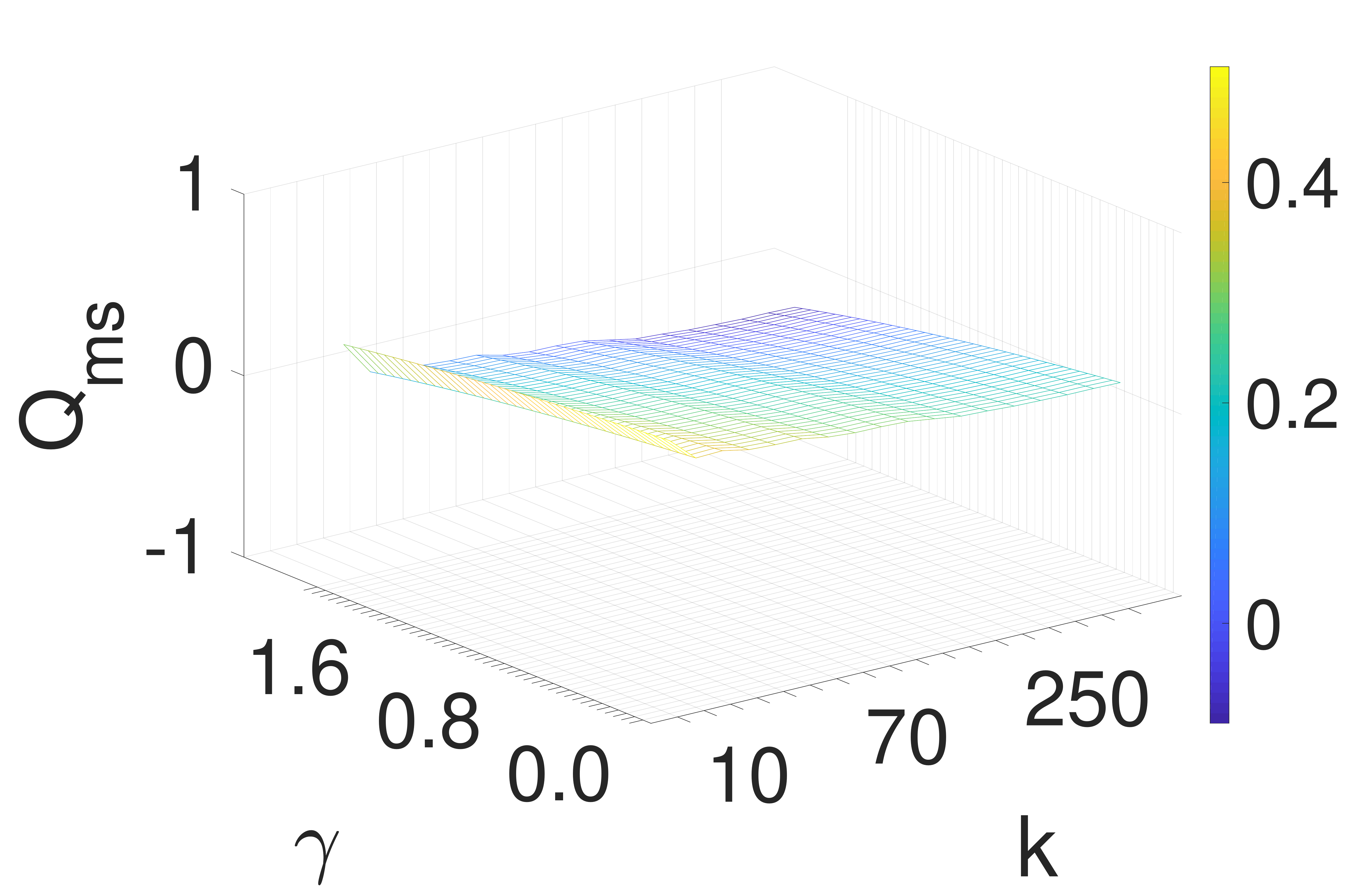} 
}\hspace{-0.4cm}
%\subfigure[London]{
%\includegraphics[scale=0.08]{./../img/london4.pdf}
%}
\subfigure[Obama]{
\includegraphics[scale=0.08]{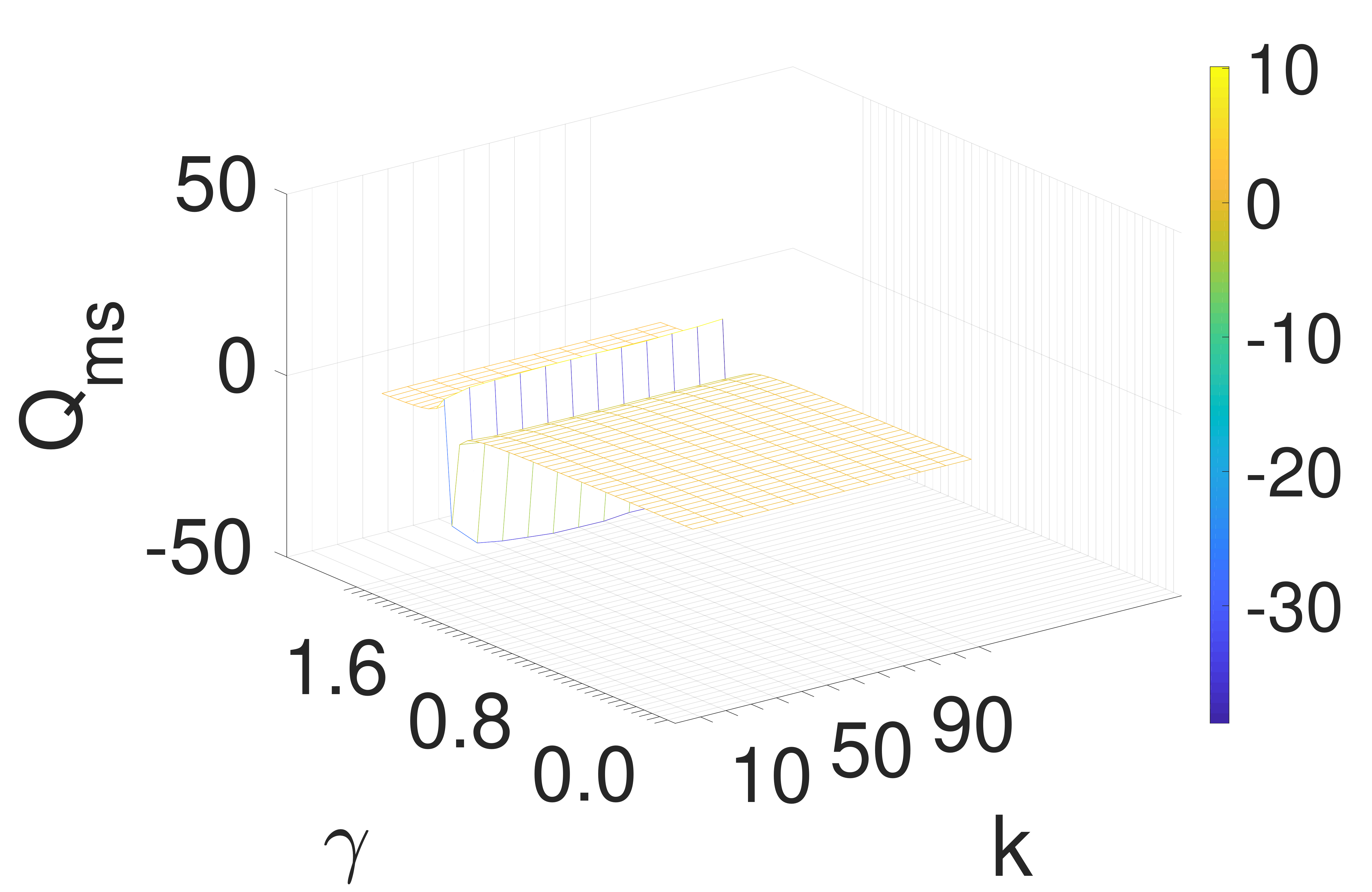} 
}\hspace{-0.4cm}
\subfigure[GH-SO-TW]{
\includegraphics[scale=0.08]{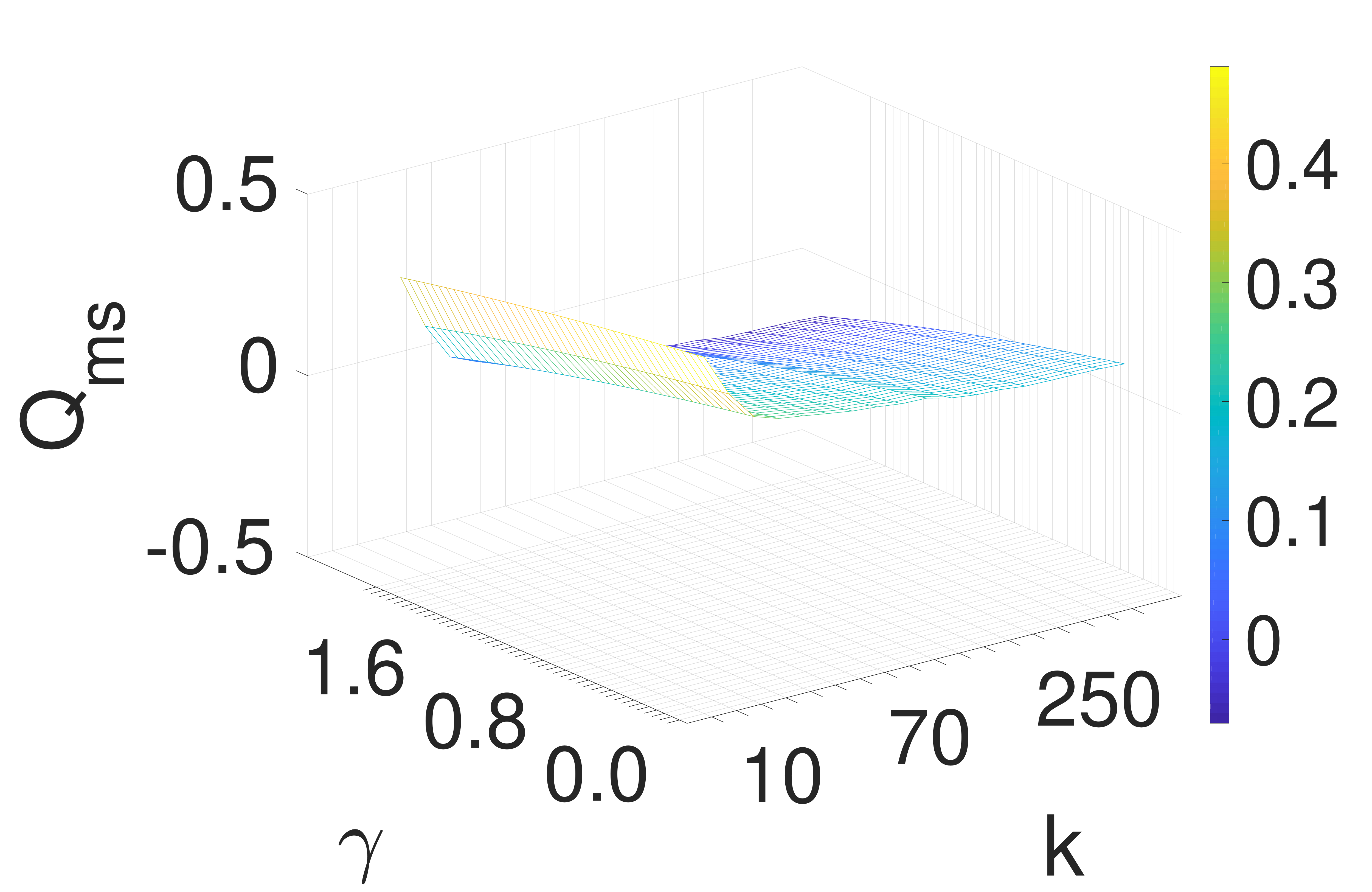}
}
\caption{Mucha et al.'s   modularity ($Q_{\textrm{ms}}$) by varying $\gamma$ with $\omega=1-\gamma$}
\label{fig:Mucha1}
\end{figure}

\begin{figure}[t!]
\centering
\subfigure[\fftwyt]{
\includegraphics[scale=0.08]{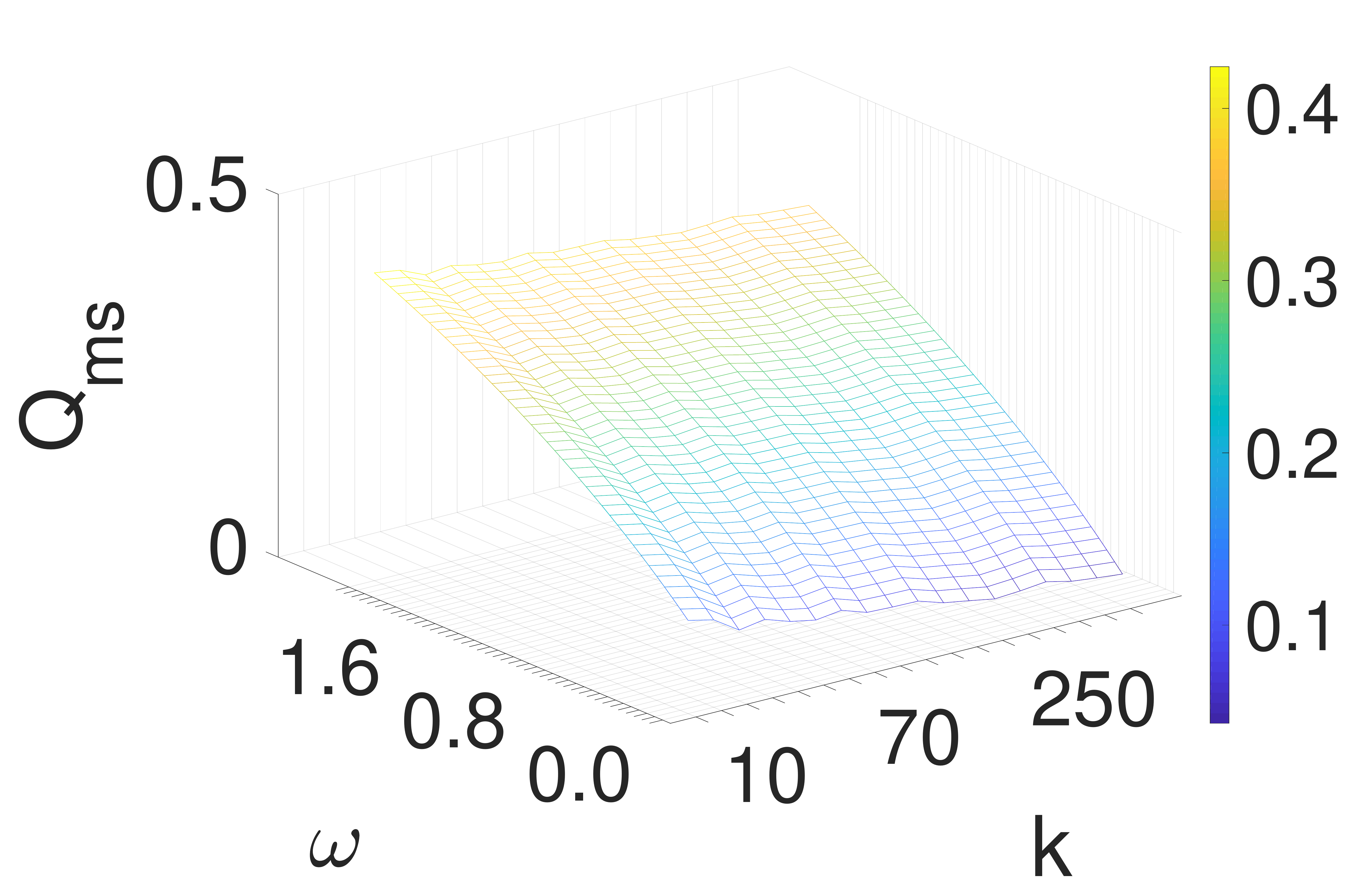}
}
\subfigure[Flickr]{
\includegraphics[scale=0.08]{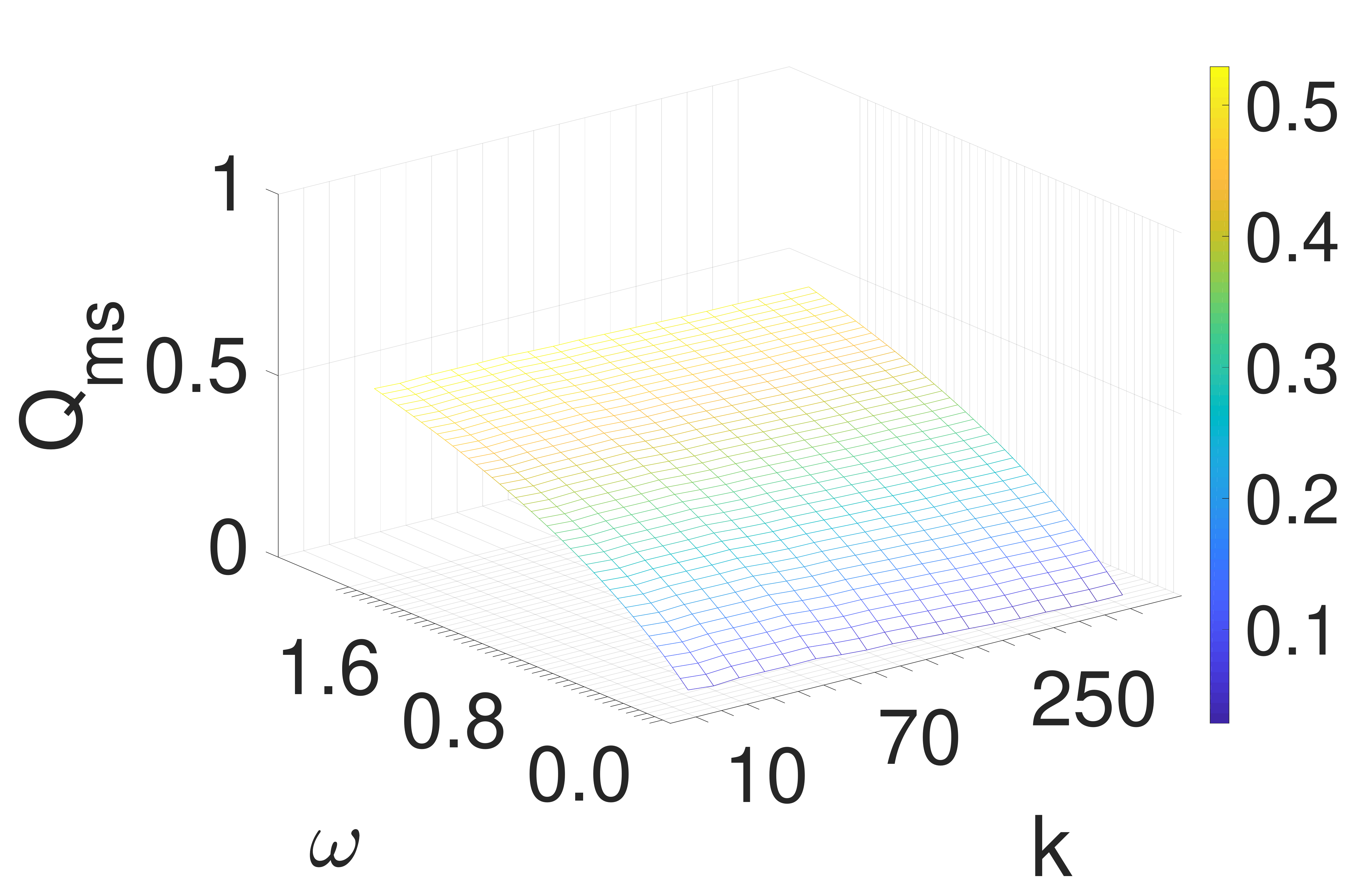}
}
\subfigure[FAO-Trade]{
\includegraphics[scale=0.08]{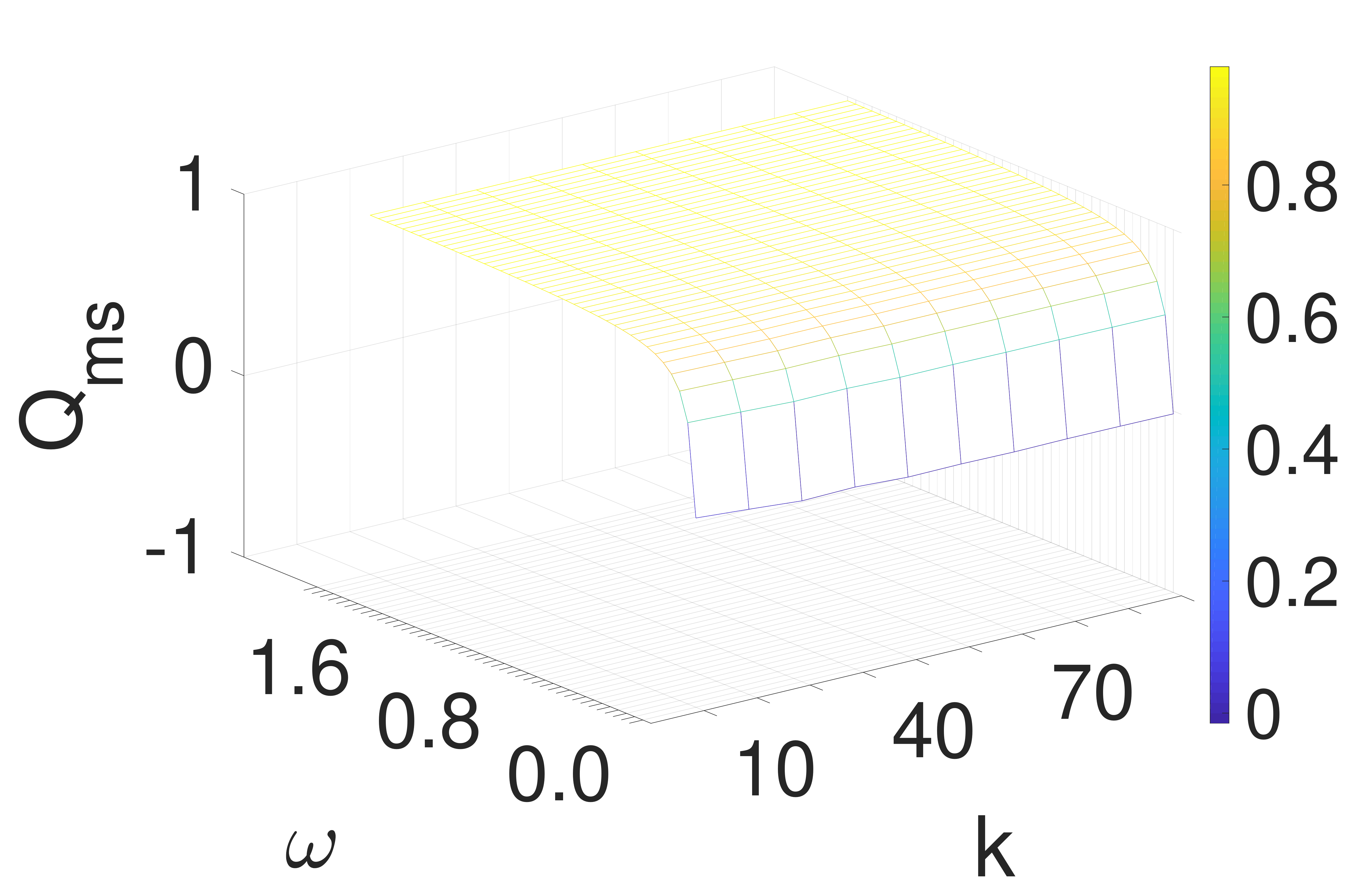}
}
%\subfigure[VC-Graders]{
%\includegraphics[scale=0.08]{./../img/vickers3.pdf}
%}
\hspace{-0.4cm}
\subfigure[AUCS]{
\includegraphics[scale=0.08]{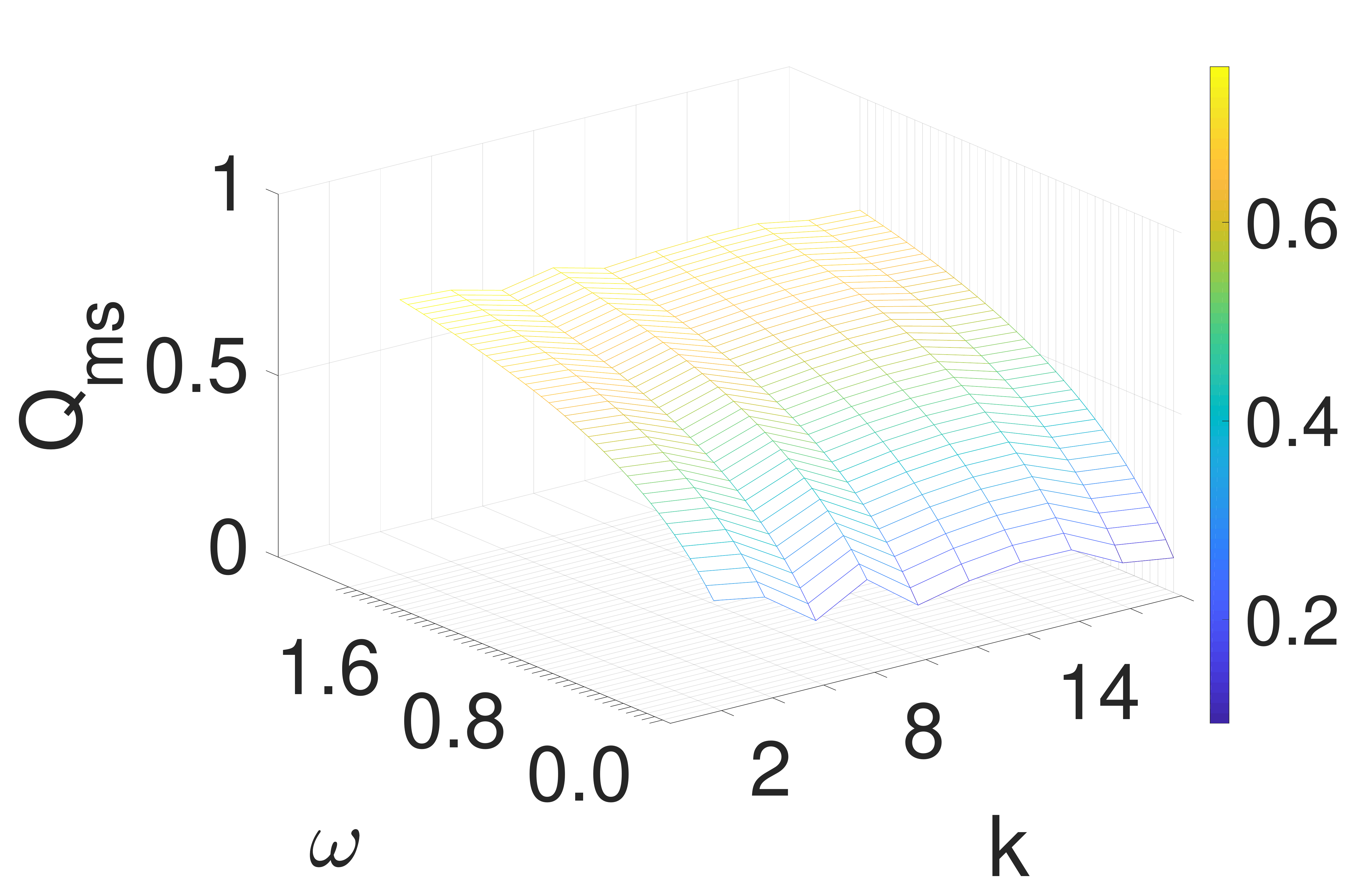}
}
\subfigure[Higgs-Twitter]{
\includegraphics[scale=0.08]{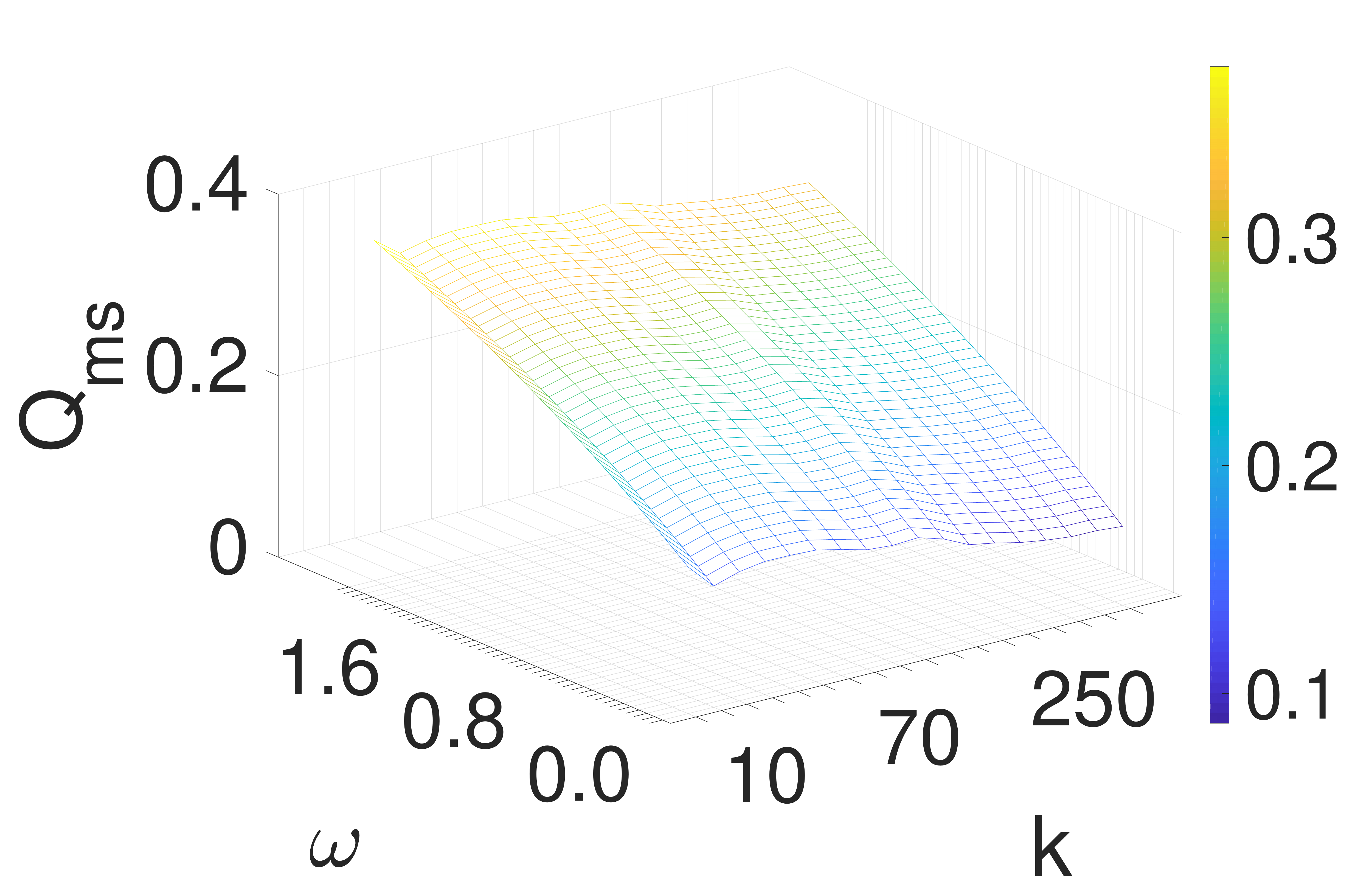}
}\hspace{-0.4cm}
%\subfigure[London]{
%\includegraphics[scale=0.08]{./../img/london3.pdf}
%}
\subfigure[Obama]{
\includegraphics[scale=0.08]{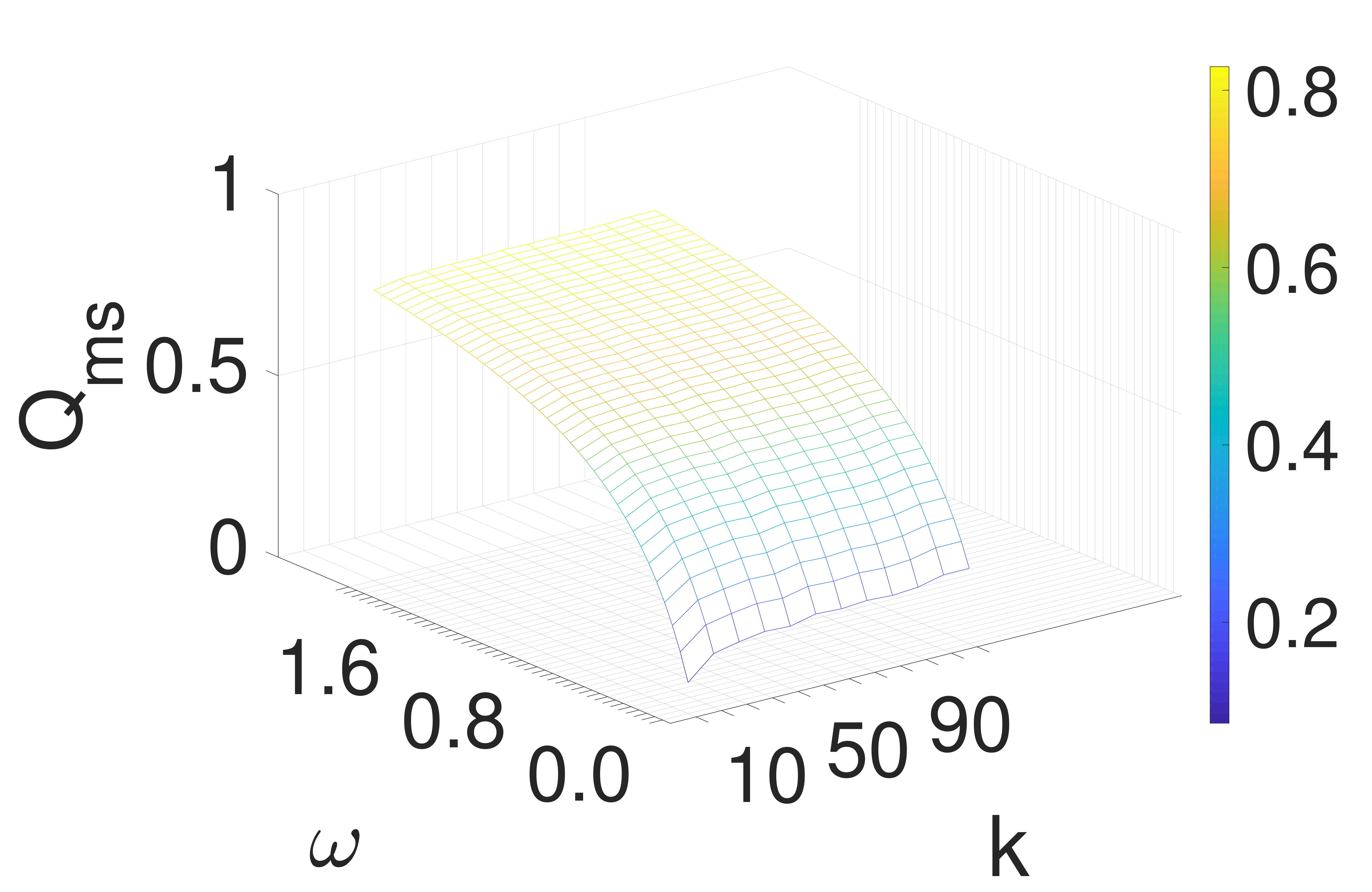}
}%\hspace{-0.4cm}
%\subfigure[GH-SO-TW]{
%\includegraphics[scale=0.08]{./../img/tw3.pdf}
%}
\caption{Mucha et al.'s   modularity ($Q_{\textrm{ms}}$) by varying $\omega$ with $\gamma=1$}
\label{fig:Mucha2}
\end{figure}

\subsection{Analysis of $Q_{\textrm{ms}}$ and comparison with $Q$} 
\label{sec:muchaanalysis}

We discuss here performance results obtained by the community detection algorithms with   $Q_{\textrm{ms}}$ as assessment criterion. We will refer to the default setting of unordered set of layers as stated in~\cite{Mucha10}. 
  
Using GL,  $Q_{\textrm{ms}}$ tends to decrease as $\gamma$ increases (while $\omega$ decreases, as it was varied with $\gamma$ as $\omega=1-\gamma$). This occurs monotonically in most datasets, within positive ranges (e.g.,    from 0.636 to 0.384 on \textit{\fftwyt},  from 0.525 to 0.391 on \textit{\ghsotw}) or including negative modularity for higher $\gamma$ (e.g., 
from 0.645 to -0.05 on \textit{Flickr}, from 0.854 to -4 on \textit{AUCS}). 
Remarkably, the simultaneous effect of  $\gamma$ and  $\omega=1-\gamma$  on  $Q_{\textrm{ms}}$ leads on some datasets (\textit{Obama}, \textit{EU-Air}, \textit{London}) to  a drastic degradation of modularity (down to much negative values) for some     $\gamma >1$, followed by a rapid increase to modularity of 1 as $\gamma$ increases closely to~2.   
 Analogous considerations hold for LART, PMM and \greedystar. For the latter method, the trend of drastic degradation of modularity followed by a rapid increase is only visible for \textit{EU-Air}. For PMM, the plots of Fig.~\ref{fig:Mucha1} show results by varying $k$, on the real-world datasets.
  Surprisingly, it appears that $Q_{\textrm{ms}}$ is  relatively less sensitive to the variation in the community structure than our $Q$. 
 % {\color{blue} 
  This is particularly visible in \textit{AUCS}, \textit{London} (not shown),  and \textit{Obama} %and \textit{FAO-Trade} 
   where $Q_{\textrm{ms}}$ shows no variation for increasing $k$. Also, it is worth noting that, for specific values of $\gamma$, $Q_{\textrm{ms}}$ may have an abrupt decrease  with very low peaks, as it happens   for % \textit{London} and 
    \textit{Obama}.  By contrast,  for \textit{FAO-Trade}, a value around 0.8 for $\gamma$ induces high modularity   which is stable for the different $k$ values.
  %}.
  
%As shown in the plots of Fig.~\ref{fig:Mucha2}, w
When varying $\omega$ within [0..2], with $\gamma=1$, $Q_{\textrm{ms}}$ tends to monotonically increase as $\omega$ increases.   This holds  consistently on all datasets (Fig.~\ref{fig:Mucha2} shows plots for some of them) with the only exception of \textit{FAO-Trade}, where $Q_{\textrm{ms}}$ is stable at 1 for $\omega >$ 0.8. Variations are always on positive intervals (e.g.,   from 0.248 to 0.621 on \textit{Flickr}, from 0.305 to 0.541 on \textit{\fftwyt}, from 0.136 to 0.356 on \textit{Higgs-Twitter}).

\vskip 0.5em
\subsubsection{\bf \em Comparison between $Q$ and $Q_{\textrm{ms}}$: qualitative evaluation on the solutions generated by the community detection methods}   
In the light of the above analysis, 
%It should be noted that, while we could not directly compare the values of $Q$ and $Q_{\textrm{ms}}$, 
 a few interesting remarks arise by observing the different behavior of $Q$ and $Q_{\textrm{ms}}$ over the same  community structure solutions, in function of the resolution and inter-layer coupling factors. From a qualitative viewpoint, the effect  of $\omega$ on  $Q_{\textrm{ms}}$ turns out to be opposite, in most cases, to the effect of our $IC$ terms on $Q$: that is, accounting more for the inter-layer couplings leads to increase    $Q_{\textrm{ms}}$, while this does not necessarily happen in $Q$. 
Less straightforward is comparing the use of a constant resolution for all layers, as done in $Q_{\textrm{ms}}$, and the use of variable (i.e., for each pair of layer and community)  resolutions, as done in $Q$. In this regard, we have previously observed that the use of a varying redundancy-based resolution factor improves $Q$ w.r.t. the setting $\gamma=1$. By coupling this general remark with the results (not shown) of an inspection of the values of $\gamma(L,C)$   in the computation of $Q$ on the various network datasets (which confirmed   that $\gamma(L,C)$ values span over its range,  in practice), we can conclude that a more appropriate   consideration of the term modeling the expected connectivity of community is realized in our $Q$ w.r.t. keeping the resolution as constant for all layers in $Q_{\textrm{ms}}$.

 \begin{figure}[t!] 
\centering
\includegraphics[height=4.8cm, width=4.8cm, keepaspectratio]{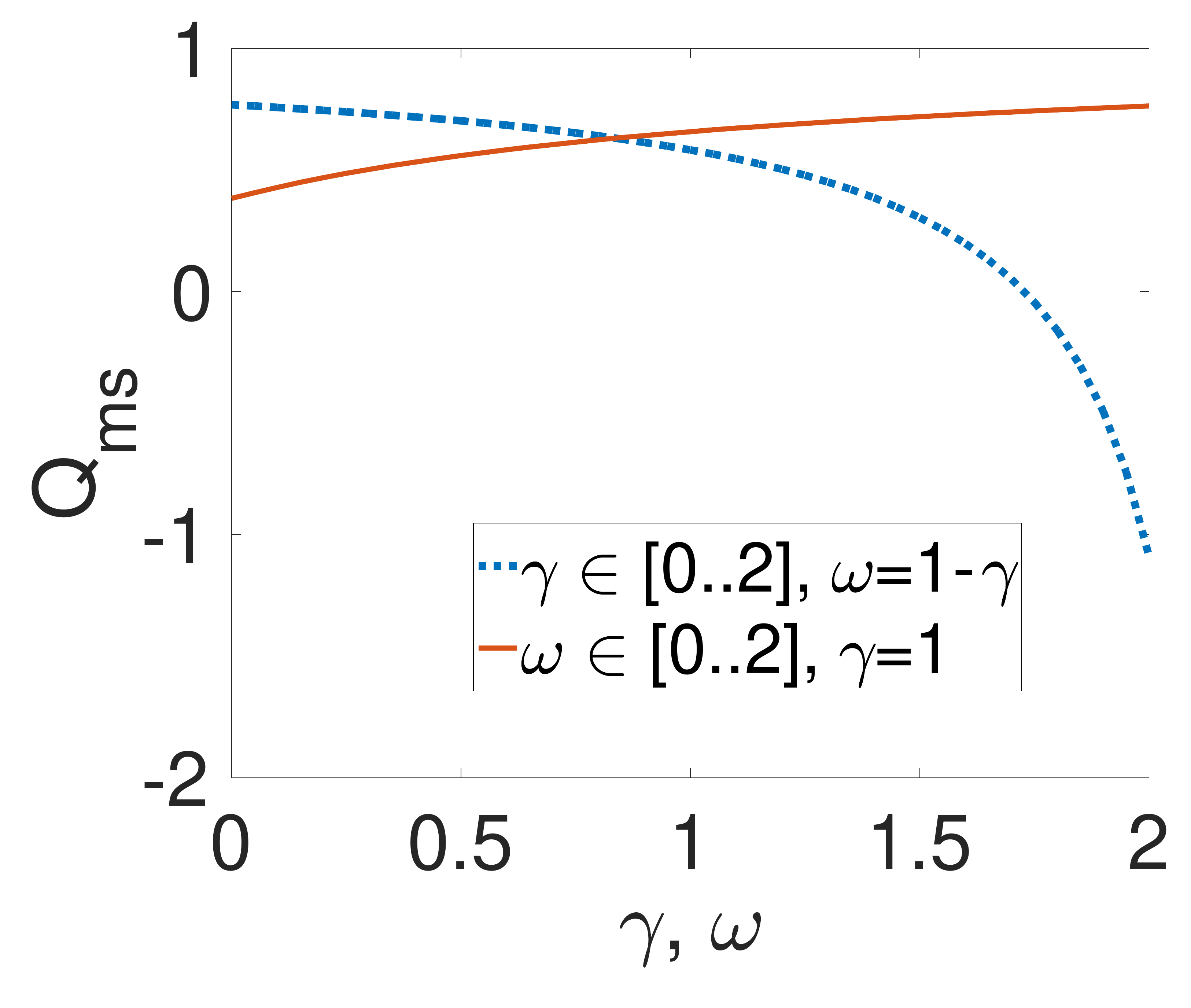}
\caption{Mucha et al.'s modularity ($Q_{\textrm{ms}}$) by varying $\gamma$ with $\omega=1-\gamma$ and by varying $\omega$ with $\gamma=1$,  on the ground-truth community structure of \textit{AUCS}.}
\label{trendMucha}
\end{figure}

\vskip 0.5em
\subsubsection{\bf \em Comparison between $Q$ and $Q_{\textrm{ms}}$: evaluation on \textit{AUCS}  ground-truth  communities}   
Let us now consider a further   stage of evaluation of $Q$ vs. $Q_{\textrm{ms}}$, which is complementary to the previous comparative analysis, with the specific purpose of assessing their behavior w.r.t. a \textit{ground-truth community structure}.  To this aim, we resorted again to the \textit{AUCS} data: in their original work~\cite{Rossi2015}, the authors filled a gap in the literature (actually, still largely open) corresponding to a lack of benchmarks for understanding multilayer/multiplex networks. In that work,  the authors also provided a ground-truth multiplex community structure for \textit{AUCS}, which reflects  affiliation of the university employees/students to research groups. Please refer to~\cite{Rossi2015} for a detailed description of how this ground-truth was obtained.

\begin{sidewaystable}[ph!]
\centering 
 \caption{Comparison between specific settings of  $Q$ and $Q_{\textrm{ms}}$ for each of the non-singleton  ground-truth communities  of   \textit{AUCS} } 
\label{tab_ground}
\begin{tabular}{|c|c|c|c|c||c|c||c|c|c|c|c|c||c|c|c|}\hline
&\!\!\#nodes/\!\!& avg.  & avg. &\!clust.\!&\multicolumn{2}{c||}{$Q_{\textrm{ms}}$}&\multicolumn{6}{c||}{$Q$}&\multicolumn{3}{c|}{$Q_{\textrm{ms}}$}
\\ \cline{6-16}
& \#edges &\!degree\!& path & coeff. &\multicolumn{2}{c||}{$\gamma=1$}&\multicolumn{2}{c|}{$\gamma=1$}&\multicolumn{4}{c||}{$\gamma(L,C)$}&$\gamma=1$&\multicolumn{2}{c|}{$\gamma=\mbox{avg}\,\gamma(L,C)$}
\\ \cline{6-16}
& & & length  & &\!$\omega=1$\!&\!$\omega=2$\!&$IC_s$&$IC_{ia}$&$ IC_s$&$IC_{ia}$&$IC^{\textrm{Adj}}_{ia}$&$IC^{\textrm{Suc}}_{ia}$ &\!\!\!$\omega\!=\!\mbox{avg}\,IC_s$\!\!\! &\!\!\!$\omega=1$\!\!\! &\!\!\!$\omega\!=\!\mbox{avg}\,IC_s$\!\!\!
\\ \hline \hline
$C_1$\!&4/8	&	3.000&	1.081&	0.523&0.076&0.083&0.037&0.052&0.038&0.054&0.073&0.069&0.064&0.078&0.067\\\hline
$C_2$\!&11/29	&4.853&	1.564&	0.635 &0.202&0.239&0.118&0.131&0.135&0.148&0.223&0.229&0.145&0.226&0.181\\\hline
$C_3$\!&6/4	&1.210&	1.148&	0.120&0.052&0.061&0.020&0.035&0.020&0.036&0.044&0.039&0.038&0.053&0.040\\\hline
$C_4$\!&5/6&2.105&	1.038&	0.413	&0.062&0.069&0.029&0.043&0.029&0.044&0.057&0.054&0.051&0.063&0.053\\\hline
$C_5$\!&3/4&2.100&	1.133&	0.600	&0.054&0.066&0.018&0.034&0.018&0.035&0.041&0.034&0.036&0.056&0.038\\\hline
$C_6$\!&6/8	&2.430&	1.208&	0.428&0.083&0.095&0.036&0.054&0.037&0.055&0.072&0.067&0.065&0.085&0.068\\\hline
$C_7$\!&7/13&3.570&	1.551&	0.659&0.123&0.141&0.057&0.074&0.058&0.075&0.103&0.101&0.096&0.128&0.103\\\hline\hline
%$\Sigma$&42/72&19.269&8.723&3.378&0.651&0.754&0.314&0.424&0.337&0.447&0.615&0.593&0.495&0.689&0.550
\multicolumn{5}{|r|}{\textit{Global modularity}} &    \textit{0.651}&\textit{0.754}&\textit{0.314}&\textit{0.424}&\textit{0.337}&\textit{0.447}&\textit{0.615}&\textit{0.593}&\textit{0.495}&\textit{0.689}&\textit{0.550}

\\\hline
\end{tabular}  
\end{sidewaystable}%  

In this context of evaluation, we analyzed  again the behaviors of  $Q$ and  $Q_{\textrm{ms}}$ under particular settings of the resolution and inter-layer coupling factors, while keeping fixed the community structure to a reference one corresponding to  ground-truth  knowledge. 
 Before going into details of such analysis, let us first provide some remarks on the trends of $Q_{\textrm{ms}}$   under its two   settings previously considered in Sect.~\ref{sec:muchaanalysis}:   varying $\gamma$ within [0..2] with $\omega=1-\gamma$, and   varying $\omega$ within [0..2] with $\gamma=1$.  
  As we observe from the results shown in 
Fig.~\ref{trendMucha}, $Q_{\textrm{ms}}$ can vary significantly depending on the setting of $\gamma, \omega$: when $\gamma=1, \omega \in [0..2]$, 
$Q_{\textrm{ms}}$ monotonically increases with $\omega$, varying within a relatively small range (i.e., from 0.38 to 0.76); however,  when $\gamma \in [0..2],  \omega=1-\gamma$, $Q_{\textrm{ms}}$ follows an inverse trend, with a more rapid decrease for $\gamma > 1.5$, and overall wider range (from 0.768 to -1.09).  Note that these considerations on the trends are consistent with the previous analysis for $Q_{\textrm{ms}}$ computed on the solution found by PMM for \textit{AUCS} (cf. Figs.~\ref{fig:Mucha1}--\ref{fig:Mucha2} (e)). 
 We shall come back later on such a high parameter-sensitivity of $Q_{\textrm{ms}}$.

Table~\ref{tab_ground} shows the global and community-specific values of $Q$ and $Q_{\textrm{ms}}$ for particular combinations of their corresponding    $\gamma$ and inter-layer coupling factors ($IC$ and $\omega$, respectively). 
Columns 2 to 5 report basic structural statistics for each of the communities, while the rest of the table is organized into three subtables: the first refers to  $Q_{\textrm{ms}}$ results, the second to our $Q$, and the third again to $Q_{\textrm{ms}}$ with $Q$-\textit{biased} settings of $\gamma, \omega$. In the latter case, we wanted to make $Q_{\textrm{ms}}$ ``closer to''  $Q$ by setting  its parameters to the values of $\gamma(L,C)$ and $IC_s$, respectively, averaged over the different communities and layers.  
Moreover, note that ground-truth communities in \textit{AUCS} are 14 in total, however we only reported  results for those (7) that contain more than one node, in order to  avoid  cluttering the table with roughly constant, zero-close modularity values that correspond to the singleton communities.

Looking at the table, communities $C_2$ and $C_7$ (resp. $C_5$, $C_3$, and $C_4$) correspond to the highest (resp. lowest) modularity values, for either modularity measure, under each of the parameter-combinations considered. 
 In general, beyond the differences in the respective values of modularity (note that  $Q_{\textrm{ms}}$ values still differ  from the corresponding $Q$ ones even for the $Q$-biased settings), the two measures appear to  behave  similarly  over the various communities.  
  To confirm this intuition, we evaluated the Pearson correlation of different pairings of $Q$ and $Q_{\textrm{ms}}$ community-specific values. 
  Results show indeed almost perfect correlation (always above 0.98).

One aspect of evaluation that we also investigated is whether disrupting the multiplex network by layer may have effect on the comparison between 
 $Q$ and  $Q_{\textrm{ms}}$, which resembles a sort of layer-oriented  \textit{resiliency} analysis. 
  More specifically, based on the structural impact due to the various layers in \textit{AUCS}~\cite{Rossi2015}, we considered  the following alternative configurations:  
  (i) we removed layer \textit{co-authorship}   (i.e., the smallest and less connected of all layers), 
  (ii) we removed layers \textit{co-authorship} and \textit{leisure}    (i.e., the ones  having the lowest number of edges), 
    (iii) we removed layer \textit{work}   (i.e., the one with the most edges), 
  (iv) we retained layers \textit{work} and \textit{lunch}   only.   
 %
%%resiliency 
%We also performed a resiliency analysis computing the values of $Q$ and Mucha et al.'s modularity $Q_{\textrm{ms}}$ for different \textit{AUCS} network configurations: (i) we delete \textit{co-authorship} and \textit{leisure} layers having the lowest number of edges, (ii) we delete \textit{work} layer having the highest number of edges, (iii) we only delete \textit{co-authorship} layer (the smallest and less connected of all layers), (iv) we keep \textit{work} and \textit{lunch} layers having the highest number of edges. 
%
 For each of such multiplex-disruption configurations, we replicated the above analysis corresponding to the full multiplex. 
  Results  (not shown)  indicated no particular differences in terms of rankings of community modularity obtained   by  $Q_{\textrm{ms}}$ and $Q$, respectively; however, we also observed a general decrease in Pearson correlation of the pairings of $Q_{\textrm{ms}}$ and $Q$ community-specific values, although the correlation remained still high, in particular always above 0.96 (e.g., when removing  layer \textit{work}, correlation was 0.965 between $Q_{\textrm{ms}}$ with $\gamma=1, \omega=\mbox{avg}\,IC_s$ and $Q$ with $\gamma=1, IC_s$, and 0.98 between $Q_{\textrm{ms}}$ with $\gamma=\mbox{avg}\,\gamma(L,C), \omega=\mbox{avg}\,IC_s$ and $Q$ with varying $\gamma(L,C), IC_s$).

To sum up, in this ground-truth evaluation,  $Q_{\textrm{ms}}$ and $Q$ exhibited consistently  similar behaviors at community level for specific settings of their respective parameters of resolution and inter-layer coupling. However, it should be emphasized that such a similarity between the two modularity values was actually achieved for either canonical settings of $\gamma, \omega$ in $Q_{\textrm{ms}}$ (i.e., $\gamma=1$ and $\omega \in \{1, 2\}$) or $Q$-biased settings of $\gamma, \omega$ (i.e., $\gamma=\mbox{avg}\,\gamma(L,C), \omega=\mbox{avg}\,IC_s$). In general, $Q_{\textrm{ms}}$ has shown to be highly sensitive to the settings of its parameters, whereas by contrast, our $Q$ has the  key advantage of  automatically determining the resolution and inter-layer coupling factors   based on the structural information of the communities in the multilayer network.

\begin{figure*}[t!]
\centering
\begin{tabular}{cc}
 {\includegraphics[height=4.5cm,width=4.8cm]{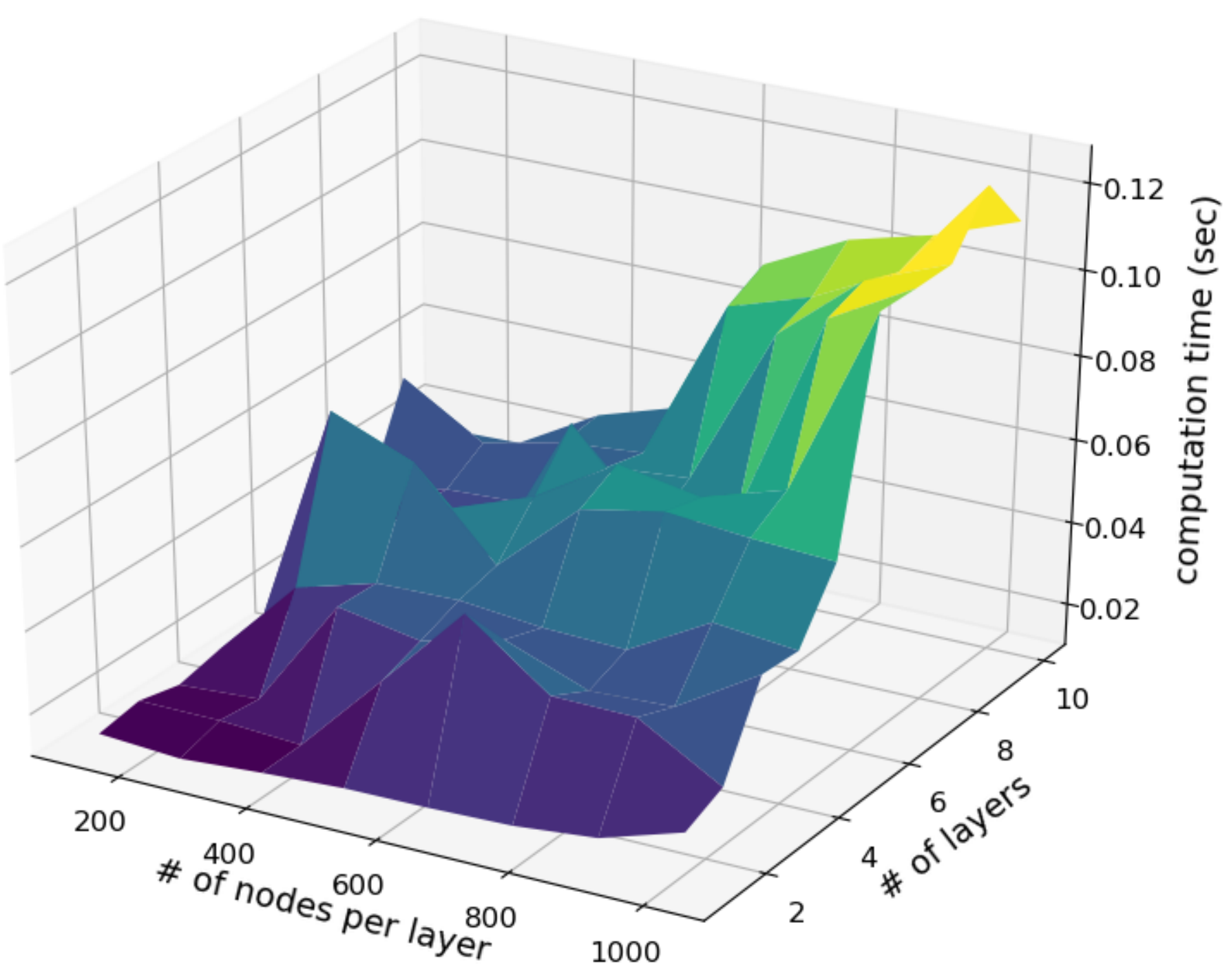}} & 
{\includegraphics[height=4.5cm,width=4.8cm]{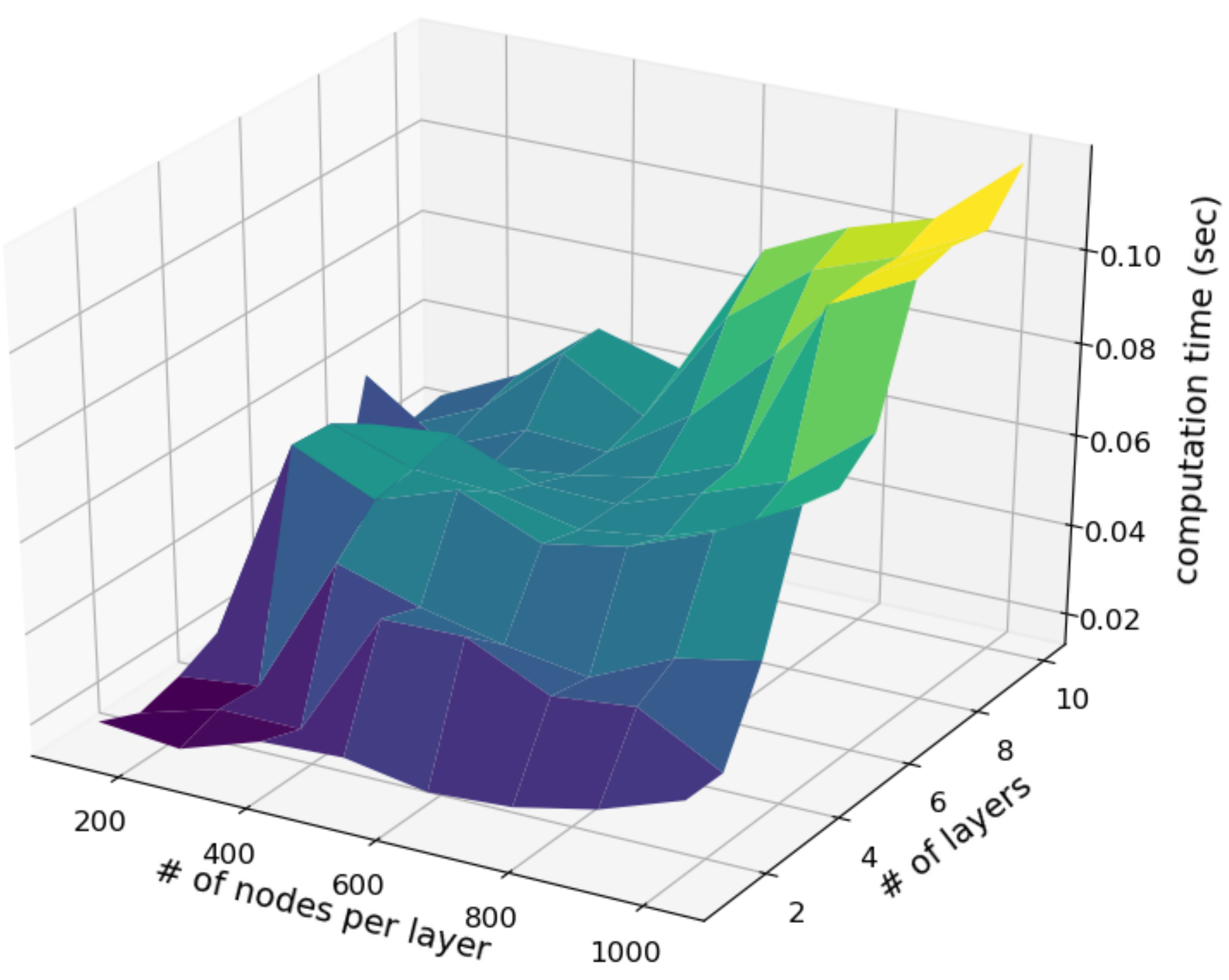}} \\
  (a) $\beta=0$ & (b) $IC_s$ \\
{\includegraphics[height=4.5cm,width=4.8cm]{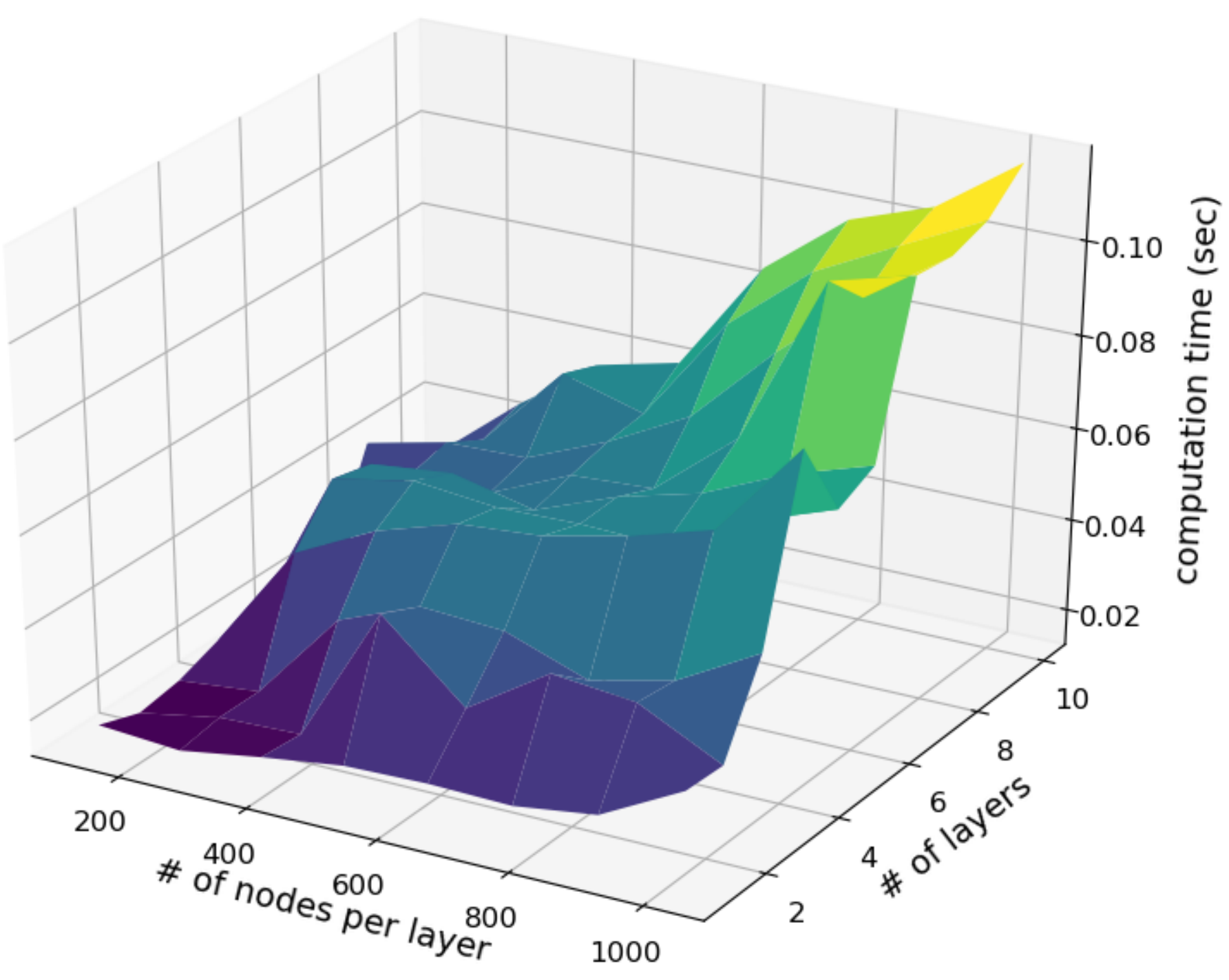}} & 
{\includegraphics[height=4.5cm,width=4.8cm]{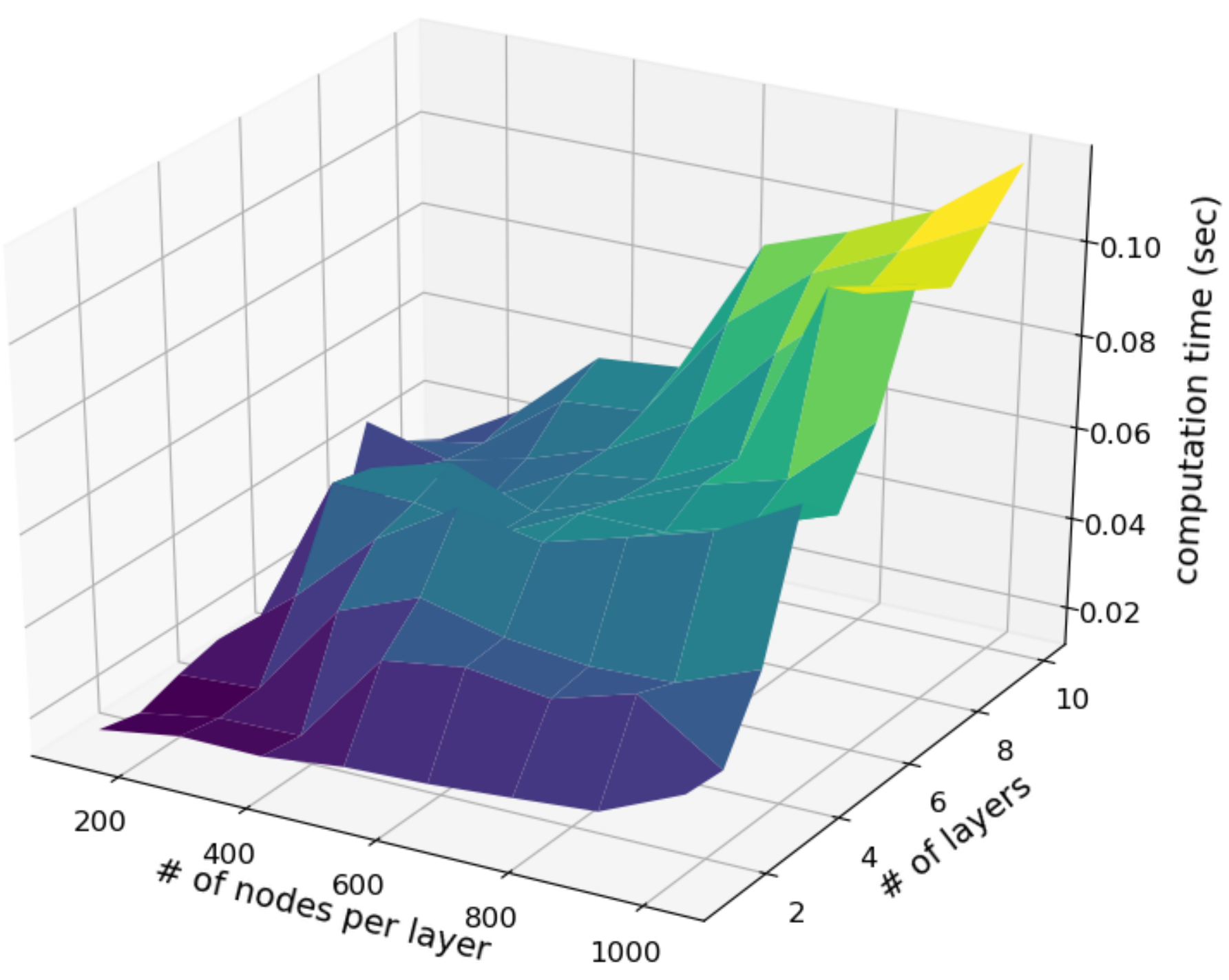}} \\
 (c)   $IC_{ia}$ &  (d)   $IC_{oa}$
  \end{tabular} 
\caption{Computation time (in seconds) of the multilayer modularity $Q$, with $\gamma=1$,   measured on the solution found by GL on the multiplex LFR network}
\label{times1} 
\end{figure*}

\begin{figure*}[t!]
\centering
\begin{tabular}{cc} 
 {\includegraphics[height=4.5cm,width=4.8cm]{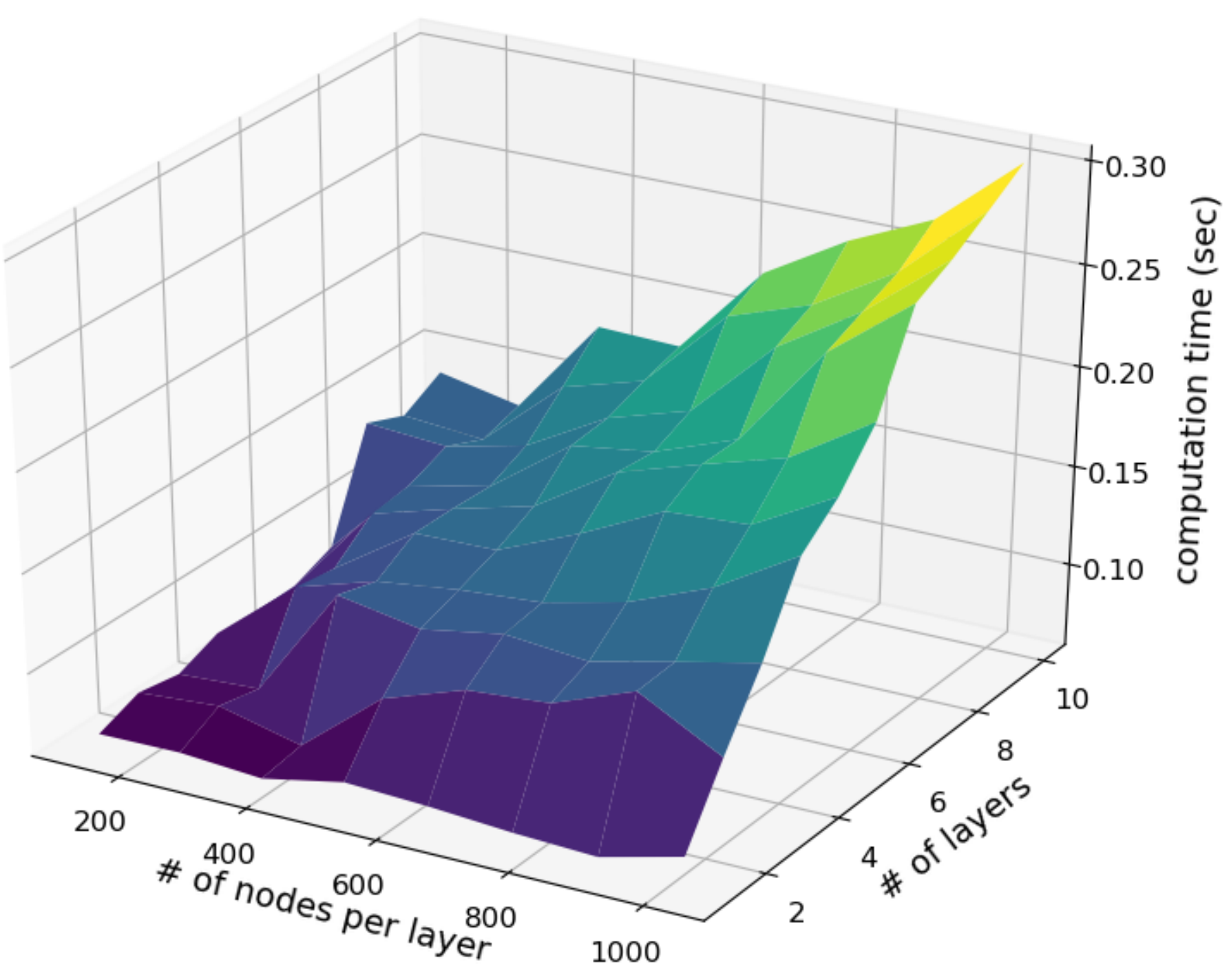}} & 
{\includegraphics[height=4.5cm,width=4.8cm]{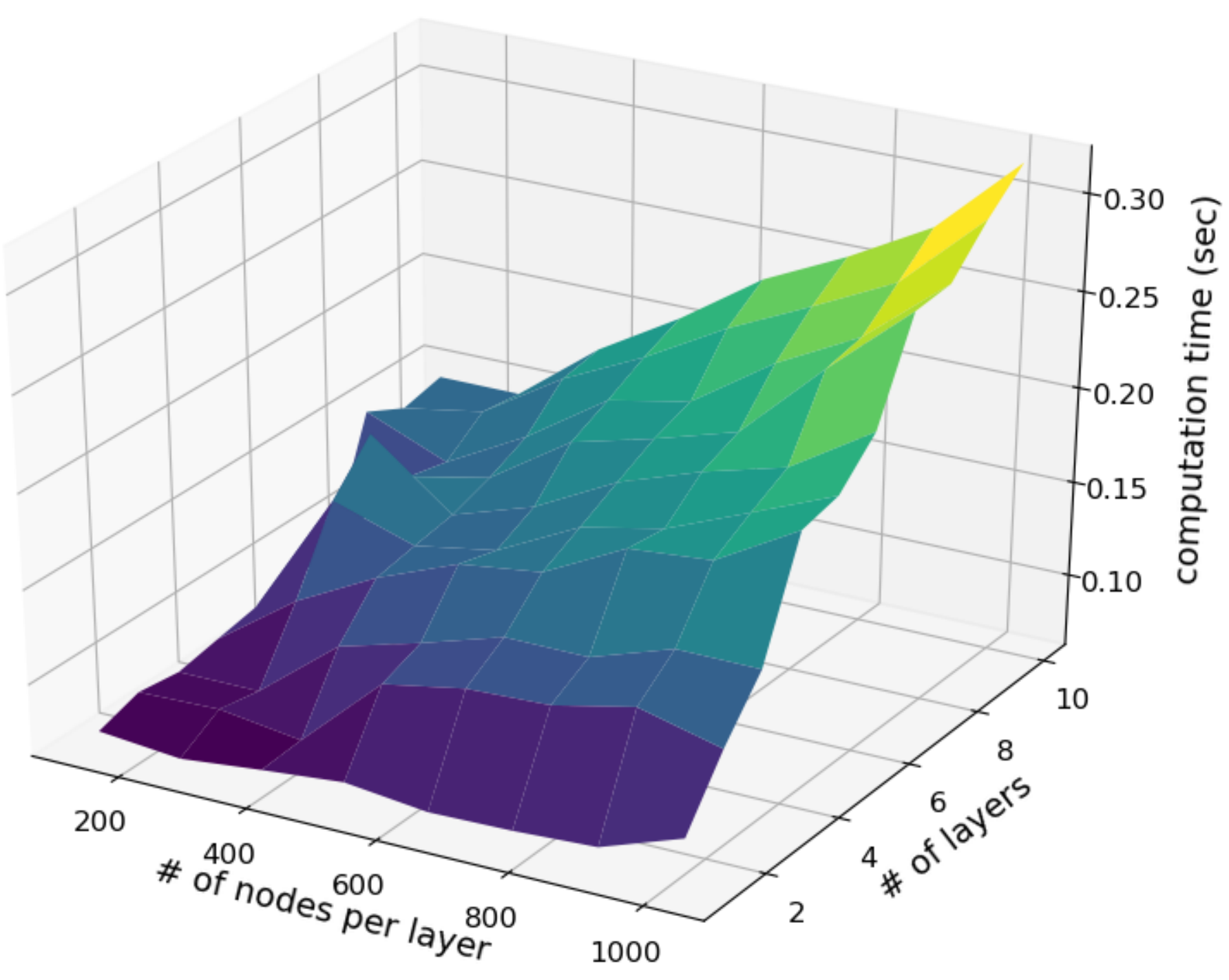}} \\ 
(a)  $\beta=0$  & (b)    $IC_s$ \\
{\includegraphics[height=4.5cm,width=4.8cm]{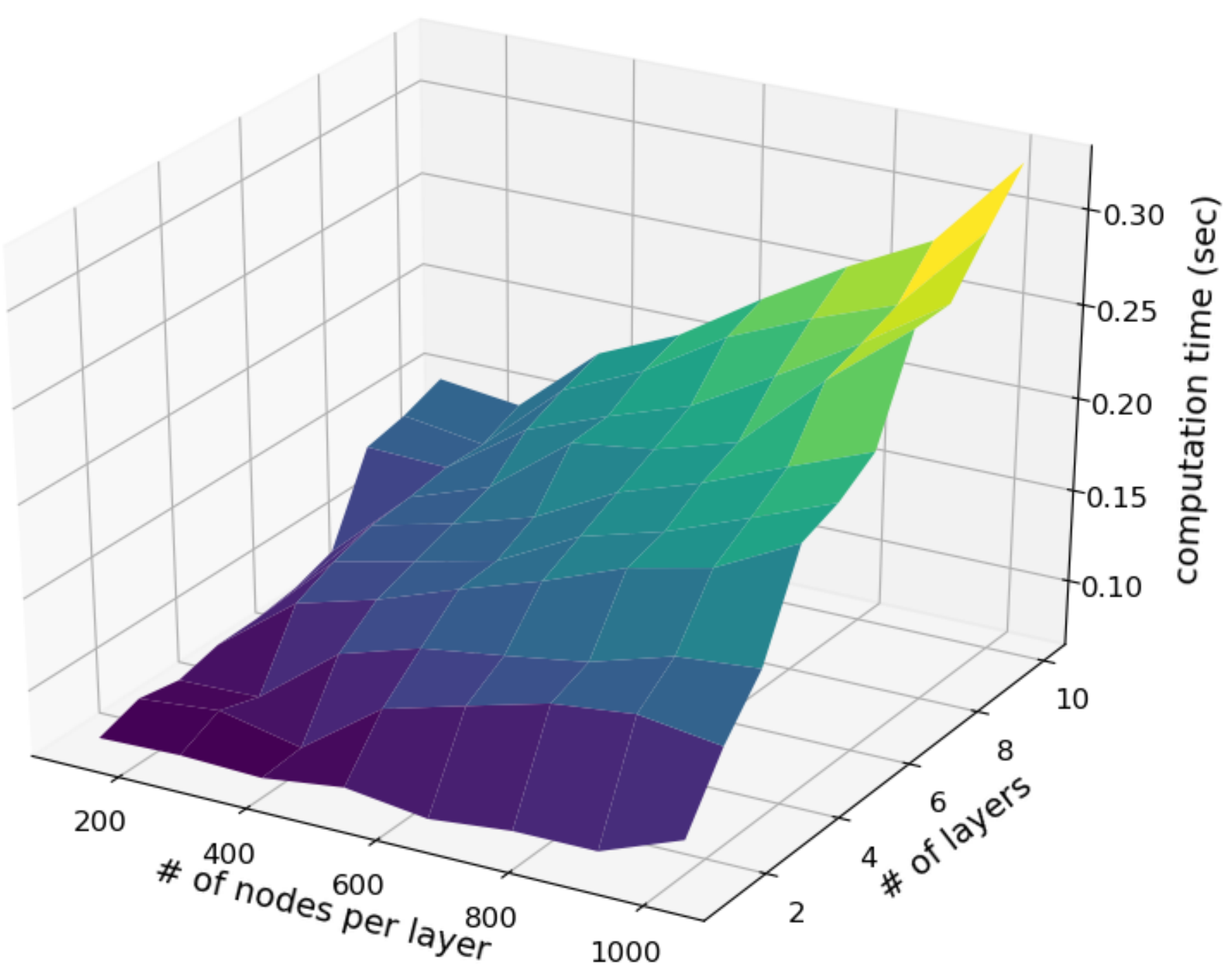}} & 
{\includegraphics[height=4.5cm,width=4.8cm]{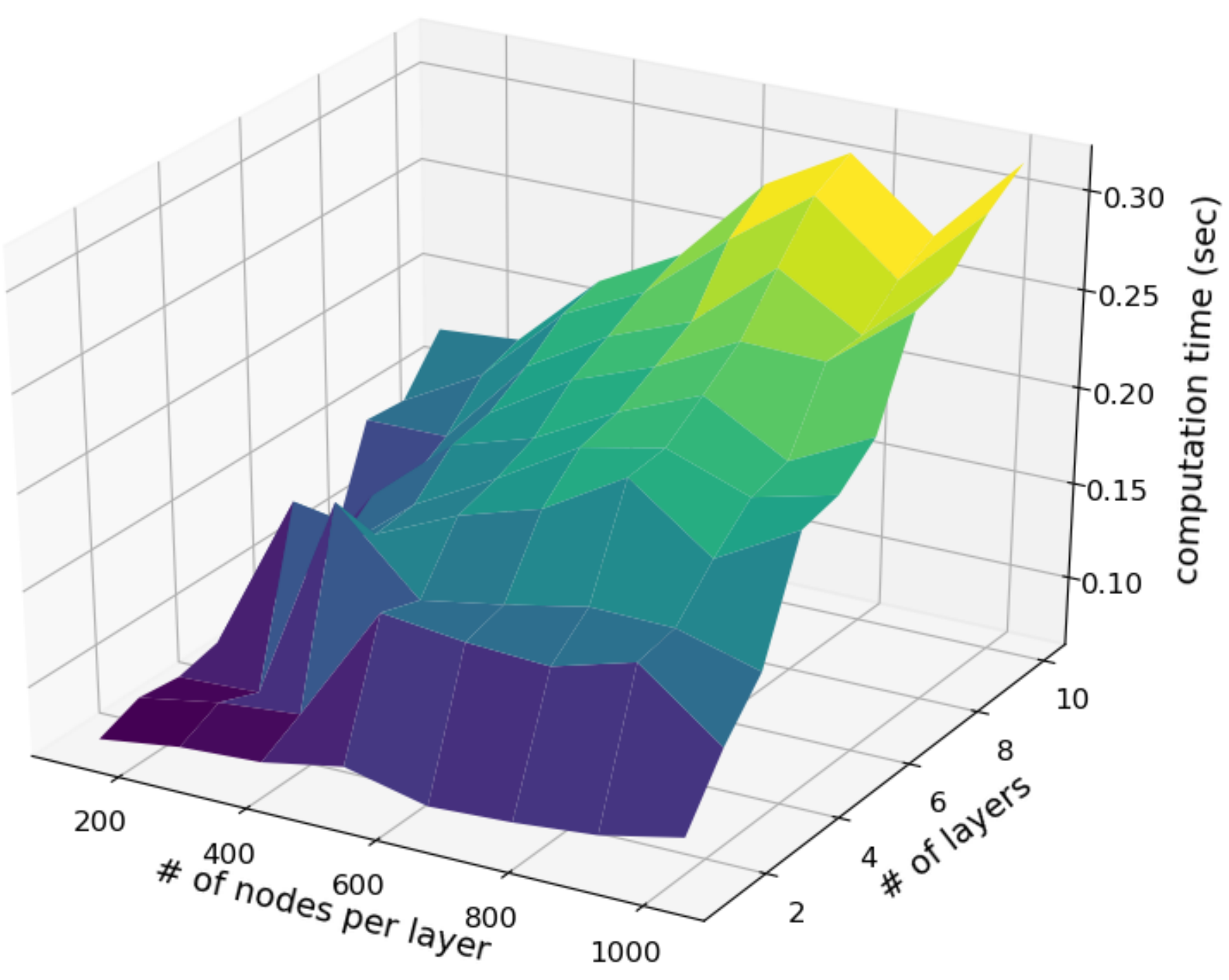}}  \\
 (c)   $IC_{ia}$ & (d)   $IC_{oa}$ 
\end{tabular}
\caption{Computation time (in seconds) of the multilayer modularity $Q$, with redundancy-based $\gamma(L,C)$, measured on the solution found by GL on the multiplex LFR network}
\label{times2} 
\end{figure*}

 \subsection{Efficiency evaluation}
We analyzed the computation time of $Q$ for the different combinations of redun\-dancy-based resolution factor $\gamma$ and inter-layer coupling factors $IC$. 
The 3-D plots in Figs.~\ref{times1}--\ref{times2} display the   time vs. the number of layers and the number of nodes per layer.   For this analysis, we referred to  the solutions found by GL for the  multilayer networks  generated through the LFR benchmark  (cf. Section~\ref{sec:snd}).  
%It is worth noting that the computation time increases as the number of network nodes and layers progressively grows. The highest peak is obtained for the network datasets of 1024 nodes and 10 layers.

%{\color{blue}
As expected, the computation time increases with both the number of layers and nodes per layer, with the latter being less evident when setting $\gamma$ fixed to 1. Also, while it is  obvious that the computation time is higher when using variable (i.e., redundancy-based) resolution factor than in the case $\gamma=1$ (with percentage increase of 50\%, for the maximum number of layers and nodes per layer), we observe   much less fluctuations in the plot surfaces  than in the case of resolution factor fixed to 1, regardless of the setting of inter-layer coupling factors. % (Figs.~\ref{times1} and \ref{times3}  vs. Figs.~\ref{times2} and \ref{times4}). 
 The reader is also referred to the {\bf \em Appendix} for further results by varying the inter-layer coupling settings.

\section{Conclusion}

We proposed a new definition of modularity for multilayer networks. Motivated by the  opportunity  of  revising the multislice modularity proposed in~\cite{Mucha10},  
 we conceived  alternative  notions of layer resolution and inter-layer coupling, which are key-enabling for  generalizations of modularity for multilayer networks.  
Using four state-of-the-art methods for multilayer community detection, synthetic multilayer networks and ten real-world multilayer networks, we provided empirical evidence of the significance of our proposed modularity. 
 
Our work paves the way for the development of new optimization methods of community detection in multilayer networks  which, by embedding our multilayer modularity, can discover     community structures having the interesting   properties relating to the proposed per-layer/community redundancy-based resolution factors and projection-based inter-layer coupling schemes.   
 In this respect, we point out that our multilayer modularity is able to cope with communities that are overlapping at entity level,  which eventually  reflect   the different roles that the same entity can play when occurring in two or more layers of the network. 
  Within this view, one benefit of adopting the multilayer network model is that the problem of computing soft community-memberships of entities can be   translated into a simpler problem of identification of crisp community-memberships of nodes within each layer.  
 Nonetheless,   a further  interesting direction would be to evaluate our multilayer modularity in contexts of node-overlapping communities. In this case, however,  one challenge to face is whether and to  what extent an   overlapping-aware multilayer modularity should be able to measure the community  overlaps    within each layer and/or across the layers. 
 Along this direction, it would be interesting to study an integration   of our multilayer modularity into recently developed works that propose probabilistic representations or stochastic generative models for overlapping community detection  in   multilayer networks~\cite{Stanley2016,Ali2019}.

\vspace{-2mm}
\section*{Acknowledgements}
 This research work has been partly supported by the Start-(H)Open POR Grant No. J28C17000380006 funded by the Calabria Region Administration, and by the NextShop PON Grant No. F/050374/01-03/X32 funded by the Italian Ministry for Economic Development.



\newpage

\appendix

\section*{Appendix}

\vspace{5mm}

\section{Analytical derivation of lower and upper bounds}

\vspace{2mm}

 {\bf Proof of Proposition 1 (Lower bound of $Q$).\ }   
Let us assume that each of the $\ell$ layer graphs in $G_{\mathcal{L}}$ has the form of a bipartite graph  $K_i(a, b)$, with $i=1, \ldots, \ell$, and sets $a$ and $b$ induce a partitioning of the set of nodes in two communities denoted as $C_1$ and $C_2$, respectively, so that $\mathcal{C}=\{C_1, C_2\}$ with $|C_1|=|C_2|=\frac{n}{2}$, and no internal links are drawn between nodes of the same community (because of the bipartite assumption).

To begin with, consider the reduction of $Q$ to its simplest form, i.e., 
  $\gamma(L,C)$ fixed to 1 for any $L,C$ and $\beta=0$. Therefore, the contribution of community $C_1$ to $Q$ is: 
%  
%\begin{equation*}
%Q(\mathcal{C}) =  \sum_{C \in \mathcal{C}} [\sum_{L \in \mathcal{L}}  - (\frac{d_L(C)}{d(V_{\mathcal{L}})})^2].
%\end{equation*}
%
%We rename the internal degree of community $C$, $d_L^{int}(C)$, as $m_i^L(C)$. Also, let $m_e^L(C)$ be the number of external edges connecting $C$ with all other communities.
%For the first community $C_1$, $Q$ can be defined as:
%
\begin{equation*}\label{eq1}
Q(C_1) =  -\sum_{L \in \mathcal{L}} \left(\frac{d_L(C_1)}{d(V_{\mathcal{L}})}\right)^2.
\end{equation*}

\noindent 
Since $d_L(C_1)=\frac{n^2}{4}$ and $d(V_{\mathcal{L}})=\frac{n^2}{4}(2\ell)=\frac{n^2 \ell}{2}$, $Q(C_1)$ is calculated as:
\begin{equation*}
Q(C_1) = -\sum_{L \in \mathcal{L}} \left(\frac{\frac{n^2}{4}}{\frac{n^2\ell}{2}}\right)^2=-\frac{1}{4\ell}.
\end{equation*}

\noindent 
The same above holds for $C_2$. Therefore, the minimum value of $Q$ when $\gamma(L,C)=1$ and $\beta=0$ is as follows:  
\begin{equation}
Q(\mathcal{C}) = -\frac{1}{4\ell}2=-\frac{1}{2\ell}.
\end{equation}

Let us     now consider $Q$ with its  redundancy-based resolution factor while keeping $\beta=0$. 
%, the multilayer modularity $Q$ can be reduced as follows:
%
%\begin{equation}
%Q(\mathcal{C}) =  \sum_{C \in \mathcal{C}} [\sum_{L \in \mathcal{L}}  - \gamma(L, C)(\frac{d_L(C)}{d(V_{\mathcal{L}})})^2].
%\end{equation}
%
 Since the internal degree of community $C_1$ is 0,  there are no redundant pairs for any community and layer, and hence  $\gamma(L,C)=2$. Consequently,  the contribution of  $C_1$ to  $Q$ is: 

\begin{equation*}
Q(C_1) =  -2 \sum_{L \in \mathcal{L}} \left(\frac{d_L(C_1)}{d(V_{\mathcal{L}})}\right)^2 = 
-2 \sum_{L \in \mathcal{L}} \left(\frac{1}{2\ell}\right)^2 = -\frac{1}{2\ell}.
\end{equation*}

\noindent 
The same above holds for $C_2$. Therefore, the lower bound of $Q$ when $\gamma(L,C)$ is variable and $\beta=0$ is as follows:  
\begin{equation}\label{eq:pluto}
Q(\mathcal{C}) =-\frac{1}{2\ell}2=-\frac{1}{\ell}.
\end{equation}

In the general form of $Q$ with both resolution and inter-layer coupling factors (i.e., varying $\gamma$ and $\beta=1$), the contribution of $C_1$ to $Q$ is:
\begin{equation*}
Q(C_1) =  \sum_{L \in \mathcal{L}} \left[-2\left(\frac{d_L(C_1)}{d(V_{\mathcal{L}})}\right)^2 + \sum_{L' \in \mathcal{P}(L)}  \frac{IC(C_1, L, L')}{d(V_{\mathcal{L}})}\right].
\end{equation*}

 In the above formula, let us indicate the terms  $X(C_1)=\sum_{L \in \mathcal{L}} -2(\frac{d_L(C_1)}{d(V_{\mathcal{L}})})^2$, and $Y(C_1)=\sum_{L \in \mathcal{L}, L' \in \mathcal{P}(L)}  \frac{IC(C_1, L, L')}{d(V_{\mathcal{L}})}$. 
If we denote with $p= \sum_{L \in \mathcal{L}}|\mathcal{P}(L)|$  the total number of valid layer-pairings, then   $d(V_{\mathcal{L}})=\frac{n^2\ell}{2}+{np}$, and $X(C_1)$ can be rewritten  as follows:

\begin{equation*}
X(C_1)=-2\sum_{L \in \mathcal{L}}\left(\frac{\frac{n^2}{4}}{\frac{n^2\ell}{2}+ {np}}\right)^2=-\frac{1}{2}\frac{n^2\ell}{(n\ell + 2p)^2}.
\end{equation*}

For the reduction of the term $Y(C_1)$, we consider the two cases of symmetric inter-layer coupling and asymmetric inter-layer coupling. 
In the first case, the minimum value for $IC_s(C_1,L_i, L_j)$ is equal to $\frac{|C_1^{(i)} \cap C_1^{(j)}|}{|V_i \cap V_j|}=\frac{1}{n}$.  %\disc{forse bisogna dire quando si ottiene questo valore. Ovvero, se non sbaglio, 1 nodo nella prima comunita' e i rimanenti n-1 nella seconda comunita'}.   
%%%%   !!!!!!!     \frac{\frac{n}{2}}{n}=\frac{1}{2}$.   
Accordingly, $Y(C_1)$ is reduced as follows:
%
% \begin{equation*}
% \begin{split}
% \beta(C_1)  = & \sum_{L \in \mathcal{L}}\sum_{L' \in \mathcal{P}(L)}\frac{\frac{1}{2}}{\frac{n^2\ell}{2}+ {np}}= \\
%  = & \sum_{L \in \mathcal{L}}\sum_{L' \in \mathcal{P}(L)} \frac{1}{n^2\ell + 2 np}=\frac{p}{n^2\ell + 2 np}.
% \end{split}
% \end{equation*}
\begin{equation*}
Y(C_1)  =  \sum_{L \in \mathcal{L}}\sum_{L' \in \mathcal{P}(L)}\frac{\frac{1}{n}}{\frac{n^2\ell}{2}+ {np}} = \frac{2p}{n^2(n\ell+2p)}
\end{equation*}

\noindent 
In the second case, $IC_a(C_1,L_i, L_j) = \frac{|C_1^{(i)} \cap C_1^{(j)}|}{|V_i \cap V_j|} \times \frac{|V_i|}{|C_1^{(i)}|}=\frac{1}{n}\times \frac{n}{\frac{n}{2}}=\frac{2}{n}$. 
%%% !!!!!!  \frac{1}{2}\times \frac{n}{\frac{n}{2}}=1$.
Accordingly, $Y(C_1)$ is reduced as follows:

\begin{equation*}
Y(C_1)=\sum_{L \in \mathcal{L}}\sum_{L' \in \mathcal{P}(L)}\frac{\frac{2}{n}}{\frac{n^2\ell}{2}+ {np}}=\frac{4p}{n^2 (n\ell   +  2p)}.
\end{equation*}

\noindent 
The  above expressions for $X$ and $Y$ hold for $C_2$. Therefore, the lower bound of $Q$ in its general form is as follows:   

\begin{equation}
\label{eq:pippo}
\begin{split}
Q(\mathcal{C}) = & 2(X(C_1)+Y(C_1)) = \\
 = & -\frac{n^2\ell}{(n\ell +2p)^2} + \frac{4(1+\eta)p}{n^2(n\ell + 2p)},
\end{split}
\end{equation}
with $\eta=0$ for $IC_s$ and $\eta=1$ for $IC_a$.

It should be noted that Eq.~\ref{eq:pluto} is a special case of  Eq.~\ref{eq:pippo} with $\beta=0$ and $d(V_{\mathcal{L}})$ discarding the contribution given by the inter-layer edges (i.e., $d(V_{\mathcal{L}}) = \frac{n^2\ell}{2}$).

 {\bf Proof of Proposition 2 (Upper bound of $Q$).\ }
Let us assume that each of the $\ell$ layer graphs in $G_{\mathcal{L}}$ has community structure $\mathcal{C}=\{C_1, C_2\}$ such that $|C_1|=|C_2|=\frac{n}{2}$ and each community is a clique with $\frac{\frac{n}{2}(\frac{n}{2}-1)}{2}$ edges. Moreover, there are no external edges connecting the communities, therefore  $d_L(C)=d_L^{int}(C)$. 
 Note that by uniformly distributing the $n$ nodes into the two communities, it can easily be shown that the  maximum of $Q$ is higher.

Analogously to the  analysis of the minimum value of $Q$, let us first consider the setting $\gamma(L,C)=1$, for any $L,C$, and $\beta=0$. 
%; therefore:
%
%\begin{equation*}
%Q(C) =   \sum_{C \in \mathcal{C}}[\sum_{L \in \mathcal{L}}  \frac{d_L^{int}(C)}{d(V_{\mathcal{L}})}  -  (\frac{d_L^{int}(C)}{d(V_{\mathcal{L}})})^2].
%\end{equation*}
 %
  Therefore, the contribution of community $C_1$ to $Q$ is: 
   
\vspace{-1mm}
\begin{equation*}
Q(C_1) =   \sum_{L \in \mathcal{L}}  \frac{d_L^{int}(C_1)}{d(V_{\mathcal{L}})}  -  \left(\frac{d_L^{int}(C_1)}{d(V_{\mathcal{L}})}\right)^2.
\end{equation*}
 
\vspace{-1mm}
\noindent 
Because $d_L^{int}(C_1)=\frac{n}{2}(\frac{n}{2}-1)$ and $d_L(\V)=\frac{n}{2}(\frac{n}{2}-1)2\ell$, the above expression is rewritten as:

\begin{equation*}
Q(C_1) =   \sum_{L \in \mathcal{L}}  \frac{\frac{n}{2}(\frac{n}{2}-1)}{\frac{n}{2}(\frac{n}{2}-1)2\ell}  -  \left(\frac{\frac{n}{2}(\frac{n}{2}-1)}{\frac{n}{2}(\frac{n}{2}-1)2\ell}\right)^2 = \frac{1}{2}-\frac{1}{4\ell}.
\end{equation*}

\noindent 
The same above holds for $C_2$. Therefore, the maximum of $Q$ when $\gamma(L,C)=1$ and $\beta=0$ is as follows:\footnote{If $C_1$ and $C_2$ would have  $n-1$ and 1 nodes, respectively, the resulting maximum value of $Q$ will be $\frac{\ell-1}{\ell}$, hence lower than what we obtain in Eq.~\ref{eq:topo}.}  
\begin{equation}\label{eq:topo}
Q(\mathcal{C}) =   2\left(\frac{1}{2}-\frac{1}{4\ell}\right)=1-\frac{1}{2\ell}=\frac{2\ell-1}{2\ell}.
\end{equation}

Consider now the setting with redundancy-based $\gamma(L, C)$ and $\beta=0$. The contribution of $C_1$ to $Q$ is calculated as: 
%
%\begin{equation*}
%Q(C_1) =   \sum_{L \in \mathcal{L}}  \frac{d_L^{int}(C_1)}{d(V_{\mathcal{L}})}  -  \gamma(L, C_1)\left(\frac{d_L^{int}(C_1)}{d(V_{\mathcal{L}})}\right)^2.
%\end{equation*}
%
%Consequently, we obtain as follows:
\begin{equation*}
Q(C_1) =   \sum_{L \in \mathcal{L}}  \frac{\frac{n}{2}(\frac{n}{2}-1)}{\frac{n}{2}(\frac{n}{2}-1)2\ell}  -  \gamma(L, C_1)\left(\frac{\frac{n}{2}(\frac{n}{2}-1)}{\frac{n}{2}(\frac{n}{2}-1)2\ell}\right)^2.
\end{equation*}

\noindent 
Since $nrp(L, C_1)=\frac{\frac{n}{2}(\frac{n}{2}-1)}{2}$, it follows that  $\gamma(L, C_1)$ is   equal to $2(1+\log_2(1+\frac{\frac{n}{2}(\frac{n}{2}-1)}{2}))^{-1}$ for each layer and community. 
 Note that the above constant quantity, hereinafter denoted as   $\gamma$, tends to be $\ll 1$, and  it is  smaller for higher number of nodes $n$. 
The contribution of $C_1$ to $Q$ is rewritten as:
 
\begin{equation*}
Q(C_1)=\frac{1}{2}-\gamma\frac{1}{4\ell}.
\end{equation*}

\noindent 
The same above holds for $C_2$. Therefore, the maximum  of $Q$ with redundancy-based $\gamma(L,C)$ and $\beta=0$ is as follows:  
\begin{equation}\label{eq:specific}
Q(\mathcal{C}) =   2\left[\frac{1}{2}-\gamma\frac{1}{4\ell}\right]=1-\gamma\frac{1}{2\ell}=\frac{2\ell-\gamma}{2\ell}.
\end{equation}
 Note that the above quantity is as much closer to 1 as  $n$ and $\ell$ are higher.

In the general setting of $Q$ with redundancy-based   $\gamma(L, C)$ and $\beta=1$, the contribution of $C_1$ to $Q$ is: 
\begin{equation*}
\begin{split}
Q(C_1) = &  \sum_{L \in \mathcal{L}} \left[  \frac{d_L^{int}(C_1)}{d(V_{\mathcal{L}})}  - \gamma(L, C) \left(\frac{d_L^{int}(C_1)}{d(V_{\mathcal{L}})}\right)^2  + \right.\\
 + & \left. \sum_{L' \in \mathcal{P}(L)}  \frac{IC(C_1, L, L')}{d(V_{\mathcal{L}})} \right].
\end{split}
\end{equation*}

 In the above formula, let us indicate the terms  $X(C_1)=\sum_{L \in \mathcal{L}}[  \frac{d_L^{int}(C_1)}{d(V_{\mathcal{L}})}  - \gamma(L, C) (\frac{d_L^{int}(C_1)}{d(V_{\mathcal{L}})})^2]$, 
 and $Y(C_1)=\sum_{L \in \mathcal{L}}$ $\sum_{L' \in \mathcal{P}(L)}  \frac{IC(C_1, L, L')}{d(V_{\mathcal{L}})}$.  
% If we denote with $p= \sum_{L \in \mathcal{L}}|\mathcal{P}(L)|$  the total number of valid layer-pairings, then   $d(V_{\mathcal{L}})=\frac{n^2\ell}{2}+{np}$, and $X(C_1)$ can be rewritten  as follows: 
 Because $d_L^{int}(C_1)=\frac{n}{2}(\frac{n}{2}-1)$, $d_L(\V)=\frac{n}{2}(\frac{n}{2}-1)2\ell+np$, and $\gamma(L, C_1)=2(1+\log_2(1+$ $\frac{\frac{n}{2}(\frac{n}{2}-1)}{2}))^{-1}$ for each layer and community, we obtain:

\begin{equation*}
\begin{split}
X(C_1)\!\!=&\sum_{L \in \mathcal{L}}  \frac{\frac{n}{2}(\frac{n}{2}-1)}{\frac{n}{2}(\frac{n}{2}-1)2\ell+np}  -  \gamma\left(\frac{\frac{n}{2}(\frac{n}{2}-1)}{\frac{n}{2}(\frac{n}{2}-1)2\ell+np}\right)^2\!\!\!= \\
 = & \frac{1}{2}\frac{(\frac{n}{2}-1)\ell}{(\frac{n}{2}-1)\ell+p}-\gamma\ell \left(\frac{1}{2}\frac{(\frac{n}{2}-1)}{(\frac{n}{2}-1)\ell+p}\right)^2,
\end{split}
\end{equation*}

\begin{equation*}
Y(C_1)=\sum_{L \in \mathcal{L}}\sum_{L' \in \mathcal{P}(L)}  \frac{IC(C_1, L, L')}{\frac{n}{2}(\frac{n}{2}-1)2\ell+np}.
\end{equation*}

For the reduction of the term $Y(C_1)$, we again consider the two cases of symmetric inter-layer coupling and asymmetric inter-layer coupling. 
In the first case, $IC_s(C_1,L_i, L_j) = \frac{|C_1^{(i)} \cap C_1^{(j)}|}{|V_i \cap V_j|}=\frac{1}{n}$. Therefore:
\begin{equation*}
Y(C_1)\!\!=\!\!\sum_{L \in \mathcal{L}}\sum_{L' \in \mathcal{P}(L)}\frac{\frac{1}{n}}{{n}(\frac{n}{2}-1)\ell+np}=\frac{p}{{n^2}(\frac{n}{2}-1)\ell+n^2p}.
\end{equation*}

In the second case, $IC_a(C_1,L_i, L_j) = \frac{|C_1^{(i)} \cap C_1^{(j)}|}{|V_i \cap V_j|} \times \frac{|V_i|}{|C_1^{(i)}|}=\frac{1}{n}\times \frac{n}{\frac{n}{2}}=\frac{2}{n}$. Therefore:

\begin{equation*}
Y(C_1)\!\!=\!\!\sum_{L \in \mathcal{L}}\sum_{L' \in \mathcal{P}(L)}\frac{\frac{2}{n}}{\frac{n}{2}(\frac{n}{2}-1)2\ell+np}=\frac{2p}{{n^2}(\frac{n}{2}-1)\ell+n^2p}.
\end{equation*}

The same above holds for $C_2$. Therefore,  the maximum value of $Q$  is as follows:
\begin{equation}\label{eq:general}
\begin{split}
Q(\mathcal{C}) = & 2(X(C_1)+Y(C_1)) = \\
 = & 2 \left[ \frac{1}{2}\frac{(\frac{n}{2}-1)\ell}{(\frac{n}{2}-1)\ell+p}-\gamma\ell \left(\frac{1}{2}\frac{(\frac{n}{2}-1)}{(\frac{n}{2}-1)\ell+p}\right)^2 + \right. \\
 + & \left.  \frac{(1+\eta)p}{{n^2}(\frac{n}{2}-1)\ell+n^2p}  \right]
\end{split}
\end{equation}
with $\eta=0$ for $IC_s$ and $\eta=1$ for $IC_a$.

  It is worth noting that Eq.~\ref{eq:specific} is a special case of  Eq.~\ref{eq:general}  with  $\beta=0$ and $d(V_{\mathcal{L}})$ discarding the contribution given by the inter-layer edges.

 \vspace{3mm}
 \section{Evaluation with ordered layers}
 
\begin{figure*}[t!]
\centering
\begin{tabular}{cc}
\includegraphics[height=2.5cm, width=4cm]{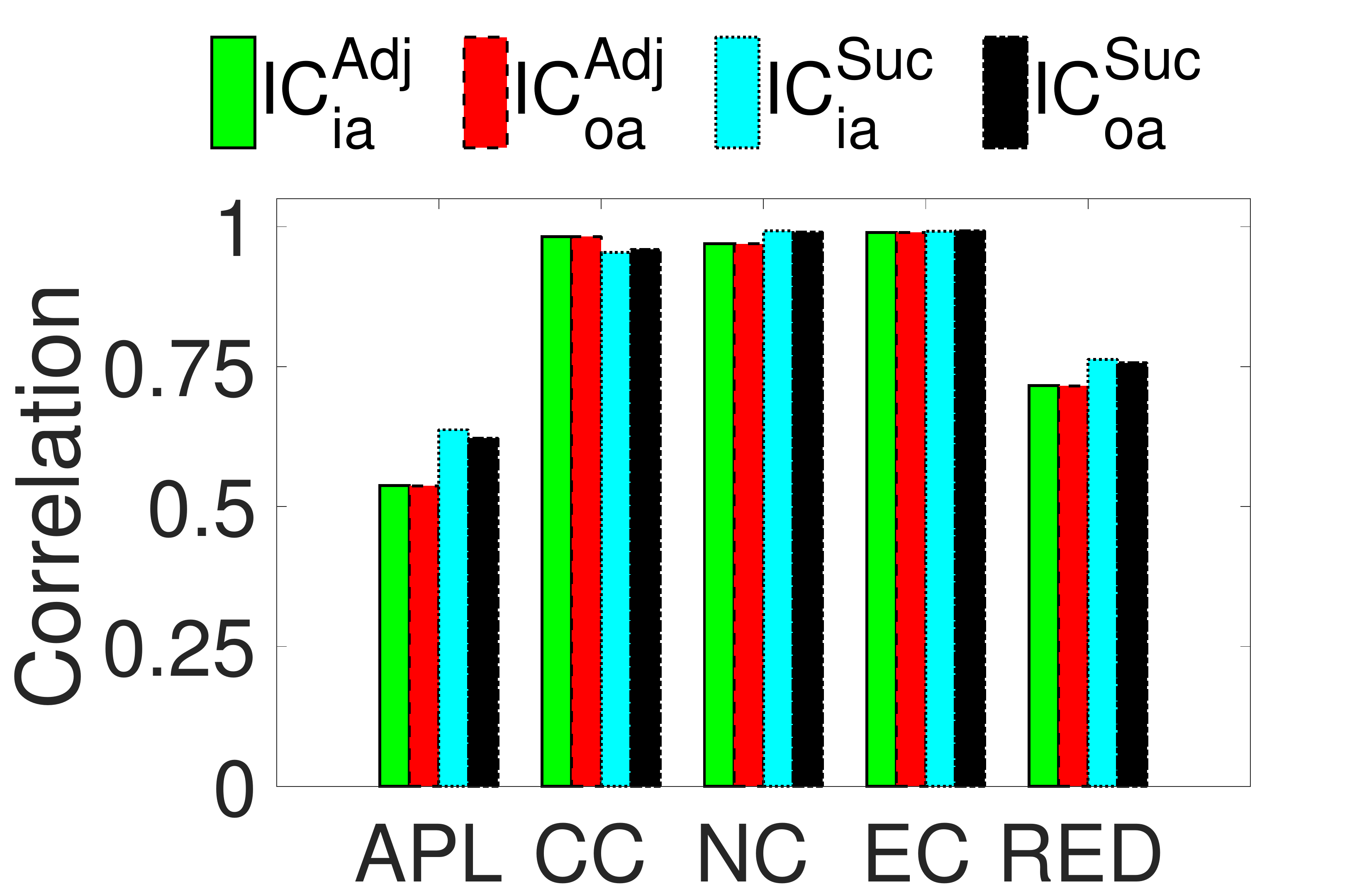}
& 
\includegraphics[height=2.5cm, width=4cm]{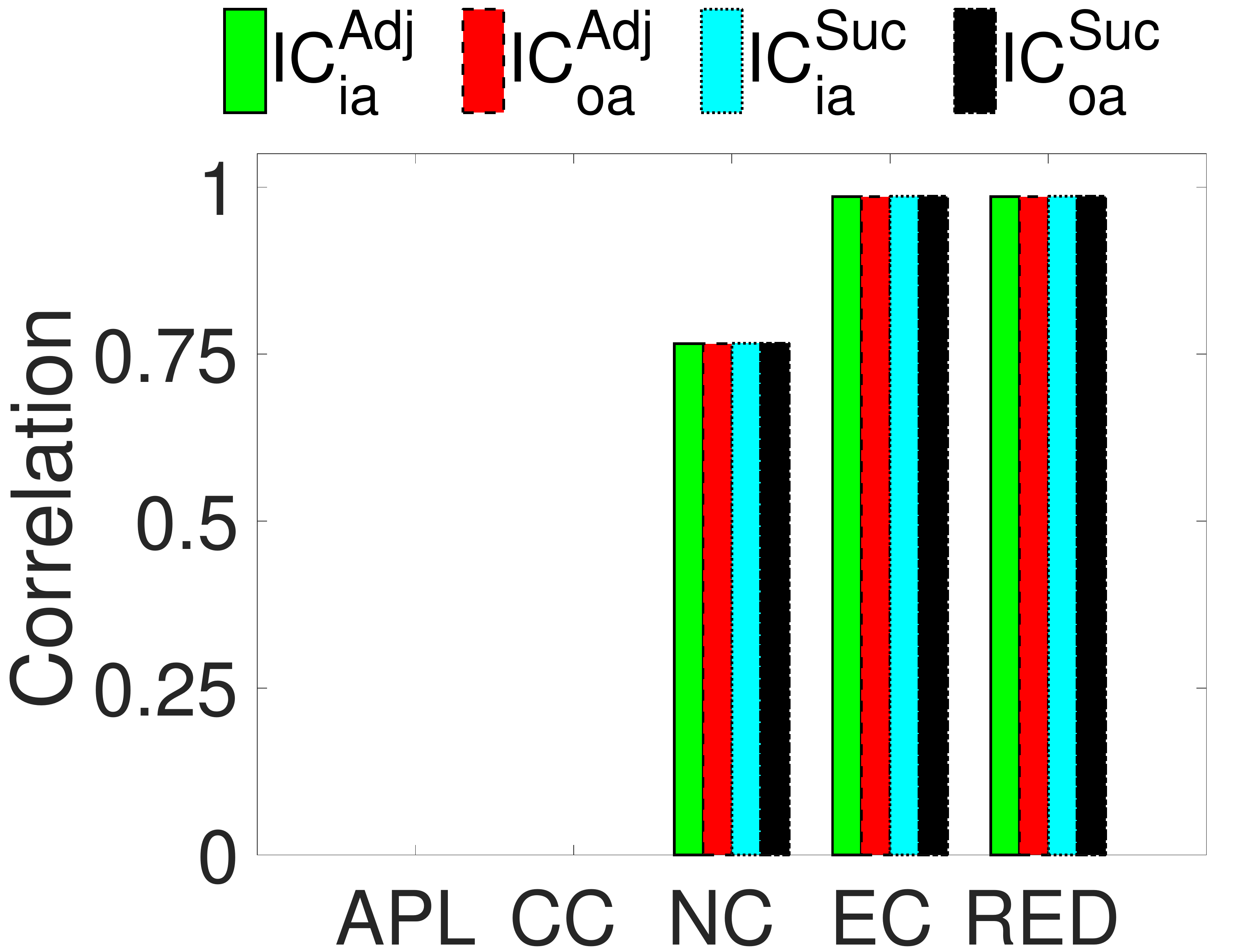}
\\ (a)   & (b)   \\
\includegraphics[height=2.5cm, width=4cm]{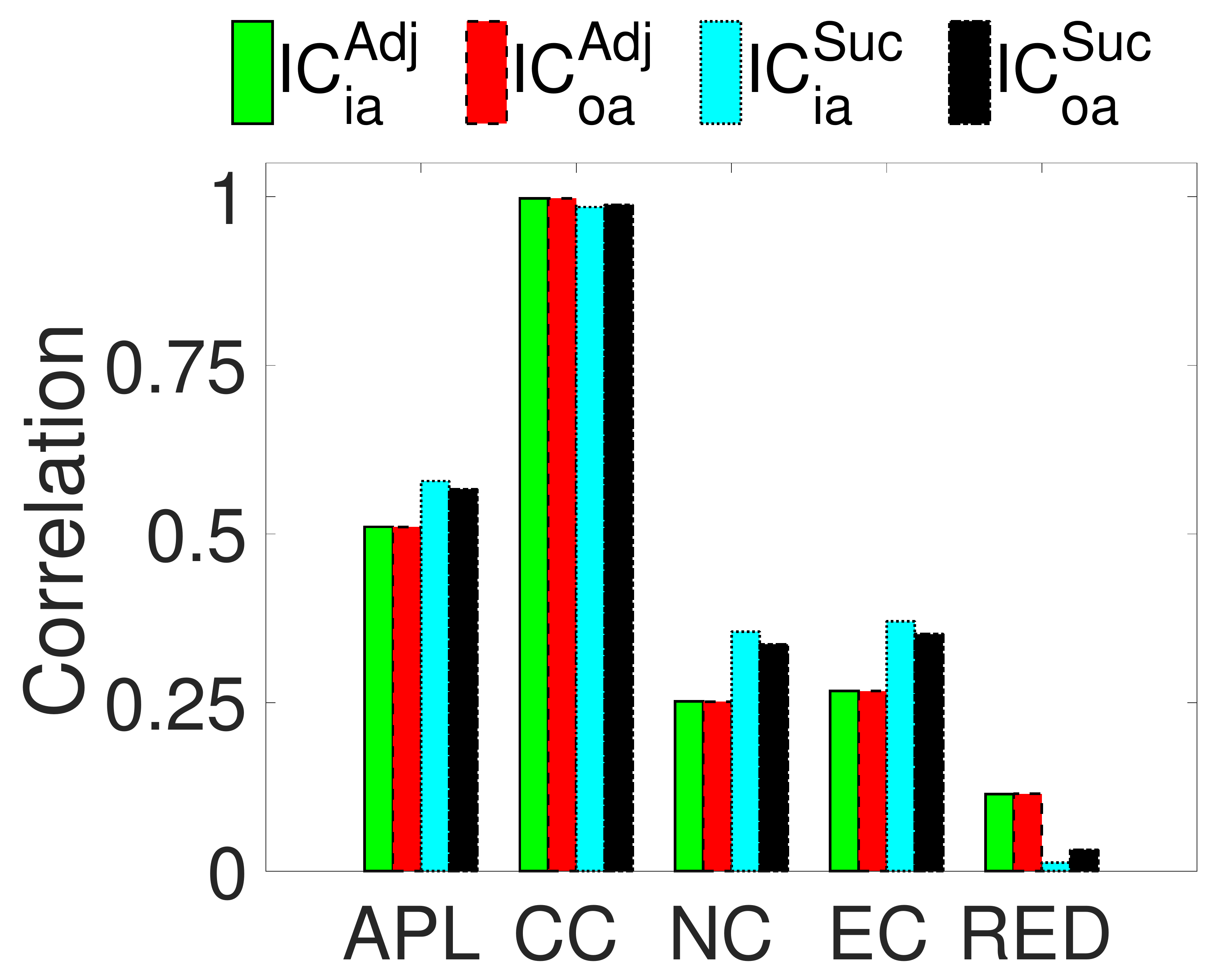}
&  
\includegraphics[height=2.5cm, width=4cm]{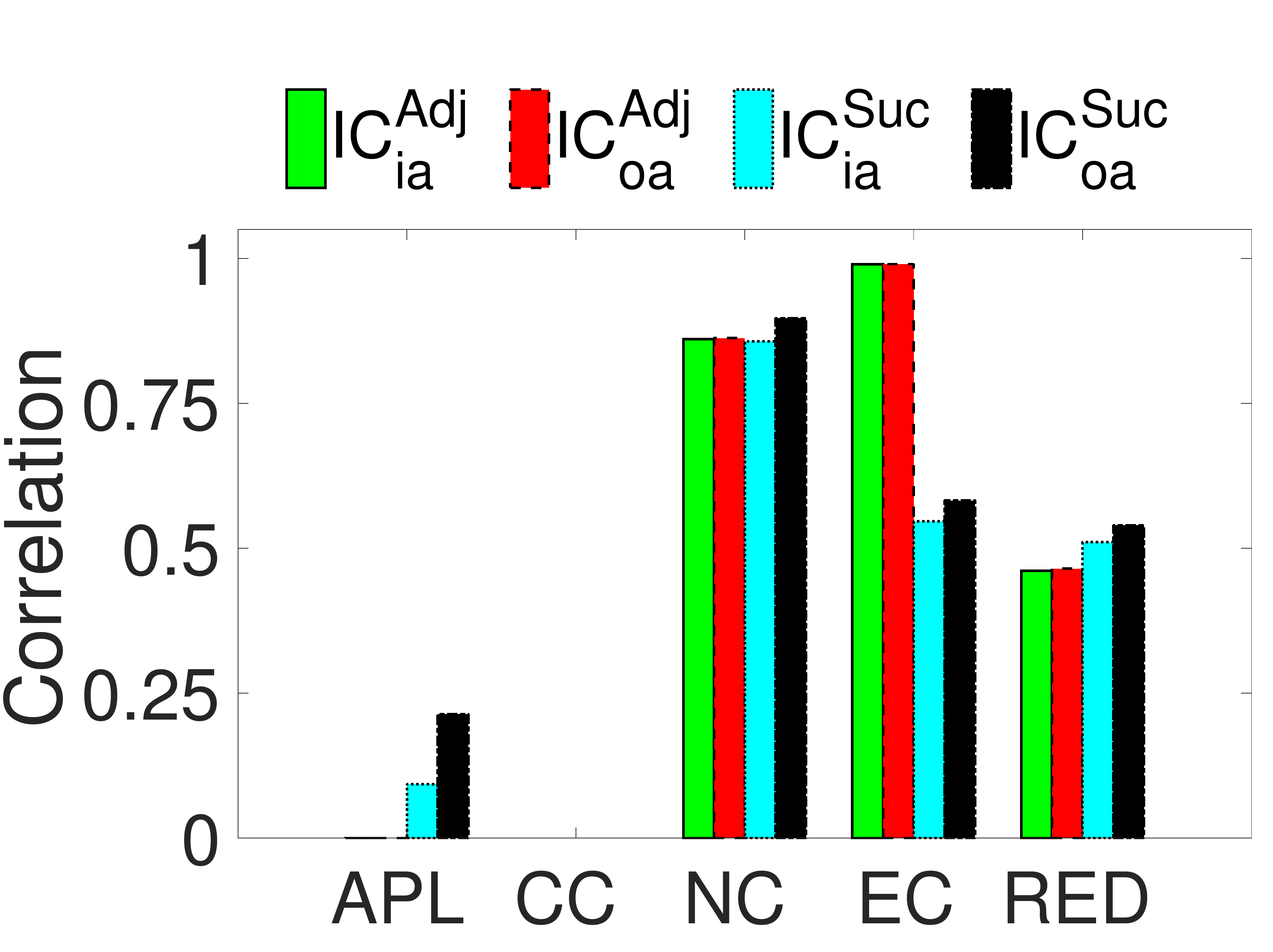}
\\ (c)   & (d)  
\end{tabular} 
\caption{Pearson correlation coefficient between average path length (APL), clustering coefficient (CC), node coverage (NC), edge coverage (EC), redundancy (RED), and the multilayer modularity $Q$ with $\gamma(L,C)$ and $IC_{ia}^{Adj}$, $IC_{oa}^{Adj}$, $IC_{ia}^{Suc}$, $IC_{oa}^{Suc}$, and descendent layer ordering,  computed on the solution found by 
(a) GL,  (b) LART, (c) PMM, and (d) \greedystar 
on the \textit{EU-Air} network. Each statistics is computed at community-level}
\label{stat2GLextra}
\end{figure*}

 Figure~\ref{stat2GLextra} provides further details on  correlation analysis using descendent layer ordering.

\begin{figure*}[ph!]
\centering
\begin{tabular}{cc}
 {\includegraphics[height=4.5cm,width=4.8cm]{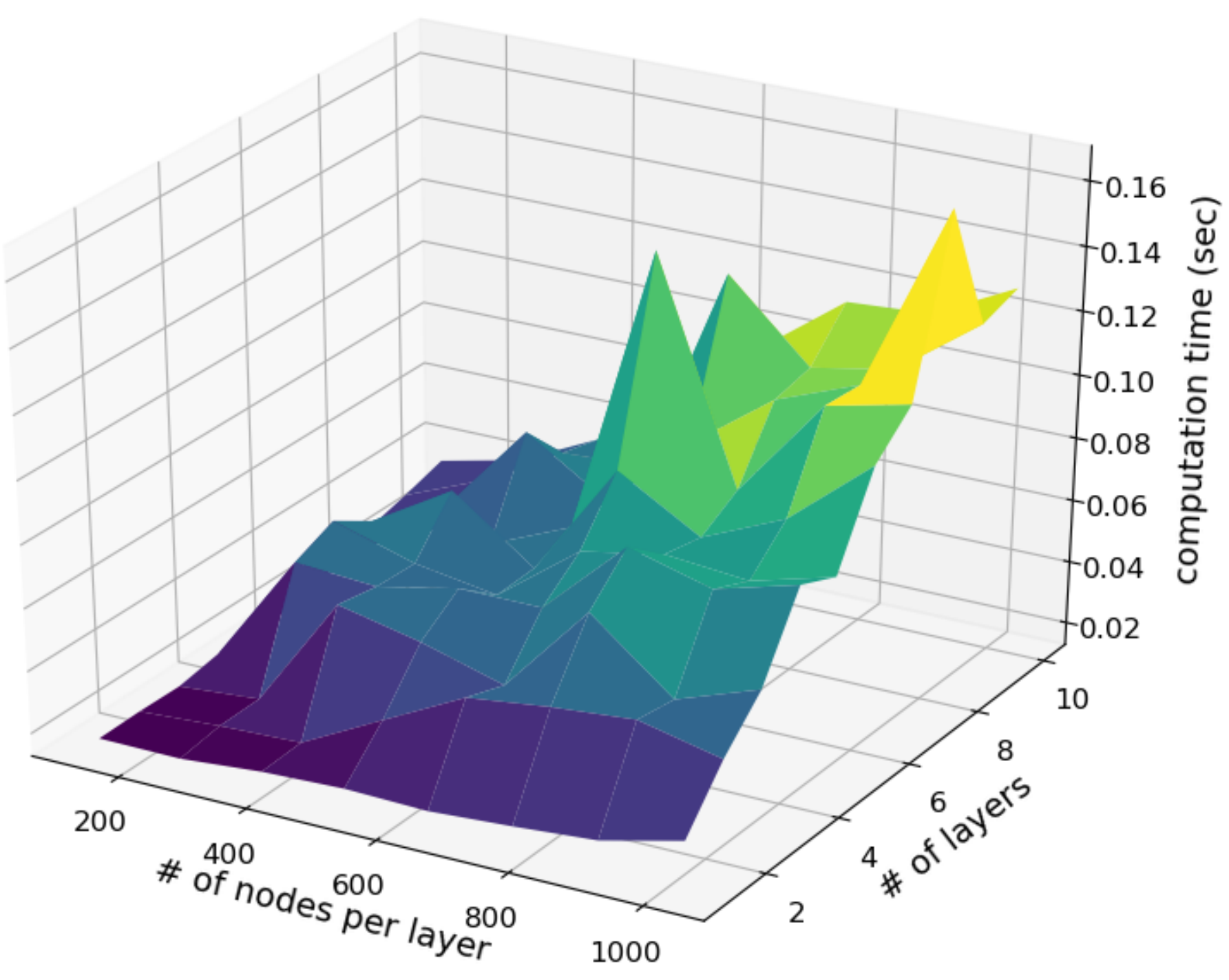}} &
{\includegraphics[height=4.5cm,width=4.8cm]{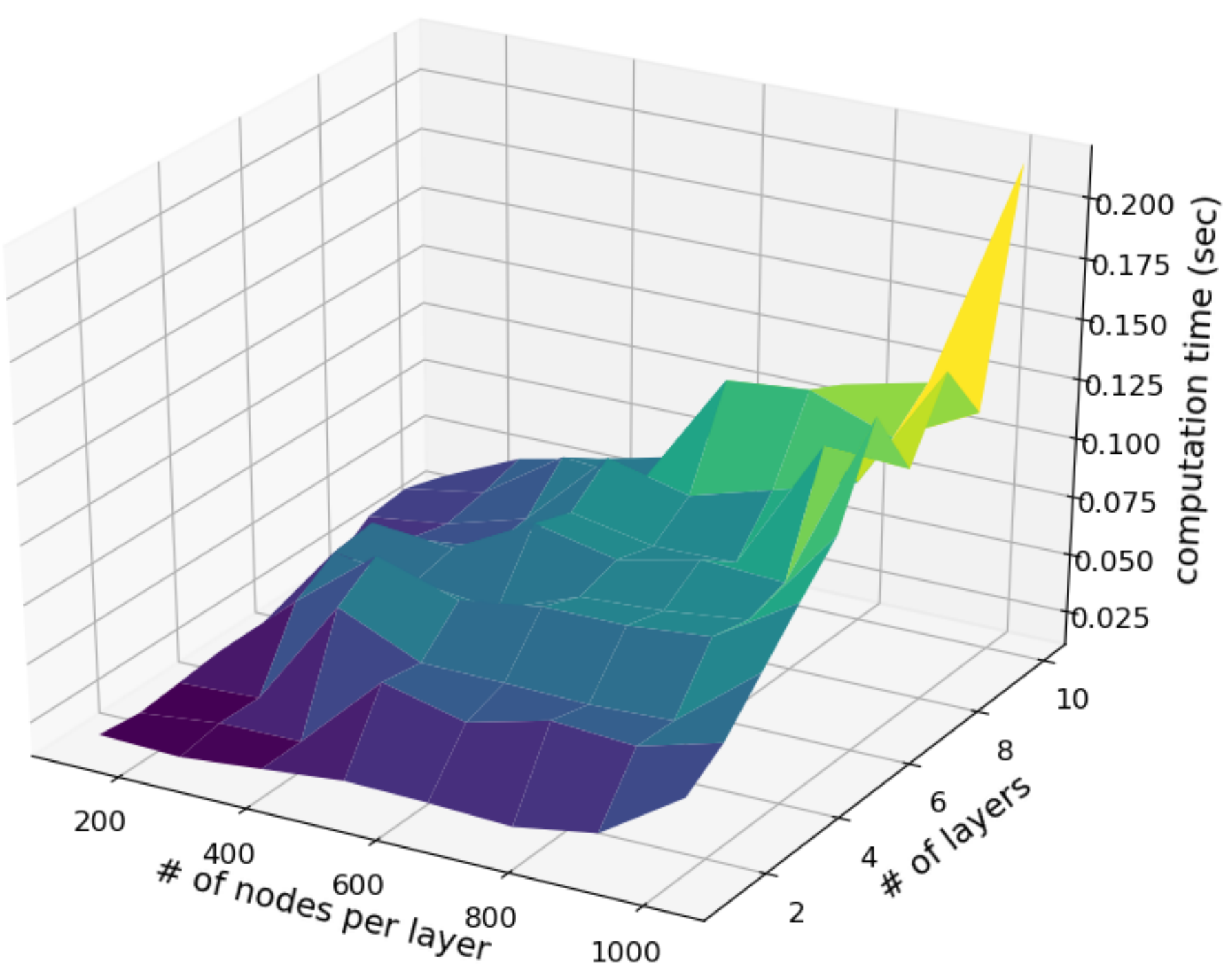}} \\  
(a)   $IC_{ia}^{Adj}$ & (b)   $IC_{oa}^{Adj}$  \\
{\includegraphics[height=4.5cm,width=4.8cm]{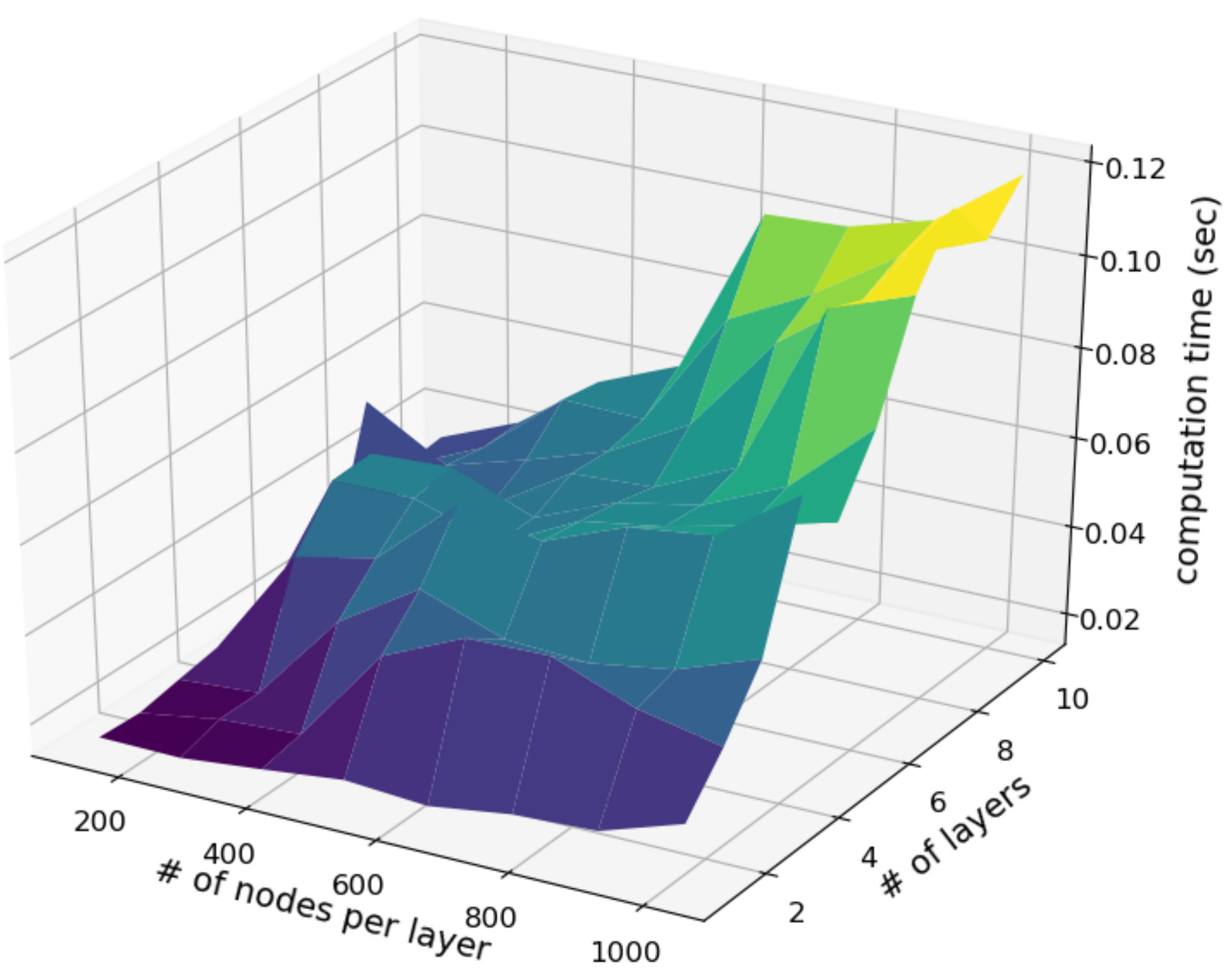}} &
{\includegraphics[height=4.5cm,width=4.8cm]{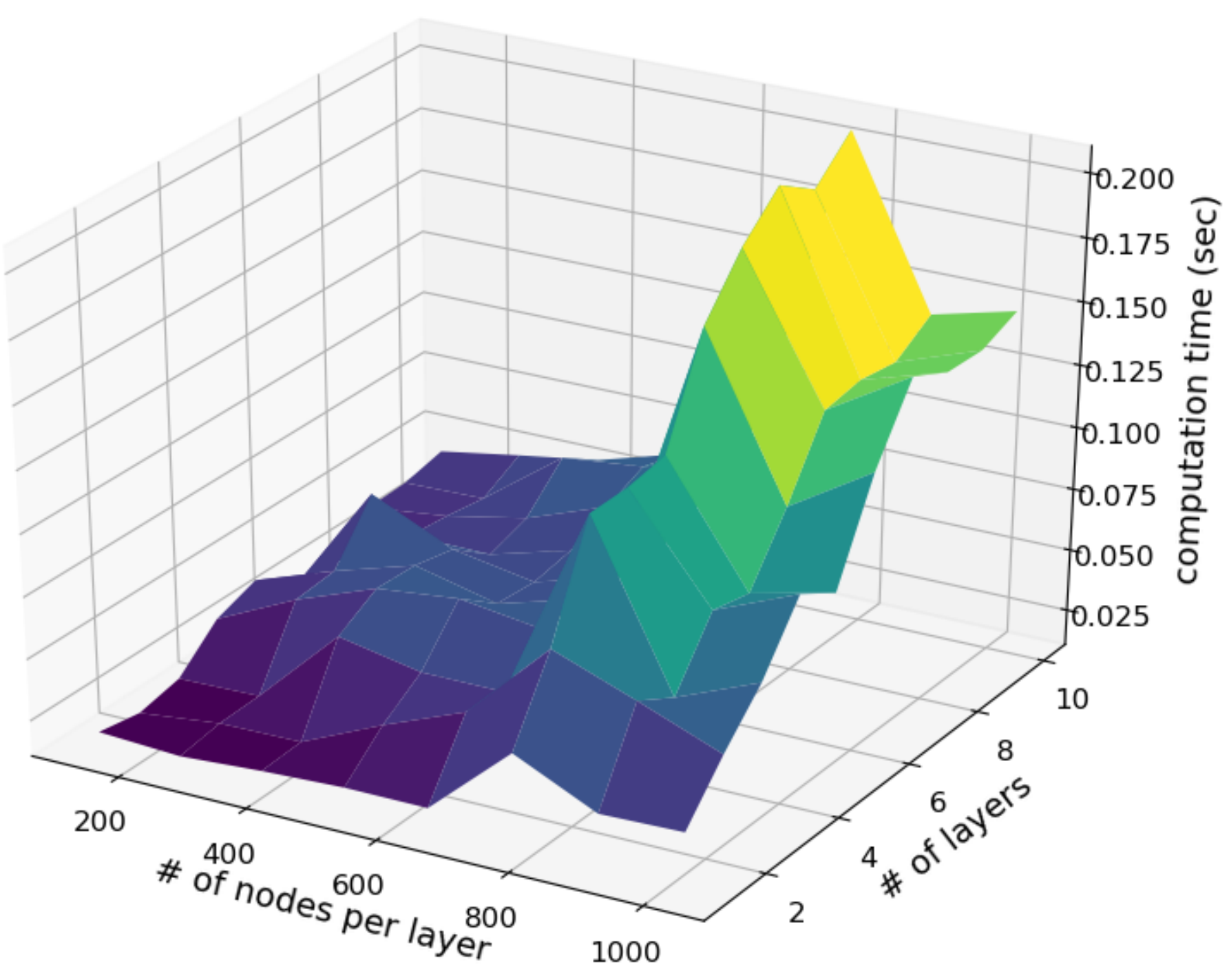}} \\
(c)   $IC_{ia}^{Suc}$ & (d)  $IC_{oa}^{Suc}$
\end{tabular} 
\caption{Computation time (in seconds) of the multilayer modularity $Q$, with $\gamma=1$, measured on the solution found by GL on the multiplex LFR network}
\label{times3}
\end{figure*}

\begin{figure*}[ph!]
\centering
\begin{tabular}{cc}
{\includegraphics[height=4.5cm,width=4.8cm]{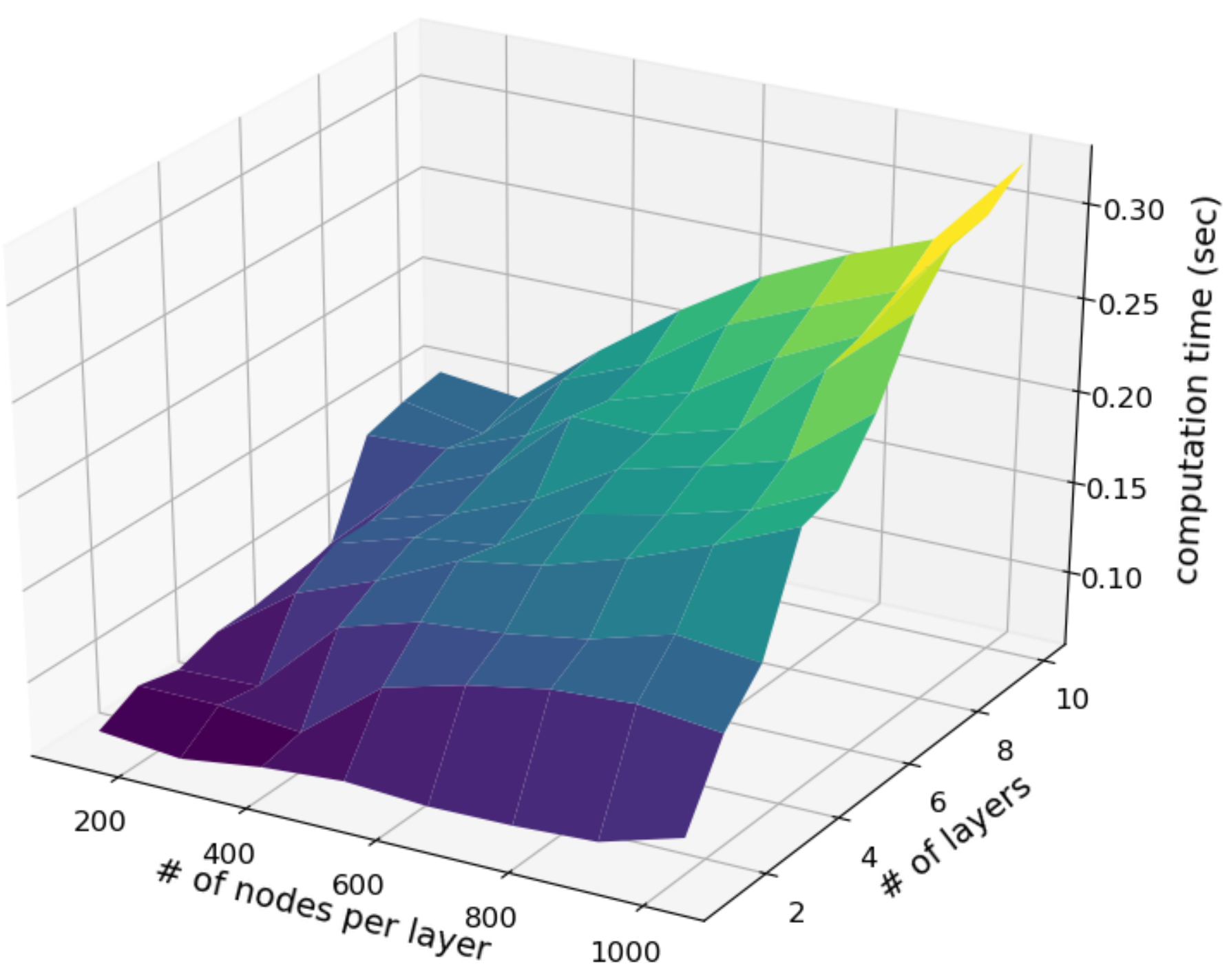}}  &
{\includegraphics[height=4.5cm,width=4.8cm]{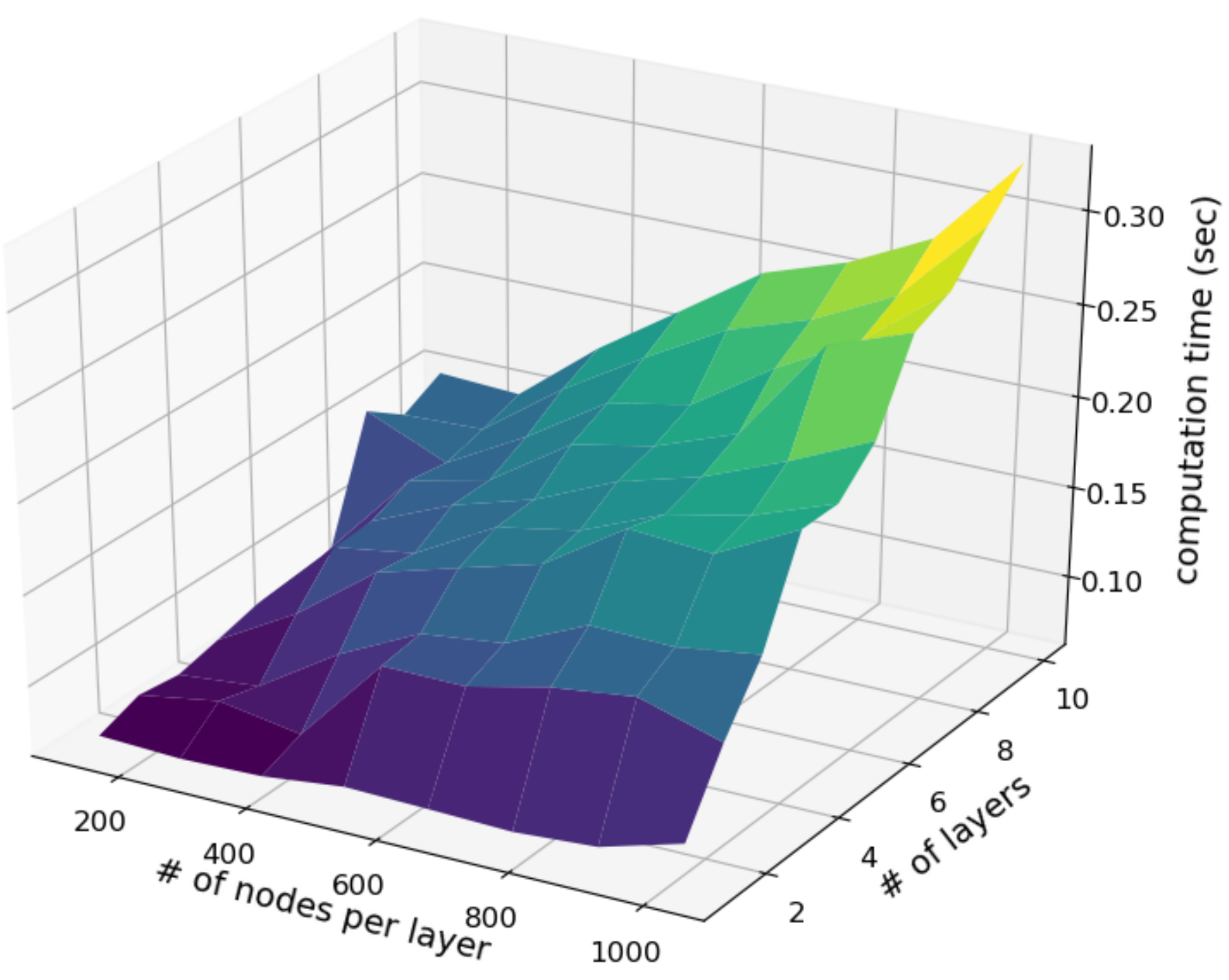}} \\
(a)   $IC_{ia}^{Adj}$ & (b)  $IC_{oa}^{Adj}$ \\
{\includegraphics[height=4.5cm,width=4.8cm]{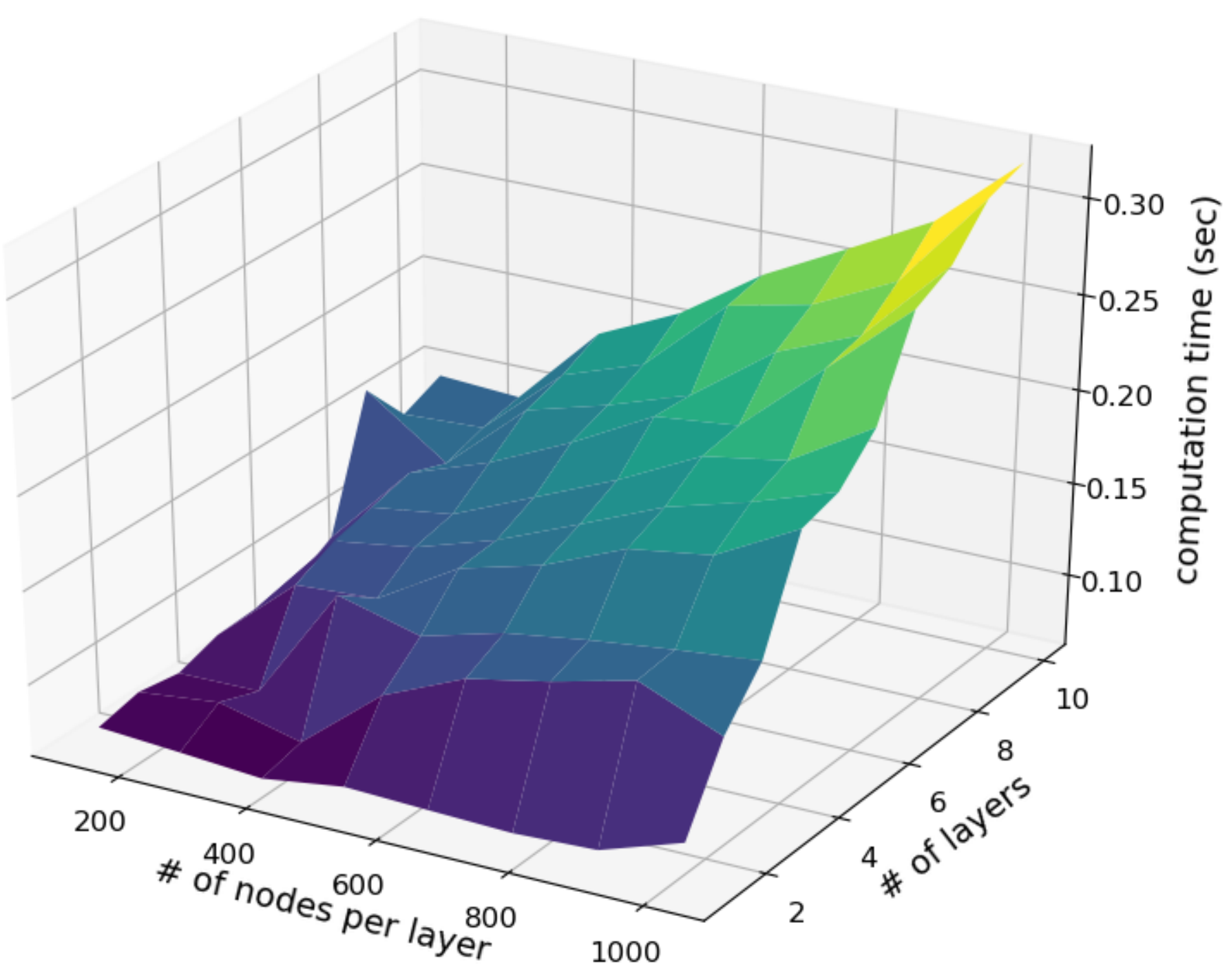}}  &
{\includegraphics[height=4.5cm,width=4.8cm]%[width=0.27\textwidth]
{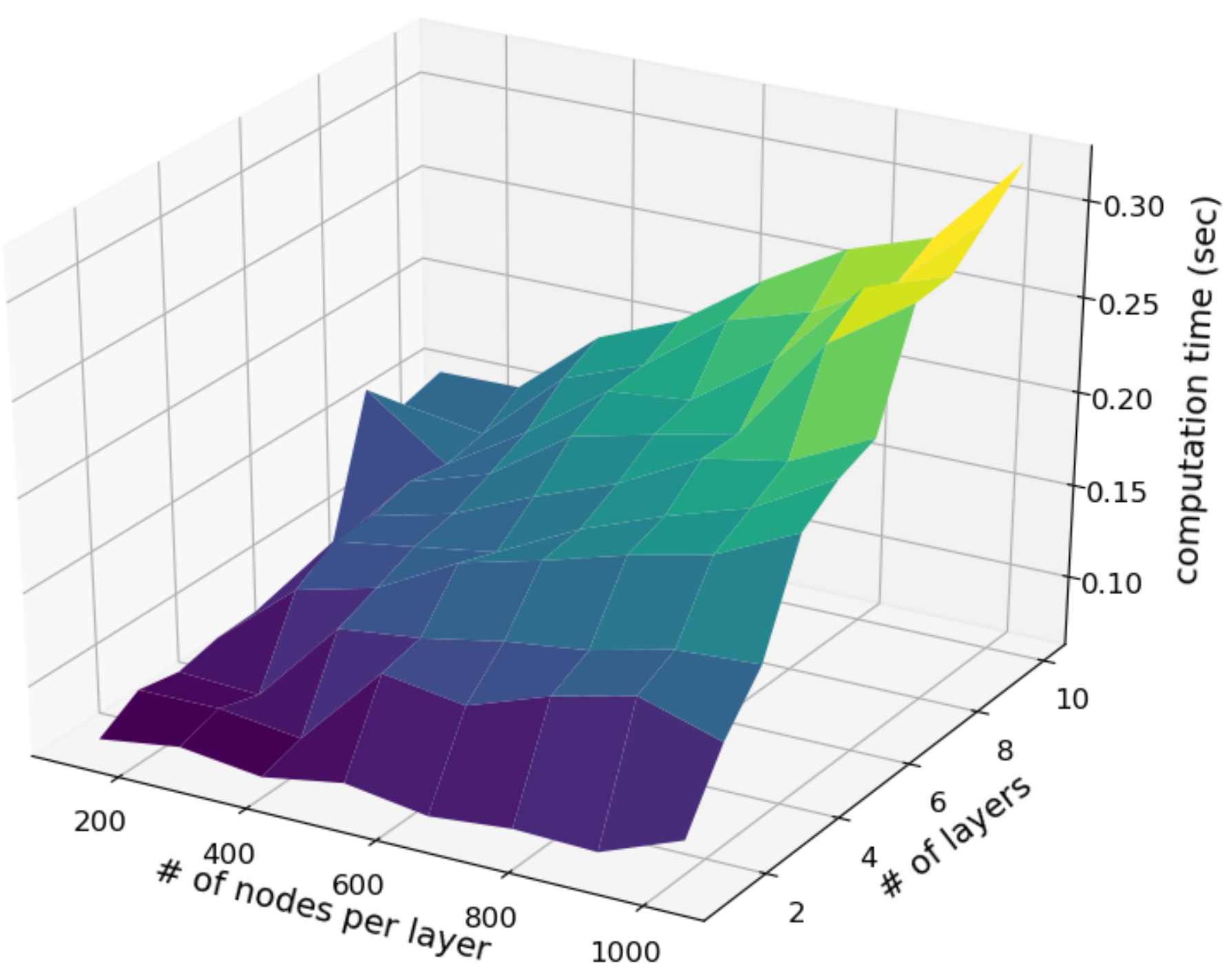}} \\
  (c)   $IC_{ia}^{Suc}$ & (d)    $IC_{oa}^{Suc}$
\end{tabular}
\caption{Computation time (in seconds) of the multilayer modularity $Q$, with redundancy-based $\gamma(L,C)$, measured on the solution found by GL on the   multiplex LFR network}
\label{times4} 
\end{figure*}

 \vspace{3mm}
\section{Efficiency results} 
Figures~\ref{times3}--\ref{times4} show the computation time of $Q$ for the different combinations of  $\gamma$ and  $IC$ factors.


\begin{thebibliography}{10}
\providecommand{\url}[1]{#1}
\csname url@samestyle\endcsname
\providecommand{\newblock}{\relax}
\providecommand{\bibinfo}[2]{#2}
\providecommand{\BIBentrySTDinterwordspacing}{\spaceskip=0pt\relax}
\providecommand{\BIBentryALTinterwordstretchfactor}{4}
\providecommand{\BIBentryALTinterwordspacing}{\spaceskip=\fontdimen2\font plus
\BIBentryALTinterwordstretchfactor\fontdimen3\font minus
  \fontdimen4\font\relax}
\providecommand{\BIBforeignlanguage}[2]{{%
\expandafter\ifx\csname l@#1\endcsname\relax
\typeout{** WARNING: IEEEtran.bst: No hyphenation pattern has been}%
\typeout{** loaded for the language `#1'. Using the pattern for}%
\typeout{** the default language instead.}%
\else
\language=\csname l@#1\endcsname
\fi
#2}}
\providecommand{\BIBdecl}{\relax}
\BIBdecl

\bibitem{AmelioT17}
A.~Amelio and A.~Tagarelli, ``{Revisiting Resolution and Inter-Layer Coupling
  Factors in Modularity for Multilayer Networks},'' in \emph{Proc. IEEE/ACM
  ASONAM}, 2017, pp. 266--273.

\bibitem{Newman04}
M.~E.~J. Newman and M.~Girvan, ``Finding and evaluating community structure in
  networks,'' \emph{Phys. Rev. E}, vol.~69, no.~2, p. 026113, 2004.

\bibitem{Newman04b}
M.~E.~J. Newman, ``Fast algorithm for detecting community structure in
  networks,'' \emph{Phys. Rev. E}, vol.~69, 2004.

\bibitem{BrandesDGGHNW08}
U.~Brandes, D.~Delling, M.~Gaertler, R.~G{\"{o}}rke, M.~Hoefer, Z.~Nikoloski,
  and D.~Wagner, ``On modularity clustering,'' \emph{{IEEE} Trans. Knowl. Data
  Eng.}, vol.~20, no.~2, pp. 172--188, 2008.

\bibitem{ChenKS14}
M.~Chen, K.~Kuzmin, and B.~K. Szymanski, ``Community detection via maximization
  of modularity and its variants,'' \emph{{IEEE} Trans. Comput. Social Syst.},
  vol.~1, no.~1, pp. 46--65, 2014.

\bibitem{Clauset04}
A.~Clauset, M.~E.~J. Newman, and C.~Moore, ``Finding community structure in
  very large networks,'' \emph{Phys. Rev. E}, vol.~70, 2004.

\bibitem{Newman05}
M.~E.~J. Newman, ``Modularity and community structure in networks,''
  \emph{Proc. Natl. Acad. Sci.}, pp. 8577--–8582,
  2005.

\bibitem{WhiteS05}
S.~White and P.~Smyth, ``A spectral clustering approach to finding communities
  in graph,'' in \emph{Proc. SDM}, 2005.

\bibitem{Reichardt06}
J.~Reichardt and S.~Bornholdt, ``Statistical mechanics of community
  detection,'' \emph{Phys. Rev. E}, vol.~74, 2006.

\bibitem{Duch05}
J.~Duch and A.~Arenas, ``Community detection in complex networks using extremal
  optimization,'' \emph{Phys. Rev. E}, vol.~72, 2005.

\bibitem{BOCCALETTI20141}
S.~Boccaletti, G.~Bianconi, R.~Criado, C.~I. del Genio, J.~G.-G. {n}es,
  M.~Romance, I.~S. {n}a Nadal, Z.~Wang, and M.~Zanin, ``The structure and
  dynamics of multilayer networks,'' \emph{Phys. Rep.}, vol. 544, no.~1,
  pp. 1--122, 2014.

\bibitem{Magnanibook}
M.~E. Dickison, M.~Magnani, and L.~Rossi, \emph{Multilayer social
  networks}.\hskip 1em plus 0.5em minus 0.4em\relax Cambridge University Press,
  2016.

\bibitem{Kivela+14}
M.~Kivela, A.~Arenas, M.~Barthelemy, J.~P. Gleeson, Y.~Moreno, and M.~A.
  Porter, ``Multilayer networks,'' \emph{J. Complex Netw.}, vol.~2,
  no.~3, pp. 203--271, 2014.

\bibitem{e20120909}
L.~G. A.~Alves, G.~Mangioni, F.~A. Rodrigues, P.~Panzarasa, and Y.~Moreno,
  ``Unfolding the complexity of the global value chain: Strength and entropy in
  the single-layer, multiplex, and multi-layer international trade networks,''
  \emph{Entropy}, vol.~20, no.~12, 2018.

\bibitem{Mucha10}
P.~J. Mucha, T.~Richardson, K.~Macon, M.~A. Porter, and J.-P. Onnela,
  ``Community structure in time-dependent, multiscale, and multiplex
  networks,'' \emph{Science}, vol. 328, no. 5980, pp. 876--878, 2010.

\bibitem{coscia2011}
M.~Coscia, F.~Giannotti, and D.~Pedreschi, ``A classification for community
  discovery methods in complex networks,'' \emph{Stat. Anal. Data
  Min.}, vol.~4, no.~5, pp. 512--546, 2011.

\bibitem{KunchevaM15}
Z.~Kuncheva and G.~Montana, ``Community detection in multiplex networks using
  locally adaptive random walks,'' in \emph{Proc. IEEE/ACM ASONAM}, 2015, pp.
  1308--1315.

\bibitem{ZhangWLY16}
H.~Zhang, C.~Wang, J.~Lai, and P.~S. Yu, ``{Modularity in Complex Multilayer
  Networks with Multiple Aspects: {A} Static Perspective},'' \emph{CoRR}, vol.
  abs/1605.06190, 2016.

\bibitem{Tagar17}
A.~Tagarelli, A.~Amelio, and F.~Gullo, ``{Ensemble-based community detection in
  multilayer networks},'' \emph{Data Min. Knowl. Discov.}, vol.~31, no.~5, pp.
  1506--1543, 2017.

\bibitem{Tagar18}
D.~Mandaglio, A.~Amelio, and A.~Tagarelli, ``{Consensus Community Detection in
  Multilayer Networks Using Parameter-Free Graph Pruning},'' in \emph{Proc. PAKDD}, 2018, pp. 193--205.

\bibitem{torreggiani2018identifying}
S.~Torreggiani, G.~Mangioni, M.~J. Puma, and G.~Fagiolo, ``Identifying the
  community structure of the food-trade international multi-network,''
  \emph{Environ. Res. Lett.}, vol.~13, no.~5, p. 054026, 2018.

\bibitem{mangioni2018multilayer}
G.~Mangioni, G.~Jurman, and M.~{De Domenico}, ``Multilayer flows in molecular
  networks identify biological modules in the human proteome,'' \emph{{IEEE}
  Trans. Netw. Sci. Eng.}, 2018.

\bibitem{alves2019nested}
L.~G. Alves, G.~Mangioni, I.~Cingolani, F.~A. Rodrigues, P.~Panzarasa, and
  Y.~Moreno, ``The nested structural organization of the worldwide trade
  multi-layer network,'' \emph{Sci. Rep.}, vol.~9, no.~1, p. 2866, 2019.

\bibitem{FortunatoB07}
S.~Fortunato and M.~Barthelemy, ``Resolution limit in community detection,''
  \emph{Proc. Natl. Acad. Sci.}, vol. 104, no.~36, 2007.

\bibitem{FORTUNATO201075}
S.~Fortunato, ``Community detection in graphs,'' \emph{Phys. Rep.}, vol.
  486, no.~3, pp. 75--174, 2010.

\bibitem{2010Tang}
L.~Tang and H.~Liu, \emph{Community Detection and Mining in Social Media}, ser.
  Synthesis Lectures on Data Mining and Knowledge Discovery.\hskip 1em plus
  0.5em minus 0.4em\relax Morgan {\&} Claypool Publishers, 2010.

\bibitem{FORTUNATO20161}
S.~Fortunato and D.~Hric, ``Community detection in networks: A user guide,''
  \emph{Phys. Rep.}, vol. 659, pp. 1--44, 2016.

\bibitem{leicht2008}
E.~A. Leicht and M.~E.~J. Newman, ``Community structure in directed networks,''
  \emph{Phys. Rev. Lett.}, vol. 100, p. 118703, Mar 2008.

\bibitem{arenas2007}
A.~Arenas, J.~Duch, A.~Fern\'{a}ndez, and S.~G\'{o}mez, ``Size reduction of
  complex networks preserving modularity,'' \emph{New J. Phys.},
  vol.~9, no.~6, p. 176, 2007.

\bibitem{newman2004}
M.~E.~J. Newman, ``Analysis of weighted networks,'' \emph{Phys. Rev. E},
  vol.~70, p. 056131, Nov 2004.

\bibitem{XIANG20124995}
J.~Xiang and K.~Hu, ``Limitation of multi-resolution methods in community
  detection,'' \emph{Physica A Stat. Mech. Appl.},
  vol. 391, no.~20, pp. 4995--5003, 2012.

\bibitem{Zhang2009}
J.~Zhang, K.~Zhang, X.~ke~Xu, C.~K. Tse, and M.~Small, ``Seeding the kernels in
  graphs: toward multi-resolution community analysis,'' \emph{New J. Phys.}, vol.~11, no.~11, p. 113003, 2009.

\bibitem{arenas2007-2}
A.~Arenas, A.~Fern\'{a}ndez, and S.~G\'{o}mez, ``Analysis of the structure of
  complex networks at different resolution levels,'' \emph{New J. Phys.}, vol.~10, no.~5, p. 053039, 2008.

\bibitem{XIANG2015127}
J.~Xiang, Y.-N. Tang, Y.-Y. Gao, Y.~Zhang, K.~Deng, X.-K. Xu, and K.~Hu,
  ``Multi-resolution community detection based on generalized self-loop
  rescaling strategy,'' \emph{Physica A Stat. Mech. Appl.}, vol. 432, pp. 127--139, 2015.

\bibitem{gomez2009}
S.~G\'{o}mez, P.~Jensen, and A.~Arenas, ``Analysis of community structure in
  networks of correlated data,'' \emph{Phys. Rev. E}, vol.~80, p. 016114, 2009.

\bibitem{traag2009}
V.~A. Traag and J.~Bruggeman, ``Community detection in networks with positive
  and negative links,'' \emph{Phys. Rev. E}, vol.~80, p. 036115, Sep 2009.

\bibitem{barber2007}
M.~J. Barber, ``Modularity and community detection in bipartite networks,''
  \emph{Phys. Rev. E}, vol.~76, p. 066102, 2007.

\bibitem{barber2008}
Y.~Xu, L.~Chen, and S.~Zou, ``Community detection from bipartite networks,'' in
  \emph{Proc. IEEE WISA}, 2013, pp. 249--254.

\bibitem{guimera2007}
R.~Guimer\`a, M.~Sales-Pardo, and L.~A.~N. Amaral, ``Module identification in
  bipartite and directed networks,'' \emph{Phys. Rev. E}, vol.~76, p. 036102,
  2007.

\bibitem{nicosia2009}
V.~Nicosia, G.~Mangioni, V.~Carchiolo, and M.~Malgeri, ``Extending the
  definition of modularity to directed graphs with overlapping communities,''
  \emph{J. Stat. Mech.}, vol. 2009, no.~03, p. P03024, 2009.

\bibitem{Berlingerio2011}
M.~Berlingerio, M.~Coscia, and F.~Giannotti, ``Finding and characterizing
  communities in multidimensional networks,'' in \emph{Proc. IEEE/ACM ASONAM},
  2011, pp. 490--494.

\bibitem{KimL15}
J.~Kim and J.~Lee, ``Community detection in multi-layer graphs: {A} survey,''
  \emph{{SIGMOD} Record}, vol.~44, no.~3, pp. 37--48, 2015.

\bibitem{Rossi2015}
L.~Rossi and M.~Magnani, ``Towards effective visual analytics on multiplex and
  multilayer networks,'' \emph{Chaos, Solitons \& Fractals}, vol.~72, pp.
  68--76, 2015.

\bibitem{Domenico2015}
M.~{De Domenico}, V.~Nicosia, A.~Arenas, and V.~Latora, ``Structural
  reducibility of multilayer networks,'' \emph{Nat. Commun.}, vol.~6,
  p. 6864, 2015.

\bibitem{cha2009www}
M.~Cha, A.~Mislove, and K.~P. Gummadi, ``{A Measurement-driven Analysis of
  Information Propagation in the Flickr Social Network},'' in \emph{Proc. WWW},
  2009, pp. 721--730.

\bibitem{kdweb2015}
G.~Silvestri, J.~Yang, A.~Bozzon, and A.~Tagarelli, ``Linking accounts across
  social networks: the case of stackoverflow, github and twitter,'' in
  \emph{Proc. Work. on Knowl. Discov. on the Web}, 2015, pp. 41--52.

\bibitem{Omodei2015}
E.~Omodei, M.~{De Domenico}, and A.~Arenas, ``Characterizing interactions in
  online social networks during exceptional events,'' \emph{Frontiers in
  Physics}, vol.~3, p.~59, 2015.

\bibitem{TangWL09}
L.~Tang, X.~Wang, and H.~Liu, ``Uncovering groups via heterogeneous interaction
  analysis,'' in \emph{Proc. IEEE ICDM}, 2009, pp. 503--512.

\bibitem{Stanley2016}
N.~{Stanley}, S.~{Shai}, D.~{Taylor}, and P.~J. {Mucha}, ``Clustering network
  layers with the strata multilayer stochastic block model,'' \emph{{IEEE}
  Trans. Netw. Sci. Eng.}, vol.~3, no.~2, pp. 95--105, April 2016.

\bibitem{Ali2019}
H.~T. {Ali}, S.~{Liu}, Y.~{Yilmaz}, R.~{Couillet}, I.~{Rajapakse}, and A.~O.
  {Hero}, ``Latent heterogeneous multilayer community detection,'' in
  \emph{Proc. ICASSP}, 2019.

\end{thebibliography}
\end{document}